\documentclass[%
 reprint,twocolumn,
superscriptaddress,
nofootinbib,
 amsmath,amssymb,
 aps,
prb,
]{revtex4-2}
\usepackage{graphicx}% Include figure files
\usepackage{dcolumn}% Align table columns on decimal point
\usepackage[normalem]{ulem}
\newcommand{\stkout}[1]{\ifmmode\text{\sout{\ensuremath{#1}}}\else\sout{#1}\fi}\usepackage{bm}% bold math
\usepackage[colorlinks=true,citecolor=blue,linkcolor=blue,urlcolor=blue]{hyperref}% add hypertext capabilities
\usepackage{physics}
\usepackage{xcolor}
\usepackage{overpic}
\usepackage[normalem]{ulem}

\begin{document}

%\preprint{APS/123-QED}

%\title{Theory of fractional Quantum Hall matter coupled to quantum light and emergent Girvin-MacDonald-Platzman polaritons}
\title{Theory of fractional quantum Hall liquids coupled to quantum light and emergent graviton-polaritons}
% Force line breaks with \\

\author{Zeno Bacciconi}%
\affiliation{SISSA --- International School for Advances Studies, via Bonomea 265, 34136 Trieste, Italy}
\affiliation{ICTP --- The Abdus Salam International Centre for Theoretical Physics, Strada Costiera 11, 34151 Trieste, Italy}
\author{Hernan B. Xavier}%
\affiliation{SISSA --- International School for Advances Studies, via Bonomea 265, 34136 Trieste, Italy}
\affiliation{ICTP --- The Abdus Salam International Centre for Theoretical Physics, Strada Costiera 11, 34151 Trieste, Italy}
\author{Iacopo Carusotto}
\affiliation{INO-CNR Pitaevskii BEC Center and Dipartimento di Fisica, Università di Trento, 38123 Povo, Italy}
\author{Titas Chanda}%
\affiliation{Department of Physics, Indian Institute of Technology Indore, Khandwa Road, Simrol, Indore 453552, India}
\affiliation{Department of Physics, Indian Institute of Technology Madras, Chennai 600036, India}
\affiliation{Center for Quantum Information, Communication and Computation (CQuICC),
Indian Institute of Technology Madras, Chennai 600036, India}
\author{Marcello Dalmonte}%
\affiliation{ICTP --- The Abdus Salam International Centre for Theoretical Physics, Strada Costiera 11, 34151 Trieste, Italy}

\date{\today}% It is always \today, today,
             %  but any date may be explicitly specified

\begin{abstract}
Recent breakthrough experiments have demonstrated how it is now possible to explore the dynamics of quantum Hall states interacting with quantum electromagnetic cavity fields. While the impact of strongly coupled non-local cavity modes on integer quantum Hall physics has been recently addressed, the effects on fractional quantum Hall (FQH) liquids -- and, more generally, fractionalized states of matter -- remain largely unexplored. In this work, we develop a theoretical framework for the understanding of FQH states coupled to quantum light. In particular, combining analytical arguments with tensor network simulations, we study the dynamics of a $\nu=1/3$ Laughlin state in a single-mode cavity with finite electric field gradients. We find that the topological signatures of the FQH state remain robust against the non-local cavity vacuum fluctuations, as indicated by the endurance of the quantized Hall resistivity. The entanglement spectra, however, carry direct fingerprints of light-matter entanglement and topology, revealing peculiar polaritonic replicas of the $U(1)$ counting. As a further response to cavity fluctuations, we also find a squeezed FQH geometry, encoded in long-wavelength correlations. By exploring the low-energy excited spectrum inside the FQH phase, we identify a new neutral quasiparticle, the graviton-polariton, arising from the hybridization between quadrupolar FQH collective excitations (known as gravitons) and light. Pushing the light-matter interaction to ultra-strong coupling regimes we find other two important effects, a cavity vacuum-induced Stark shift for charged quasi-particles and a potential instability towards a density modulated stripe phase, competing against the phase separation driven by the Stark shift. Finally, we discuss the experimental implications of our findings and possible extension of our results to more complex scenarios.

\end{abstract}
\maketitle

%\tableofcontents

\section{Introduction}
The possibility of controlling quantum matter properties via cavity embedding has sparked a lot of interest in recent years \cite{GarciaVidal_2021science_cavityrev, Mivehvar2021, Schlawin_2022apr_cavityrev,Bloch_2022nat_cavityrev}. Vacuum fluctuations of strongly confined electromagnetic modes have been proposed as handles on various phenomena \cite{Bartolociuti_prb2018_magnetotransport, Curtis_2019prl_superconductivity,Andolina_2024prb_amperean,Li_2020prl_quantumlight,Arwas_2023prb_transport,Pupillo_2017prl_transport,Chiocchetta_natcomm2021,Dmytruk_commphys2022,Ashida_2020prx_ferroelectricity,Lenk_2022prb_dmft,Tokatly_2021prb_gyrotropic,Rokaj_2022prb_hodstadter,Debernardis_prb2022_intersubband,andolina_2024dicke_heat,chiriacò2023thermal,Guerci_2020prl_superradiant}, and pioneering experiments have demonstrated the non-trivial role of cavity quantum electrodynamics (QED) set-ups in shaping matter properties \cite{AppuglieseFaist_science2022,JarcFausti_2023,ParaviciniBagliani_2018nature_magnetotransport,enkner_2023_vonklitzing,Orgiu_2015natmat_transport}. A particularly intriguing framework is that of topological phases, whose traditional many-body understanding faces fundamentally new questions due to the non-local nature of the cavity degree of freedom. 

On this point, a recent breakthrough experiment ~\cite{AppuglieseFaist_science2022} has shown that transport properties in the integer quantum Hall (IQH) regime can be affected by a split-ring cavity in the ultra-strong coupling regime, even in the absence of any driving. This effect has been proposed to arise from cavity mediated hoppings~\cite{Ciuti_prb2021} involving cyclotron transitions to higher Landau levels (LLs) assisted by vacuum photons or, more recently, as a consequence of cavity losses~\cite{Rokaj_prl2023_losses}.  While significant progress has been made in the understanding of IQH states coupled to quantum light \cite{Ciuti_prb2021,Arwas_2023prb_transport,Rokaj_2022prb_hodstadter,Rokaj_prl2023_losses}, its effect on the much richer physics of fractional quantum Hall (FQH) matter \cite{tsui_1982prl_FQH,Laughlin1983ansatz} remains largely unexplored.

%The FQH effect \cite{tsui_1982prl_FQH} is a fundamentally different state of matter which, differently from its integer counterpart, genuinely arise from many-body correlations. 

Fundamentally different from its integer counterpart, the FQH effect \cite{tsui_1982prl_FQH} can only arise from genuine many-body correlations, responsible to lift the degeneracy in the partially filled Landau level and open a bulk gap. 
In FQH phases, new collective degrees of freedom emerge, such as the paradigmatic fractionally-charged anyonic quasi-particles \cite{Nakamura_2020nat_anyons}, one of the smoking guns of topological order, and  neutral magnetoroton modes \cite{GMP1986,Pinczuk_prl1993collectiveobs}, also dubbed as gravitons \cite{Liang_nature2024_graviton} at long wavelengths given the link to a more recent geometric description of FQH correlations \cite{Haldane_prl2011_quantummetric,son2013_newtoncartan_graviton,GolkarSon_2016jhep_graviton,Liou_prl2019_graviton,Liu2018quench,Ippoliti_prb2018_geometryfluxattach,Nguyen2022mganetorotons,Balram_2022prx_fqhgravitons,pubalram_2024_nematic}. Another hallmark of topologically ordered phases is their peculiar many-body entanglement structure \cite{LiHaldane_prl2008,Regnault2017_lectnotes}, which naturally connects to their spectral properties and has proven to be a valuable tool for the classification of quantum phases of matter. In the face of this, it is tempting to ask what is the fate of such rich and profound phenomena under the action of quantum light?

In this work we address this question by putting forward a theory describing FQH liquids coupled to a cavity QED environment. The starting point of our analysis is a careful modeling of lowest Landau Level (LLL) electrons coupled to quantum light. In particular, we introduce a projected lowest Landau level (LLL) model written in the Dipole gauge and justified in terms of energy scale separation. This recovers the projected minimal coupling procedure of Ref. \cite{Dmytruck_prb2021} with the addition of a novel vacuum-induced Stark shift effect, proportional to the local value of the cavity field fluctuations. This highlights the role of cavity field gradients which, in agreement with Kohn's theorem \cite{Kohntheorem}, are essential to non-trivially couple to electronic correlations within the LLL\cite{footnote1,haldane2023_quadrupole}.

\begin{figure}
    \centering
    \begin{overpic}[width=.99\linewidth]{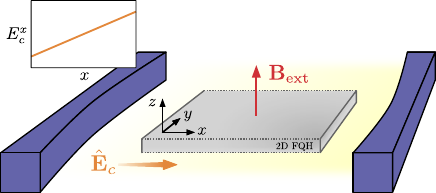}
    \put(90,40){(a)}
    \end{overpic}\\ \vspace{.5cm} 
    \begin{overpic}[width=0.4\linewidth]{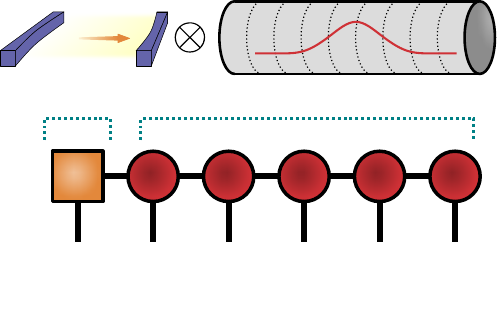}
    \put(-5,70){(b)}
    \end{overpic}
    \begin{overpic}[width=0.5\linewidth]{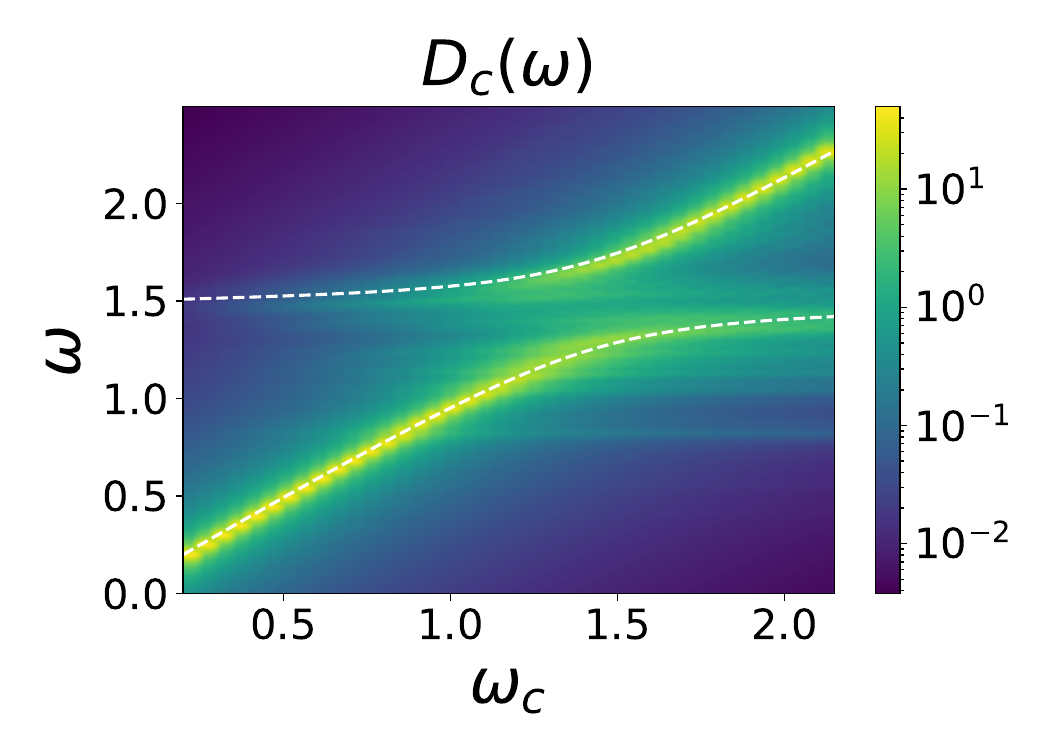}
    \put(5,70){(c)}       
    \end{overpic}
    \caption{(a) Illustration of the FQH bar placed inside the cavity. The electrons are confined to move in the 2D plane under the action of a perpendicular magnetic field $\bm{B}_\mathrm{ext}$. The in-plane component of the cavity electric field pierces the principal edges of the strip perpendicularly. Periodic boundary conditions on $y$ are depicted as dotted edges, while open boundary conditions on $x$ are depicted as a sharp cut. The inset shows the mode profile with a uniform gradient used in most of the discussion. (b) MPS ansatz used to describe the hybrid light-matter system. Orange square represent the cavity degree of freedom while the red circles represent LLL electronic orbitals on a cylinder geometry. (c) Cavity density of states $D_c(\omega)$ as a function of cavity frequency $\omega_c$ revealing graviton-polaritons. Dashed lines represent an analytical effective model for the polariton frequencies with no fitting parameters.}
    \label{fig:sketch}
\end{figure}

%In order to get salient features, we investigate the $\nu=1/3$ Laughlin state coupled to a cavity with a constant gradient and no losses, schematically depicted in Fig.~\ref{fig:sketch}.
As a proof of principle, we investigate a simplified scenario given by the $\nu=1/3$ Laughlin state coupled to a cavity with a smoothly varying gradient (constant or linearly varying), and no losses. The setting is schematically depicted in Fig.~\ref{fig:sketch}(a).
The full quantum dynamics of the system (light and matter) is studied numerically with a novel hybrid tensor network ansatz shown in Fig.~\ref{fig:sketch}(b) that combines the success in representing FQH states \cite{Zalatel_prl2013,zalatel_2012prb_mps} using matrix product states (MPS)~\cite{Schollwock2011, Orus2014, Paeckel2019}, with cavity-matter correlations in 1D \cite{Bacciconi_2023scipost,bacciconi2023topological,ChiriacòChanda_prb2022,Passetti_2023prl_dmrg}. This allows us to investigate in an unbiased way properties of both ground and low-energy excited states, using algorithms based on the density-matrix-renormalization group (DMRG)~\cite{White1992, White1993} and on the time-dependent variational principle (TDVP)~\cite{Haegeman2011, Haegeman2016}. Exact diagonalization (ED) is also used in order to test our findings. 

We show that the topological properties of the Laughlin state remain stable against the introduction of the non-local mode, as signaled by the quantized Hall resistivity, which we calculate by adapting the original flux insertion argument \cite{Thouless_1994_fluxinsertion}. However, entanglement properties are drastically affected: the entanglement spectrum features different series of eigenvalues featuring typical $U(1)$ counting of FQH states \cite{LiHaldane_prl2008}, with each series corresponding to an approximately quantized value of the cavity photon, and with an inter-series separation that we denote as \textit{polariton entanglement gap}. This reveals a collective coupling with matter excitations at zero transferred momenta which we also detect in spectral properties. These matter excitations are revealed to be gravitons, i.e., the long-wavelength part of the magnetoroton dispersion, and, once they hybridize with cavity photons, become graviton-polaritons. They give rise to a typical polariton doublet, which can be used in spectroscopy experiments as a smoking gun for the strong-coupling regime. Through a simple effective model that almost matches quantitatively with the finite-size numerical simulations, see Fig.~\ref{fig:sketch}(c), we also provide analytical predictions for the collective Rabi frequency in terms of the graviton quadrupole moment.

Another major consequence of the cavity mode on the ground state is the squeezing of the FQH metric~\cite{Haldane_prl2011_quantummetric}, that is revealed by a striking change in the long-wavelength correlations and by the spatially anisotropic profile of electron correlation holes. Only in regimes where the cavity field gradients are strong at the single-particle level cavity-mediated interactions can drive an instability of the FQH liquid to a stripe phase, reminiscent of other transitions mediated by different anisotropy sources \cite{Kumar2022prb_stripes}, accompanied by the softening of the full magnetoroton dispersion. However, before such transition occurs, spatially dependent cavity-induced vacuum Stark shifts become important and potentially generate phase separation. Their role in the FQH phase is to renormalize the energy cost of charged quasi-particles. 

\paragraph*{Summary of results.---} Before describing the structure of the work, we summarize here our main findings:
\begin{itemize}
    \item A microscopic QED theory describing the coupling of quantum light to electrons in the LLL, pointing out that the relevant light-matter coupling is expressed in terms of field gradients;
    %\item the formulation of a hybrid-MPS architecture to treat quantum Hall samples coupled to cavity modes up to relatively large volumes.
    \item Resilience of the quantized Hall resistivity upon the introduction of the non-local cavity-mode degree of freedom, with the latter imprinting an anisotropic FQH geometry in the ground state;
    \item A new entanglement structure in hybrid quantum Hall states, where the role of quantum light is to introduce a ``band" of chiral Luttinger liquids multiplets, each with an approximately quantized photon number;
    \item The prediction of graviton-polariton modes, describing the hybridization of the typical magnetoroton mode with quantum light in the setup we consider;
    \item Cavity vacuum-induced Stark shifts for charged quasi-particle excitations, controlled by the same collective energy scale which controls the splitting of graviton-polaritons.
\end{itemize}
The paper is organized as follows.

In Sec.~\ref{sec:model} we describe the system under study, starting from a brief recap on LLs and the microscopic derivation of the LLL light-matter model based on energy scale separation between LL, ensuring a proper treatment of the ultra-strong coupling regime. We discuss the role of cavity field gradients, and detail the numerical methods we use to solve the full cavity-matter problem.

In Sec.~\ref{sec:topological_order} we present a detailed study of the nature of the FQH topological order in the presence of a strongly coupled non-local cavity degree of freedom. We do so by looking at two key markers, viz., the entanglement spectrum structure and the transverse Hall resistivity. 

In Sec.~\ref{sec:spectral} we investigate bulk spectral properties of the hybrid light-matter FQH state. We start by reviewing the magnetoroton spectrum, i.e., the low-energy gapped neutral excitations on top of FQH states. Then we present numerical evidence for the effect of the cavity degree of freedom on the full magnetoroton dispersion, from the formation of the hybrid graviton-polariton to the lowering of the magnetoroton minimum. We then construct an effective model which builds on top of the Girvin-MacDonald-Platzman (GMP) treatment of magnetorotons, and is able to capture analytically the salient features of the graviton-polaritons that we observe numerically.

In Sec.~\ref{sec:stripe}, we extend our study of other ground state properties in the ultra-strong coupling regime. We first find that in the FQH phase strong cavity fluctuations can squeeze the emergent FQH geometry. Only at the single-particle strong coupling regime we find an instability towards a stripe phase, which however can be realized only if single-particle cavity-induced terms are compensated or neglected. By reintroducing them we show how vacuum-induced Stark shift are governed by a collective energy scale and, in the FQH phase, can renormalize the energy cost of charged quasi-particles.

In Sec.~\ref{sec:experimental_disc}, we discuss the connection of our findings to realistic experimental scenarios taking as a reference a split-ring resonator \cite{AppuglieseFaist_science2022}. While energy scales match, we highlight how carefully designed resonators with strong field gradients are needed to reach strong coupling to FQH physics. We also comment on the possible role of cavity screening of Coulomb interactions neglected in our treatment. 

To conclude in Sec.~\ref{sec:conclusions}, we summarize our results and draw a more general picture on how confined electromagnetic modes can be used to both probe and control correlations in the LLL. 

\section{The model}
\label{sec:model}

In this section we start by reviewing the LL physics (Sec. \ref{sec:LL}) and the split-ring cavity set-up (Sec. \ref{sec:cavity_setup}). In Sec. \ref{sec:howtoLL} we discuss the light-matter coupling in a full model without LL truncation. Them in Sec. \ref{sec:LLL_trunc} we present the first result of this work, namely the derivation of a QED Hamiltonian for LLL electrons coupled to a cavity. After discussing some of the physical implications of the obtained model (Sec. \ref{sec:gradients}), we spell out the full Hamiltonian used in the rest of the paper (Sec. \ref{sec:hamiltonian}) and describe the MPS ansatz used to solve it (Sec. \ref{sec:num_meth}).

\subsection{Landau levels}
\label{sec:LL}
We consider a collection of $N_e$ spinless electrons with mass $m_e$ and elementary charge $q=|e|$ living on the $(x,y)$ plane under a strong magnetic field $\boldsymbol{B}_\mathrm{ext}=(0,0,B)$ in the $z$-direction, Fig. \ref{fig:sketch}(a), represented in the so called Landau gauge by an external vector potential $\boldsymbol{A}_\mathrm{ext}= (0,xB,0)$. We choose to work with periodic boundary conditions along $y$, implying a cylinder geometry with a finite circumference $L_y$. The single-particle eigenstates can be written as \cite{tong2016lectures}:
\begin{equation}
    \psi_{k,n}(x,y)= \frac{1}{\sqrt{L_y}}e^{iky}\phi_{n}(x-X_k),\qquad k=\frac{2\pi}{L_y}j_k,
\end{equation}
with $n=0,1,2,...,$ being the LL index and $j_k \in\mathbb{Z}$ labeling the momentum along $y$. Here $\phi_{n}$ are eigenfunctions of the harmonic oscillator with characteristic length equal to the magnetic length $l_B=\sqrt{\frac{\hbar }{eB}}$, frequency equal to the cyclotron frequency $\omega_B=\frac{eB}{m_e}$, and centered around $X_k=k l_B^2$. In presence of an external potential $W$, which can account for both disorder and confining potential, the single-particle Hamiltonian in the LL basis takes the following general form:
\begin{equation}
    \hat{\mathcal{H}}_{0}=\sum_{n,k} \left(n+\frac{1}{2}\right)\omega_B \,\hat{n}_{k,n}
  + \sum_{k,k',n,n'} W_{k,k'}^{n,n'} \hat{c}^\dagger_{k,n}\hat{c}_{k',n'}
\end{equation}
where $\hat{c}_{k,n}$ is the annihilation operator for the orbital $(k,n)$, $\hat{n}_{k,n}=\hat{c}^\dagger_{k,n}\hat{c}_{k,n}$, and $W_{k,k'}^{n,n'}$ are the matrix elements of the external potential. Importantly, we focus on the LLL ($n=0$) where the single-particle Hamiltonian then reads:
\begin{equation}
    \hat{H}_{0}= \hat{\Pi} \hat{\mathcal{H}}_{0}\hat{\Pi} = \sum_{k,k'} W^{0,0}_{k,k'} \hat{c}^\dagger_k\hat{c}_{k'},
\end{equation}
where $\hat{\Pi}$ is a projector onto the LLL, and we dropped the LL index on the fermionic operator. In the following we are going to focus on $W=0$, except when explicitly stated. 

The other important ingredient are two-body interactions represented by a central potential $V(|\boldsymbol{r}_1-\boldsymbol{r}_2|)$. In the continuum this read:
\begin{align}
    \hat{\mathcal{H}}_{int}= &\int \dd^2\boldsymbol{r}_1 d^2\boldsymbol{r}_2 \hat{\psi}^\dagger(\boldsymbol{r}_1)\hat{\psi}^\dagger(\boldsymbol{r}_2)\cdot\nonumber\\ &\cdot V(|\boldsymbol{r}_1-\boldsymbol{r}_2|) \hat{\psi}(\boldsymbol{r}_2)\hat{\psi}(\boldsymbol{r}_1)
\end{align}
with $\hat{\psi}(\boldsymbol{r})=\sum_{k,n}\psi_{k,n}(\boldsymbol{r})\hat{c}_{k,n}$. Neglecting LL mixing and projecting onto the LLL, the interaction term can be written as:
\begin{equation}\label{eq:Hint}
    \hat{H}_\mathrm{int}= \sum_{k_1,k_2,k_3,k_4}V_{k_1,k_2,k_3,k_4}\hat{c}^\dagger_{k_1}\hat{c}^\dagger_{k_2} \hat{c}_{k_3}\hat{c}_{k_4},
\end{equation}
where $V_{k_1,k_2,k_3,k_4}$ are the matrix elements of $V(r)$ in the LLL. In order to keep the analysis simple, we adopt the first Haldane pseudopotential \cite{Rezayi_prb1994}, i.e., the shortest range fermionic interaction, for which the Laughlin wavefunction is an exact zero-energy ground-state. Its matrix element on the cylinder are \cite{Rezayi_prb1994}:
\begin{align}\label{eq:matrix-elements-Hint}
    V_{k_1,k_2,k_3,k_4}&= V_0\delta_{k_1+k_2}^{k_3+k_4} \frac{\sqrt{2\pi}}{L_y} \left[(k_1-k_3)^2-(k_2-k_3)^2 \right]\nonumber \\ &\times e^{-\frac{1}{2}(k_1-k_3)^2} e^{-\frac{1}{2}(k_2-k_3)^2},
\end{align} 
where the energy scale of the interaction is set to $V_0=1$, and the $1/L_y$ factor guarantees the interaction term to be extensive.
In absence of disorder, both total number of particles $\hat{N}=\sum_k \hat{n}_k$ and total momentum along the $y$ direction $\hat{K}_y=\sum_k k\,\hat{n}_k$ are conserved and can be fixed to $N_e$ and $K_y$. We will consider a finite cylinder in the $x$ direction by truncating in the orbital space, e.g., $j_k=-(M-1)/2,\dots,(M-1)/2$. This give rise to a cylinder of circumference $L_y$ and length in the OBC direction of $L_x=2\pi M/L_y$. To fix the correct filling, we fix $M$ with the condition $M=3N_e-2$. Such choice also guarantees a unique ground state at $K_y=0$, which is the $\nu=1/3$ Laughlin state. 

\subsection{Cavity set-up}
\label{sec:cavity_setup}

We consider a single-mode cavity model, inspired by the split-ring resonator used in Ref. \cite{AppuglieseFaist_science2022}. Other cavity resonances are expected to appear higher in energy and are thus going to be neglected. The split-ring mode can be understood in terms of an LC resonance \cite{DeBernardis_pra2018,BlazquezRabl_prl2023} at frequency $\omega_c=1/\sqrt{LC}$, where $L$ and $C$ are the effective inductance and capacitance, respectively. By shrinking the capacitor region, one can reach impressive enhancement of the vacuum electric field fluctuations, and in some cases, enter the strong-coupling regimes, even at the single-electron level \cite{Keller_nanolett2017_strongcoupling}. The free quantized Hamiltonian for the LC resonator can be written in terms of the electric and magnetic field energy density as:
\begin{equation}
    \hat{H}_c=\int \dd^3r\bigg[\frac{\epsilon_0}{2}\hat{\boldsymbol{E}}_c^2 +\frac{1}{2\mu_0}(\boldsymbol{\nabla} \times\hat{ \boldsymbol{A}_c})^2 \bigg]
    =\omega_c \hat{a}^\dagger \hat{a},
\end{equation}
where $\hat{\boldsymbol{E}}_c$ is the electric field and $\hat{\boldsymbol{A}}_c$ is the vector potential. 

In the region of interest, i.e., where the Hall bar is placed, these are expanded on a single quantized bosonic mode as:
\begin{align}\label{eq:vector_potential}
    \hat{\boldsymbol{A}}_c &=  A_c \boldsymbol{f}_c(\hat{a}+\hat{a}^\dagger), \\ 
     \hat{\boldsymbol{E}}_c &= i E_c \boldsymbol{f}_c (\hat{a}-\hat{a}^\dagger),  
\end{align}
with $\boldsymbol{f}_c$ the dimensionless mode function. Here, we assume in fact to have a negligible cavity magnetic field component $\hat{\boldsymbol{B}}_c=\boldsymbol{\nabla} \times \hat{\boldsymbol{A}}_c \simeq0$ implying $\boldsymbol{\nabla}\times\boldsymbol{f}_c \simeq 0$. The intensity of cavity vacuum fluctuations is:
\begin{align}\label{eq:Ec_veff}
E_c=\omega_c A_c = \sqrt{\hbar\omega_c/(2\epsilon_0 V_\mathrm{mode})}
\end{align}
 and is expressed as a function of the effective mode volume $V_\mathrm{mode}$. With this choice, we have $\int \dd^3r\, \boldsymbol{f}_c \cdot \boldsymbol{f}_c=V_\mathrm{mode}$.

We focus on the specific case of vanishing electric field component in the $y$ direction and generic field along $x$. The in-plane part of the mode function can be written as:
\begin{equation}\label{eq:mode_function_x}
    \boldsymbol{f}_c^\parallel(\boldsymbol{r}) = f_c(x) \boldsymbol{u}_x,
\end{equation}
with $f_c(x)$ being a generic function and $\boldsymbol{u}_x$ being the unit vector in the $x$ direction. In the following we will also use a shorthand notation for the cavity field $E_c(x)=E_c f_c(x)$.

In the ideal case of infinite parallel mirror plates \cite{DeBernardis_pra2018} living in the $yz$-plane, we have $f_c(x)=f_c$ independent of $x$. However, relevant field gradients are expected \cite{AppuglieseFaist_science2022}, and can be controlled to some extent. Note that the gradient of the cavity mode function in 3D is constrained by Gauss's law such that, in the case of an uniform dielectric within the capacitor plates, we have $\boldsymbol{\nabla}\cdot \boldsymbol{f}_c = 0$. This does not constrain the in-plane gradients which can indeed be finite. The split-ring set-up naturally implements an electric field perpendicular to the edges of the Hall bar \cite{AppuglieseFaist_science2022}, modeled here as the open edges of the cylinder in our configuration, Fig. \ref{fig:sketch}(a). We remark that this particular choice of cavity configuration differs from that considered in Ref. \cite{winterZilberberg_arxiv2024} where the electric field is exactly parallel to edge of the system.

\subsection{Quantum light-matter coupling}\label{sec:howtoLL}

In the continuum model one can implement minimal substitution, $\hat{\boldsymbol{p}}\to \hat{\boldsymbol{p}}-e \hat{\boldsymbol{A}}_c$, as an unitary transformation $\hat{\mathcal{U}}$ \cite{Dmytruck_prb2021}. In the case of purely electric field ($\nabla \times \hat{\boldsymbol{A}_c}\simeq 0$), the unitary operator $\hat{\mathcal{U}}$ depends on the cavity mode function via the cavity electric pseudopotential $\chi(\boldsymbol{r})$:
\begin{equation}\label{eq:chix_def}
    \chi(\boldsymbol{r})=e\int_{\boldsymbol{r}_0}^{\boldsymbol{r}}\dd\boldsymbol{r}' \cdot A_c \boldsymbol{f}_c(\boldsymbol{r}')\;.
\end{equation}
which satisfies $\nabla \chi(\boldsymbol{r})=e A_c\boldsymbol{f}_c(\boldsymbol{r})$. This allows us to define an unitary operator $\hat{\mathcal{U}}$ which implements the light-matter coupling:
\begin{equation}
    \hat{\mathcal{U}}= \exp \left[i(\hat{a}+\hat{a}^\dagger)\hat{\mathcal{P}} \right] \;,
\end{equation}
where $\hat{\mathcal{P}}$ is the polarization-like dimensionless electronic operator summarizing the coupling to the mode function
(from now on, in short, polarization):
\begin{equation}
    \hat{\mathcal{P}}=\int \dd^2\boldsymbol{r} \, \chi(\boldsymbol{r}) \hat{\psi}^\dagger (\boldsymbol{r}) \hat{\psi}(\boldsymbol{r}) \;.
\end{equation}
The so-called Coulomb gauge Hamiltonian can then obtained by applying the unitary $\hat{\mathcal{U}}$, which shifts the electronic momenta $\hat{\boldsymbol{p}}\to \hat{\boldsymbol{p}}-e \hat{\boldsymbol{A}}_c$, on the bare electron part:
\begin{align}
\hat{\mathcal{H}}^{coul}&=\hat{\mathcal{U}}^\dagger \left(\hat{\mathcal{H}}_0+\hat{\mathcal{H}}_{int}\right)\hat{\mathcal{U}}+\hat{H}_{c}=\nonumber\\ &=\hat{\mathcal{H}}_0+ \hat{\mathcal{H}}_{int} +\hat{H}_c + \hat{\mathcal{H}}_{para}+\hat{\mathcal{H}}_{dia} 
\end{align}
with the light-matter coupling now expressed in terms of a paramagnetic and diamagnetic terms:
\begin{align}\label{eq:colomb_paramagnetic}
    &\hat{\mathcal{H}}_{para}=\int \dd^2\boldsymbol{r}\hat{\psi}^\dagger (\boldsymbol{r}) \frac{e}{2m}\left( -i\nabla\cdot\hat{\boldsymbol{A}}_c-i\hat{\boldsymbol{A}}_c\cdot\nabla  \right)\hat{\psi}(\boldsymbol{r})\\
    &\hat{\mathcal{H}}_{dia}=\int \dd^2\boldsymbol{r}\hat{\psi}^\dagger (\boldsymbol{r}) \frac{e^2}{2m}\left( \hat{\boldsymbol{A}}_c\cdot\hat{\boldsymbol{A}}_c  \right)\hat{\psi}(\boldsymbol{r})
\end{align}
Here the light-matter coupling is expressed via currents, hence momentum of electrons.

The Dipole gauge Hamiltonian can then be obtained by applying the inverse unitary transformation $\hat{\mathcal{T}}=\hat{\mathcal{U}}^\dagger$ \cite{Dmytruck_prb2021}, also known as Power-Zienau-Wolley (PZW), to the full Coulomb gauge Hamiltonian  $\hat{\mathcal{H}}^{coul}$ (and not just to the bare electron part as done above):
\begin{align}
 &\hat{\mathcal{H}}^{dip}=\hat{\mathcal{U}}\hat{\mathcal{H}}^{coul}\hat{\mathcal{U}}^\dagger= \hat{\mathcal{H}}_0+\hat{\mathcal{H}}_{int} +\hat{\mathcal{U}}\hat{H}_c\hat{\mathcal{U}}^\dagger \\ \label{eq:dipoleH_full}
     &\qquad=\hat{\mathcal{H}}_0+\hat{\mathcal{H}}_{int}+ \hat{H}_c +\hat{\mathcal{H}}_{DP} +\hat{\mathcal{H}}_{PP} \;;
\end{align}
where the light-matter coupling is encoded in two new terms:
\begin{align}  \label{eq:DP_full}
\hat{\mathcal{H}}_{DP}&=i\omega_c(\hat{a}-\hat{a}^\dagger)\hat{\mathcal{P}} \\  \label{eq:PP_full}\hat{\mathcal{H}}_{PP}&=\omega_c\hat{\mathcal{P}}^2
\end{align}
with the latter often referred to as self-polarization term. Here the light-matter coupling is expressed through the polarization (position) rather than through currents (momentum) of electrons as in the Coulomb Hamiltonian. We also remark that, despite the nomenclature, no dipole approximation ($\hat{\boldsymbol{E}}_c$ uniform) has to be performed. Indeed we can verify that the polarization operator carry information about all multi-poles of the charge distribution. Expanding the cavity field up to first order around $\boldsymbol{r}=0$ and substituting it into $\chi(r)$, we get:
\begin{align}
    &\hat{\mathcal{P}}\simeq e \frac{E_c}{\omega_c} \boldsymbol{f}^\parallel_c(0)\cdot   \int \dd^2r\; \boldsymbol{r} \,\hat{\psi}^\dagger(\boldsymbol{r})\hat{\psi}(\boldsymbol{r}) +\\ \nonumber
    &+ e  \frac{E_c}{\omega_c} \sum_{a,b}\left[\partial_{a}  (\boldsymbol{f}^\parallel_c(0)\cdot\boldsymbol{u}_b )\right]  \int \dd^2r\; \frac{r_a r_b}{2}\,\hat{\psi}^\dagger(\boldsymbol{r})\hat{\psi}(\boldsymbol{r})\; ,
\end{align}
where $a$ and $b$ run over $\{x,y\}$. This clearly includes both dipole (first line) and quadrupole (second line) contributions. In particular for a cavity mode as that of Eq. \ref{eq:mode_function_x} only the $(a,b)=(x,x)$ contribution of the quadruople will contribute to the in-plane dynamics. Note that the the $z$ direction is absent from the discussion as the well thickness is assumed to be infinitesimal.

In order to specialize to the case of Landau Levels we just need to expand the polarization operator on the single particle basis introduced in Sec. \ref{sec:LL} as:
\begin{align}
    \hat{\mathcal{P}}= \sum_{k,k'}\sum_{n,n'}\chi_{k,k'}^{n,n'} \hat{c}^\dagger_{k,n}\hat{c}_{k',n'}
\end{align}
where the electric pseudopotential (or, equivalently, polarization) matrix elements are defined as:
\begin{equation}\label{eq:matrix_element}
    \chi^{n,n'}_{k,k'}= \int \dd ^2r\, \psi_{k,n}^*(\boldsymbol{r}) \,\chi(\boldsymbol{r})\, \psi_{k',n'}(\boldsymbol{r})\;.
\end{equation}
For our specific choice of cavity mode ($\boldsymbol{f}^\parallel_c(\boldsymbol{r})=f_c(x)\boldsymbol{u}_x$) and neglecting second order derivatives of the cavity field, we can write a more explicit expression:
\begin{align}\label{eq:matrix_element_Kcons}
    \chi_{k,k'}^{n,n'}&= \frac{l_B}{\sqrt{2}}\chi'(X_k)\left[\sqrt{n}\delta_{n,n'+1}+(n\leftrightarrow n')\right]\delta_{k,k'}+\nonumber\\
&+   \frac{l_B^2}{4}\chi''(X_k)\left[\sqrt{n(n-1)}\delta_{n,n'+2}+(n\leftrightarrow n')\right]  \delta_{k,k'}+\nonumber\\
&+\left[\chi(X_k) + \frac{l_B^2}{4} \chi''(X_k) (2n+1)\right]\delta_{n,n'}\delta_{k,k'}.
\end{align}  
Here $l_B$ is the magnetic length, $X_k$ is the center of the orbital with momentum $k$, and $\chi'$ and $\chi''$ represent the first and second order derivatives of the cavity electric pseudopotential, giving respectively dipole $\chi'\sim E_c$ and quadrupole $\chi'' \sim E_c'$ contributions. Higher order derivatives of the cavity field just give further cyclotron transition. There are in general two rather different kind of terms: {\it i)} those contained in the first two lines which drive inter-LL cyclotron transition; {\it ii)} those contained in the last one affect the intra-LL dynamics. In the rest of the discussion we are often going to shorten the notation of the LLL matrix elements of interest as $\chi^{0,0}_{k,k'}=\chi_k\delta_{k,k'}$.

We now want to remark that gauge transformations change the physical meaning of both cavity and matter operators. One should instead focus on physical observables, which are gauge invariant. For example, the cavity electric field is expressed differently in the two gauges:
\begin{align}\label{eq:Efield_CvsD}
    &\text{Coulomb}: \;\; \hat{\boldsymbol{E}}_c = iE_c(\hat{a}-\hat{a}^\dagger) \boldsymbol{f}_c,\\
    &\text{Dipole}: \;\; \hat{\boldsymbol{E}}_c = E_c \left[i(\hat{a}-\hat{a}^\dagger) - 2\hat{\mathcal{P}}\right] \boldsymbol{f}_c.
\end{align}
In this sense the no-go theorems \cite{AndolinaNoGO_prb_2019,AndolinaNoGO_epj2022}, which forbid a macroscopic coherent occupation of the cavity in the Coulomb gauge, constrain the groundstate coherent occupation in the Dipole gauge to be:
\begin{equation}\label{eq:acoh_Dipole}
    \langle \hat{a}\rangle_C=0\;\; \Longrightarrow\;\; \langle \hat{a}\rangle_D=i \langle\hat{\mathcal{P}}\rangle_D  ,
\end{equation}
where $\langle\dots\rangle_{C(D)}$ denotes the expectation value in the Coulomb (Dipole) gauge. We also note that there is an extra freedom in the choice of an overall constant in $\hat{\mathcal{P}}$, the origin of our system, which guarantees us that we can always find a basis where $\langle\hat{a}\rangle_D=0$ also in the Dipole gauge.

\subsection{LLL truncation}\label{sec:LLL_trunc}
Simple LLL models have been incredibly succesful in describing FQH systems. This relies on the fact that at high magnetic fields $B$ the cyclotron frequency $\omega_B\sim B$ dominates the interaction energy scale $V\sim e^2/\epsilon l_B\sim \sqrt{B}$ (Coulomb energy). Corrections in $V/\omega_B$ have also been extensively studied \cite{Sodemann_prb2013_LLmixing,Zalatel_prb2015_LLmixing,Rezayi_prl2017_LLmixing} and are often required for more quantitative comparison with experiments.

Performing a LLL truncation in the cavity QED set-up under consideration should however done with care. Truncations of light-matter interactions in ultra-strong coupling regimes have been shown to be problematic \cite{DeBernandis_pra2018_2,DiStefanoNori_2019natphys}. In this regard, many efforts have been devoted to devising controlled, effective low-energy models for truncated electronic systems that are strongly coupled to quantized electromagnetic modes \cite{Dmytruck_prb2021,LiEckstein_prb2020}.

Here we start from the single mode cavity QED model coupled to the full continuum of Landau Level electrons and perform a Lowest Landau Level truncation of the light-matter interaction terms based on controlled energy scale separation. As we will see, this procedure recovers results that can be obtained by a truncated minimal substitution prescription described in Ref. \cite{Dmytruck_prb2021}, up to a re-normalization of single particle terms. Numerical insights on the effect of the LLL truncation performed here can be found in App. \ref{app:LLL}.

We first require a cavity frequency $\omega_c\ll \omega_B$, which is usually verified in the experimental situations under consideration \cite{AppuglieseFaist_science2022}. Then also the light-matter interaction energy scale should be controlled. This however depends on the desciption we adopt, either Dipole or Coulomb, where the interaction is wrtitten in terms of different operators. In the Dipole gauge we have that cavity-mediated inter-LL transitions (see Eq. \eqref{eq:matrix_element_Kcons}) are controlled by the energy scale:
\begin{equation}
    \epsilon_{lm}^{dip}\sim e l_B E_c \;.
\end{equation}
Contrary, in the Coulomb gauge the light-matter coupling is expressed via the momentum of electrons (Eq. \eqref{eq:colomb_paramagnetic}) which carry a factor $1/m$ and hence a cyclotron frequency $\omega_B=eB/m$, leading to an energy scale for the light-matter interaction term of:
\begin{equation}
    \epsilon_{lm}^{coul}\sim e l_B E_c \frac{\omega_B}{\omega_c}\;.
\end{equation}
Even though the two energy scales are different we remind that the total Hamiltonians, before truncation, are equivalent up to global unitary transformations. It is clear however that performing a LLL truncation is more suitable in the Dipole gauge, where inter-LL matrix elements do not scale with $\omega_B$.
We also remark that at the theoretical level the LLL limit ($\omega_B=eB/m\to \infty$) should be taken with the mass of the electrons $m\to 0$ rather than with the magnetic field $B\to \infty$. This allows us to have a non-zero length scale for the system $l_B$ which also controls other intra-LL matrix elements (see Eq. \eqref{eq:matrix_element_Kcons}).

\paragraph*{Dipole gauge.---} Given that $\omega_B$ dominates on all other terms in the Dipole gauge, we can directly truncate $\hat{\mathcal{H}}^{dip}$:
\begin{align}\label{eq:Hdip_trunc}
    \hat{H}^{dip}&= \hat{\Pi}_{LLL} \hat{\mathcal{H}}^{dip} \hat{\Pi}_{LLL} =\nonumber\\&=
    \hat{H}_{int}+\hat{H}_c+\hat{H}_{DP}+\hat{H}_{PP}\;,
\end{align}
where in general we use $\hat{O}_{\alpha}=\hat{\Pi} \hat{\mathcal{O}}_\alpha \hat{\Pi}$ as a short notation for the LLL projected version of an operator. Equation \eqref{eq:Hdip_trunc} can be understood as the zeroth order expansion in $1/\omega_B$ of the full model in Eq. \eqref{eq:dipoleH_full}, with neglected second order corrections akin to what usually happens for standard FQH systems \cite{Rezayi_prl2017_LLmixing,Sodemann_prb2013_LLmixing}. Expanding the two projected light-matter interaction terms we get the following form:
\begin{align}
    &\hat{H}_{DP}=i\omega_c(\hat{a}+\hat{a}^\dagger) \hat{P}\\
    &\hat{H}_{PP}=\omega_c\hat{P}^2 + \sum_k \mu_k\hat{c}^\dagger_{k,0}\hat{c}_{k,0}
\end{align}
where $\hat{P}=\hat{\Pi} \hat{\mathcal{P}}\hat{\Pi}$ is the truncated polarization operator:
\begin{align}
    \hat{P}=\sum_k \chi_{k,k}^{0,0} \hat{c}^\dagger_{k,0}\hat{c}_{k,0}\;, 
\end{align}
and $\mu_k$ a cavity mediated single particle potential:
\begin{align}\label{eq:muk_full}
    \mu_k&= \omega_c \left(\chi^{0,1}_{k,k}\chi^{1,0}_{k,k} +\chi^{0,2}_{k,k}\chi^{2,0}_{k,k}\right)=\nonumber\\
    &= \frac{\left(e l_B E_c(X_k) \right)^2 }{2\omega_c}+  \frac{\left(e l_B^2 \partial_x E_c(X_k) \right)^2 }{8\omega_c}\;.
\end{align}
The structure of the light-matter interaction is then very similar to the one of the one of the full model (Eq. \eqref{eq:DP_full} and \eqref{eq:PP_full}) except for the single particle contribution $\mu_k$ in the self-polarization term which arises from the fact that:
\begin{equation}
\hat{\Pi}\,\hat{\mathcal{P}}^2\,\hat{\Pi}\neq\left(\hat{\Pi}\,\hat{\mathcal{P}}\,\hat{\Pi}\right)^2 \;.
\end{equation}
As a general comment we note that this term is important only for inhomogenues cavity electric fields, reducing to a global energy shift in the homogeneous case. In the above equation we can clearly distinguish dipole and quadrupole contributions, with higher order missing because of the assumption of negligible higher order derivatives of the electric field.

From a physical perspective it is possible to understand this term as a renormalization of the LLL vacuum energy ($\frac{1}{2}\omega_B$) via a spatially dependent \textit{depolarization shift} \cite{Todorov_prb2012} of the cyclotron frequency. Another equivalent interpretation is via a spatial dependent renormalization of the electronic mass, which enters in $\omega_B=eB/m$. Given the dependence of this single particle term on local vacuum fluctuations of the cavity electric field and the dipole/quadrupole moment of the cyclotron transition, we dub it vacuum-induced Stark shift, in analogy with other well known Stark shifts effects in quantum optics.  

\paragraph*{Coulomb gauge.---} As we discussed above, truncating in the Coulomb gauge is problematic because of the energy scale governing inter-LL transitions. However one can still define a \textit{Coulomb gauge} Hamiltonian using a truncated unitary transformation \cite{Dmytruck_prb2021}:
\begin{align}
    \hat{U}=\hat{\Pi}\;\hat{\mathcal{U}}\;\hat{\Pi}=\exp\left[ i(\hat{a}+\hat{a}^\dagger)\sum_{k,k'}\chi_{k,k'}^{0,0} \hat{c}^\dagger_{k,0} \hat{c}_{k',0} \right]
\end{align}
to get:
\begin{align}
    \hat{\Tilde{H}}^{coul}&= \hat{U} \hat{H}^{dip} \hat{U}^\dagger=\\
    &=\hat{U}\left(\hat{H}_0 + \hat{H}_{int}\right)\hat{U}^\dagger + \hat{H}_c + \sum_k \mu_k \hat{c}^\dagger_k\hat{c}_k\;
\end{align}
Note that we have used a tilde on the coulomb gauge Hamiltonian to stress that it differs from a bare truncation of the full one $\hat{\Tilde{H}}^{coul} \neq \hat{\Pi} \hat{\mathcal{H}}^{coul}\hat{\Pi}$. Interestingly the effect of the truncated unitary transformation $\hat{U}$ on LLL electron operators correspond to a simple Peierls phase dressing:
\begin{align}
    \hat{c}_k\to \hat{U} \hat{c}_k \hat{U}^\dagger = e^{i(\hat{a}+\hat{a}^\dagger)\chi_k}\hat{c}_k\; ,
\end{align}
also found for tight-binding models \cite{Dmytruck_prb2021}.

\subsection{Role of gradients}
\label{sec:gradients}

In a clean system, gradients are fundamental to couple the cavity field to electrons within the LLL. This is a consequence of the celebrated Kohn's theorem \cite{Kohntheorem}, whose corollary is that a uniform field can only couple to the cyclotron mode \cite{Rokaj_prl2023_losses}, generating transitions among different Landau levels. In view of that we now consider a further simplified cavity mode by considering a linear expansion $f_c(x)=C_0 +C_1 x$ and leave the discussion of more complicated modes for Sec. \ref{sec:experimental_disc}. For this particular shape of the mode function, the LLL matrix elements are:
\begin{align}\label{eq:chi-k-full}
    \chi_k= \frac{eE_c}{\omega_c}\left(\frac{l_B^2}{4} C_1+C_0(kl_B^2-x_0)+\frac{1}{2}C_1(kl_B^2-x_0)^2\right),
\end{align}
with $x_0$ the origin for the integration in Eq. (\ref{eq:chix_def}). In the Coulomb gauge $\hat{\Tilde{H}}^{coul}$, the matrix elements of the coupling can be readily understood by the dressing of $\hat{H}_\mathrm{int}$ and $\hat{H}_0$. For the interaction term $\hat{H}_\mathrm{int}$, we have that the four-body terms $\hat{c}^\dagger_{k_1}\hat{c}^\dagger_{k_2}\hat{c}_{k_3}\hat{c}_{k_4}$ get dressed with the following phase factor:
\begin{align}\label{eq:4body_phase}
    &\exp\left[i(\hat{a}+\hat{a}^\dagger)(\chi_{k_1}+\chi_{k_2}-\chi_{k_3}-\chi_{k_4} )\right].
\end{align}
However, from momentum conservation along $y$, we have that $k_1+k_2=k_3+k_4$, and for the explicit expression of $\chi_k$ the dressing phase can be written as:
\begin{align}\label{eq:Peierls_phase_gradient}
    \exp\left[i\frac{e E_c C_1 l_B^4}{\omega_c}(\hat{a}+\hat{a}^\dagger)(k_1-k_4)(k_4-k_2)\right].
\end{align}
The other important cavity-induced effect is the presence of a vacuum-induced Stark shift $\mu_k$ which, in the case of a simple gradient and up to overall constants, reads: 
\begin{align}
    \mu_k= \frac{1}{\omega_c}\left[e l_B E_c(C_0+ C_1 k l_B^2)\right]^2
\end{align}
At this point, we highlight two important properties:
\begin{itemize}
    \item First, as anticipated from the Kohn's theorem \cite{Kohntheorem}, the constant part of the electric field ($C_0$) is decoupled;
    \item Second, an uniform gradient of the electric field generates a uniform light-matter coupling plus a non-uniform single particle potential.
\end{itemize}
The latter follows by the fact that in Eq. \eqref{eq:Peierls_phase_gradient} only differences in momenta, hence relative distance on $x$, appear. Alternatively, when sticking to the dipole gauge formulation, one can see that, single-particle potentials aside, the constant part of the field ($C_0$) only couples to conserved quantities, such as the number of electrons and momentum along $y$. In contrast, the gradient ($C_1$) couples to the $xx$-component of the quadrupole moment operator, $\hat{Q}^{xx}\propto\sum_{kk'}\int\dd^2r\,x^2\psi_{k}^*(\bm{r})\psi_{k'}(\bm{r})\hat{c}^\dagger_{k} \hat{c}_{k'}$, associated to electrons in the LLL \cite{haldane2023_quadrupole}.

Another way to avoid Kohn's restriction and actually couple the LLL to a constant electric field is via the presence of an external potential for the electrons. Either a confining potential~\cite{NardinCarusotto_pra2023_nonlinearedge} or disorder will do the job, realizing however quite different scenarios. A confining potential on $x$ will have a dominant effect at the edges where its variation are stronger while disorder will contribute to bulk properties \cite{footnote2}. These effects can be important in actual experimental realization but go beyond the scope of this work.

\subsection{Hamiltonian}\label{sec:hamiltonian}
In order to better understand the many-body physics at play we will first study the effect of cavity mediated interactions ($\chi_k$) and only later reintroduce cavity induced single particle potentials ($\mu_k$). We will thus label the model Hamiltonians according to the terms present in it, $\hat{H}_{\chi}$ and $\hat{H}_{\chi,\mu}$ for generic cavity modes $f_c(x)$. 

In particular we first focus on the effect of uniform electric field gradients by choosing $f_c(x)=C_0+ C_1 x$ and setting both cavity vacuum-induced Stark shifts ($\mu_k$) and other external potentials ($W$), or equivalently their sum, to zero. We choose to work in the Dipole gauge where the toy model Hamiltonian reads:
\begin{align}\label{eq:Htot_chi}
\hat{H}_\chi &=\hat{H}_\mathrm{int} +\omega_c\hat{a}^\dagger\hat{a} + i\omega_c g(\hat{a}-\hat{a}^\dagger)\sum_k \Big(\frac{k^2}{2}-\kappa_0\Big) \hat{n}_k \nonumber \\
&+ \omega_c g^2 \bigg[\sum_k\Big(\frac{k^2}{2}-\kappa_0\Big) \hat{n}_k  \bigg]^2,
\end{align}
with $\hat{H}_\mathrm{int}$ the interaction Hamiltonian for the first Haldane pseudopotentials, Eq. \eqref{eq:Hint}, $\kappa_0$ the overall constant in the definition of $\chi_k$ related to the choice of origin, and
\begin{equation} \label{eq:g_expression}
    g= \frac{e C_1 E_c l_B^2}{\omega_c}
\end{equation}

a dimensionless coupling constant proportional to the electric field gradient $\partial_x E_c(x)=E_c C_1$. The energy scale of the interaction and the magnetic length are all set to unity, $V_0=l_B=1$, and $\omega_c=1$, unless specified otherwise. The latter choice is motivated by the fact that we need to focus on a limited parameter space, and is not related to any fine tuning.

For a more direct comparison of different cavity frequencies $\omega_c$, we will sometimes rescale the coupling $g$ and use $g\sqrt{\omega_c}$ as a tuning parameter for the light-matter coupling. This choice indeed guarantees that, following Eq.~\eqref{eq:Ec_veff} and  Eq.~\eqref{eq:g_expression}, the coupling ($g\sqrt{\omega_c}$) does not carry hidden dependencies on $\omega_c$, but only depends on the magnetic length $l_B$ and the cavity mode via its overall effective volume and its spatial variations. In Sec. \ref{sec:spectral_eff} it will also become clear that this choice guarantees a fixed physical Rabi coupling between gravitons and cavity photons (see Eq.~\eqref{eq:rabi_frequency}).

We are now going to reintroduce the cavity induced potential $\mu_k$, although for a slightly different profile of the cavity electric field which keeps inversion symmetry $x\to -x$. Inspired by the cavity field reported in Ref. \cite{AppuglieseFaist_science2022}, we choose a parabolic profile $f_c(x)= C_0 + \frac{1}{2}C_2 x^2$. In this case, to allow for a direct comparison with the uniform gradient case, we will require $|\partial_x E_c(x)|\in[0,2g]$ between $x\in[-\frac{L_x}{2},\frac{L_x}{2}]$ and $E_c(\pm L_x/2)=E_c(0)$, obtained by fixing $C_2=g /(L_x/2)$ and $C_0=\frac{1}{2}C_2(L_x/2)^2 $. Within this choice, the Hamiltonian can be written as:
\begin{align}\label{eq:Htot_chimu}
    \hat{H}_{\chi,\mu}= \hat{H}_\chi + \frac{\omega_c g^2 L_x^2}{16}\sum_k \hat{n}_k\left[ 1+ \left(\frac{k}{L_x/2}\right)^2\right]^2 +const
\end{align} 
where now $\chi_k$ is changed accordingly to the new mode function.

The single-particle term above is in fact exactly the same found in cold atom gases trapped by optimal means: there, the microscopic reason for such spatially dependent Stark shifts stems from a finite waist of the lattice laser beam. As in the latter physical setting, getting rid of this term seems challenging: however, owing to its simple functional form, it is relatively easier to understand their physical effects. We can then borrow a considerable understanding on how these terms affect the system dynamics from the abundant cold atom literature on the topic.

In particular, the above mentioned terms have long been investigate in the context of the local density approximation (LDA), which describes a given physical system as a collection of subsystems, each one at its own value of the effective local potential $\mu$. Such approximation works particularly well in the presence of incompressible phases: those occupy finite region of each sample, and are by definition stable to local potentials (even more so if their nature is topological). A paradigmatic example of the relevance and validity of such approximation is the observation of Mott insulator phases with cold atoms in optical lattices~\cite{greiner2002quantum}, characterized by typical wedding-cake like structures~\cite{preiss2015quantum}. 
A comprehensive review of such phenomenology for the specific case of fractional quantum Hall states is presented in Ref.~\cite{cooper2008rapidly}, and additional methods to exploit space dependent structures are described in Ref.~\cite{roncaglia2011rotating}. Based on the arguments reported above and extensively review in literature~\cite{cooper2008rapidly,dalibard2011colloquium}, below, we will focus most of our ground state analysis on $\hat{H}_{\chi}$, and present a further discussion on the effects of $\mu$ in Sec. \ref{sec:qp_ren} and App. \ref{app:LLL}.

{\it Observables. -} We also introduce some relevant observables that we use throughout the paper. The expressions we provide below are valid for the Dipole gauge, where cavity operators are dressed rather than matter ones. First, we have the real-space charge density:
\begin{equation}\label{eq:density_def}
    \hat{n}(x)=\frac{1}{L_y}\sum_k \hat{n}_k |\phi_0(x-X_k)|^2,
\end{equation}
which is independent on the position $y$ because of the translational invariance. In order to get information about correlations in the $y$ direction we use the density-density correlations from the two-particle correlator $G^{(2)}$ defined as:
\begin{equation}\label{eq:g2_def}
    G^{(2)}(\boldsymbol{r}_1,\boldsymbol{r}_2)= \langle \hat{\psi}^\dagger (\boldsymbol{r}_1)\hat{\psi}^\dagger (\boldsymbol{r}_2)\hat{\psi}(\boldsymbol{r}_2)\hat{\psi}(\boldsymbol{r}_1)\rangle.
\end{equation}
Another important quantity is the guiding center density operator \cite{GMP1986,KumarHaldane_prb2022}, which in second quantization reads:
\begin{equation}\label{eq:guidingcenter_def}
    \hat{\Bar{\rho}}(\boldsymbol{q})= \frac{1}{M}e^{-iq_xq_y/2} \sum_k e^{-iq_x k} \hat{c}^\dagger_k\hat{c}_{k+q_y}.
\end{equation}
From this we define the connected guiding center static structure factor:
\begin{equation}\label{eq:staticSq_def}
    S(\boldsymbol{q}) =\bra{\mathrm{GS}}\delta\hat{\Bar{\rho}}(-\boldsymbol{q})\delta\hat{\Bar{\rho}}(\boldsymbol{q}) \ket{\mathrm{GS}}, 
\end{equation}
with $\delta \hat{\Bar{\rho}}=\hat{\Bar{\rho}}-\bra{\mathrm{GS}}\hat{\Bar{\rho}}\ket{\mathrm{GS}}$, and its dynamical counterpart:
\begin{align}\label{eq:dynamicalSq}
    S(\omega,\boldsymbol{q})&=\frac{1}{M}\sum_n |\bra{n}\delta \hat{\Bar{\rho}}(\boldsymbol{q})\ket{\mathrm{GS}}|^2\delta_\eta(\omega -E_n+E_{\mathrm{GS}}) \nonumber\\
    &-\{\omega\to -\omega\},
\end{align}
with $\ket{n}$ being a many-body eigenstate with energy $E_n$, and $\eta$ being a broadening parameter that should be sent to zero. 
Regarding the cavity, we use its density of states as a way to probe polaritons at finite frequency:
\begin{align}\label{eq:Acavity_def}
    D_{c}(\omega) =& \sum_n |\bra{n}\hat{a}^\dagger \ket{\mathrm{GS}}|^2\delta_\eta(\omega -E_n+E_\mathrm{GS}) \nonumber\\
    -&\sum_n |\bra{n}\hat{a} \ket{\mathrm{GS}}|^2\delta_\eta(\omega +E_n-E_\mathrm{GS}) ,
\end{align}
which can be obtained from the retarded cavity Green's function as $D_{c}(\omega)=-\frac{1}{\pi}\mathrm{Im}\,G^R_c(\omega)$, with $G^R_c(t)=i\theta(t)\langle [\hat{a}(t),\hat{a}^\dagger(0) ]\rangle$. We remark that in the ultra-strong coupling regime a precise calculation for the outcome of transmission/reflection experiments should also take into account anomalous correlations \cite{Deliberato_pra2016_uscspectralfunction,CiutiCarusotto_pra2006_inputoutpuspectral} and use the gauge invariant electric field, rather than the gauge dependent cavity operator $\hat{a}$ \cite{SettineriNori_prr2021_gaugespectral}. We leave these refinements for a future work.

\subsection{Numerical methods}\label{sec:num_meth}
In order to study the strongly-coupled light-matter system, we perform DMRG simulations for the ground state, and a combination of TDVP and ED for spectral functions. DMRG methods have been extensively used in the context of FQH systems to find the ground state of microscopic Hamiltonians in an unbiased way \cite{Zalatel_prl2013,Misguich_iop2021,He2021_cdwdmrg}. The cylinder geometry, in particular, allows for a very direct mapping of the LLL orbitals spanned by a single quantum number $k$ to a quasi-1D chain with long range interactions. Each electronic orbital $k$ is mapped onto a different MPS site, keeping track of both momentum and charge quantum numbers.  In the case of the first Haldane pseudopotential the range is finite and depends on the circumference of the cylinder $O(L_y)$. The price to pay in order to use a 1D MPS ansatz is twofold: (i) The MPO representation of the interaction Hamiltonian requires a large bond dimension; and (ii) the MPS bond dimension needs to grow exponentially with $L_y$. 

\begin{figure}
    \centering
    \includegraphics[width=\linewidth]{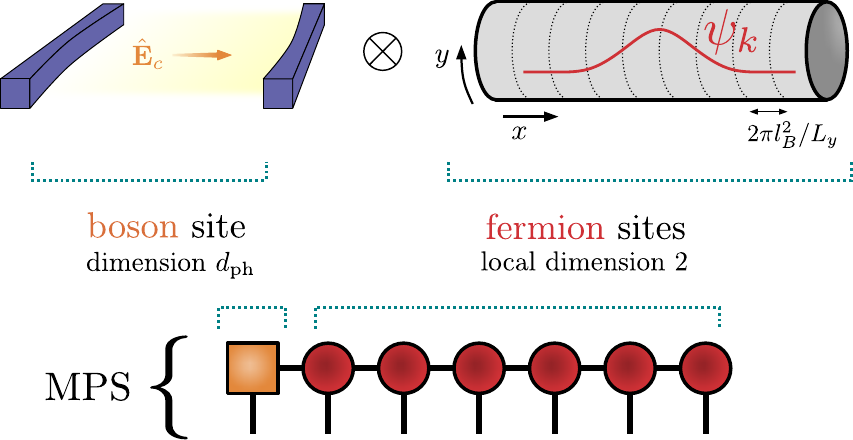}
    \caption{Cartoon for the hybrid MPS ansatz. The cavity photon is placed at the first site of the MPS, while the rest of the MPS sites represent the LLL wavefunctions $\psi_k$. Each electronic site (red circles) represent an orbital $k$ of the LLL with $k=2\pi j_k/L_y$ and $j_k=-(M-1)/2, \dots ,(M-1)/2$.}
    \label{fig:model}
\end{figure}

\subsubsection{Hybrid MPS ansatz}
Here we introduce a hybrid cavity-matter finite MPS ansatz where the photon is placed at the beginning of the MPS, Fig.~\ref{fig:model}, also used in Refs. \cite{ChiriacòChanda_prb2022,bacciconi2023topological,Bacciconi_2023scipost,Passetti_2023prl_dmrg}. Empirically we observed that this MPS ansatz is still efficient enough in representing the non-local correlations of the cavity mode, provided that the bond dimension $m$ is large enough. On a more speculative ground, we also argue that a single collective state, crucial in carrying light-matter correlations, can be represented as a finite bond dimension MPS. Thus adding a few of these collective states only results in a finite computational cost. The light-matter interactions are also easy to represent as an MPO due to their infinite range nature \cite{crosswhite_prb2008_longrangeMPO}. Here the choice of the Dipole gauge is actually important to avoid the dressing of the 4-body operators in the only matter part, which can become quite expensive, differently from the dressing of 2-body operators \cite{Bacciconi_2023scipost}. Moreover we make use of the freedom in the choice for the overall constant in the light-matter matrix elements $\kappa_0$, Eq. \eqref{eq:Htot_chi}, such that the constraint given in Eq. \eqref{eq:acoh_Dipole} gives zero coherent occupation of the photon $\langle \hat{a}\rangle_D=0$. This guarantees a good convergence with the truncation of the cavity Hilbert space at $d_\mathrm{ph}=64$. Note that the total number of electrons and total momentum along $y$ are still good quantum numbers that we conserve in our simulations. We limit the bond dimension of the MPS up to $m=1600$ which allows us to keep the truncation error always below $10^{-6}$.

Apart from ground state properties via DMRG, we also investigate excited states and dynamical properties using different methods. Regarding the excited states, it is possible to directly get good results from the local effective Hamiltonians constructed during DMRG runs, as discussed in Ref. \cite{Mila_prb2017_excited}. In particular, local targeting of the excited states has been found to be quite accurate for critical systems in 1D \cite{Mila_prb2017_excited}, whose success owes to the delocalized nature of the low-lying spectrum. As discussed in more depth in Sec. \ref{sec:spectral_num}, we find that this method gives good qualitative results and is even able to capture mixed light-matter polariton states (see Appendix \ref{app:excited_appendix} for more details on convergence). We further study dynamical properties via time evolution with TDVP (two-site updates), or directly using Lanczos methods in ED~\cite{ED_spectralfunctions}. For these TDVP runs we limit the bond dimension of the MPS to $m=200$ which still guarantees good convergence for small circumferences at a reduced computational cost.

Interestingly, the TDVP algorithm can also be used to implement change of gauges via the unitary $\hat{U}$. In particular we divide $\hat{U}$ in many steps and apply them sequentially as commonly done in TDVP evolutions. The ``Hamiltonian" of this gauge change is:
\begin{equation}
    \hat{h}_\mathrm{gauge}= (\hat{a}+\hat{a}^\dagger)\sum_k\chi_k\hat{n}_k,
\end{equation}
so that:
\begin{equation}
    \hat{U}=\exp(-i\hat{h}_\mathrm{gauge}),
\end{equation}
and, in this units, the evolution time is $\tau=1$. 
The fact that the MPS representation of the many-body ground state in a different gauge remains efficiently compressible (i.e., with a small enough bond dimension) is not guaranteed a priori and has to be checked. We find this to be the case in the FQH phase. During the change of gauge we keep the bond dimension of the MPS constant.

\section{Topological order in cavity}\label{sec:topological_order}

The presence of a genuine non-local degree of freedom makes the present setting not immediately classifiable within the framework of topological phases of matter. In the context of cavity mediated topology, a lot of attention has been dedicated to regimes where the cavity degree of freedom can be integrated out, both in the quantum materials context \cite{Dmytruk_commphys2022,Chiocchetta_natcomm2021,Ciuti_prb2021,dag_arxiv2023_topology_graphene} and in cold atom set-ups \cite{piazza_prl2017_topology_peierls,
Chanda_quantum2021selforganized, Chanda_PRB_2022,
Cuadra_prl2023_atomtopology}. Integrating out the non-local mode generally produces effective long-range interactions which simplify the picture of a mixed cavity-matter system and require the inspection of a matter-only model. This neglects by construction light-matter entanglement and gives a direct interpretation of cavity mediated topology in terms of ``standard" topology. 

Here, we are interested in the opposite situation, where light-matter entanglement cannot be neglected -- and, as we show below, does carry key signatures of topological order.

Few works \cite{bacciconi2023topological,Nguyen_prl2023,santos_arxiv2023_ssh,nguyen_arxiv2024_ssh} have recently investigated the questions above in the context of symmetry protected topological (SPT) phases. In particular, Ref.~\cite{bacciconi2023topological}  pointed out that, in the case of Majorana fermions, respecting the symmetry that protects the topology is fundamental. The FQH effect instead belongs to a fundamentally different set of topological states which display topological order -- that is, where order is intrinsically related to emergent gauge theories and entanglement. Addressing the interplay of topological order and quantum light is thus a completely distinct challenge with respect to the above mentioned SPTs.

As a paradigm of FQH, we are going to study the effect of the non-local cavity degree of freedom on the topological properties of a $\nu=1/3$ Laughlin state. In particular we focus on two markers, first on the Hall resistivity (Sec. \ref{sec:flux_ins}) and second on the entanglement properties (Sec. \ref{sec:ent_spectrum}). They represent two key aspects of topological order: quantization of transport properties, and fingerprints in the entanglement structure of the state. A final discussion on the resilience of topological properties is given at the end of the section. Throughout this section we are going to neglect single particle terms and use the toy model hamiltonian $\hat{H}_\chi$ ( Eq. \eqref{eq:Htot_chi}).

\subsection{Hall resistivity via flux insertion}\label{sec:flux_ins}

\begin{figure*}
    \centering
    \raisebox{.8cm}{\begin{overpic}[width=0.30\linewidth]{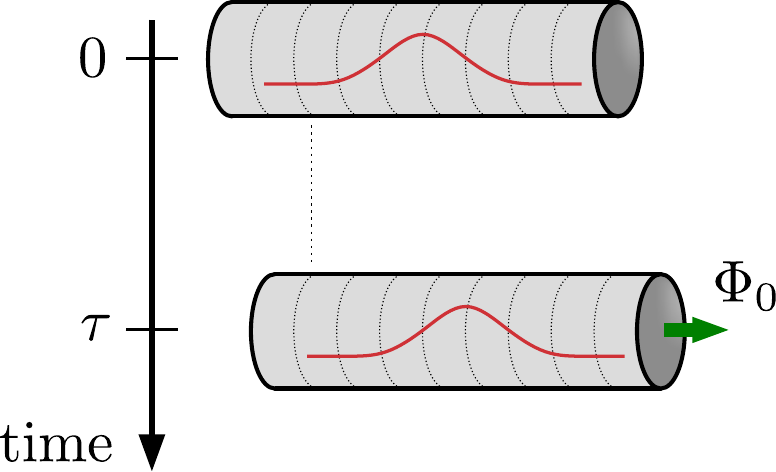}
    \put(0,60){(a)}
    \end{overpic}}
    \quad
    \begin{overpic}[width=0.30\linewidth]
    {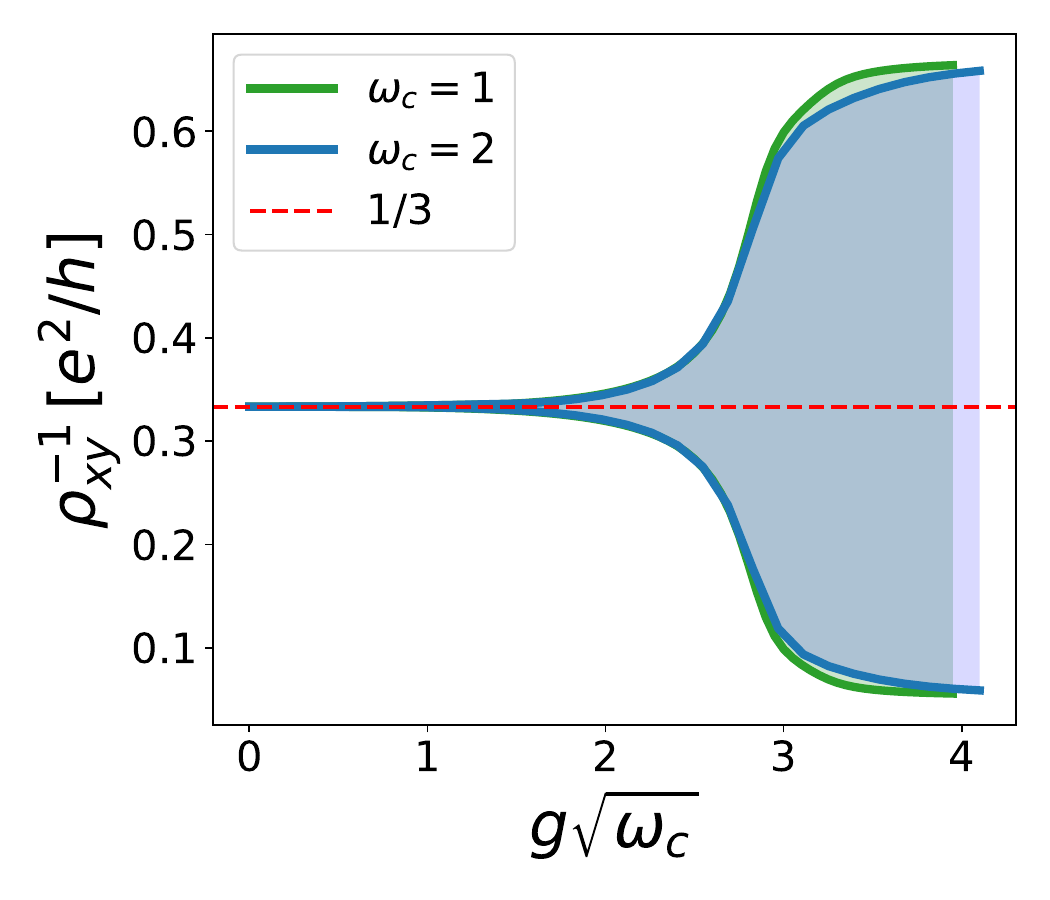}
    \put(25,20){(b)}
    \end{overpic}
    \quad
    \begin{overpic}[width=0.30\linewidth]{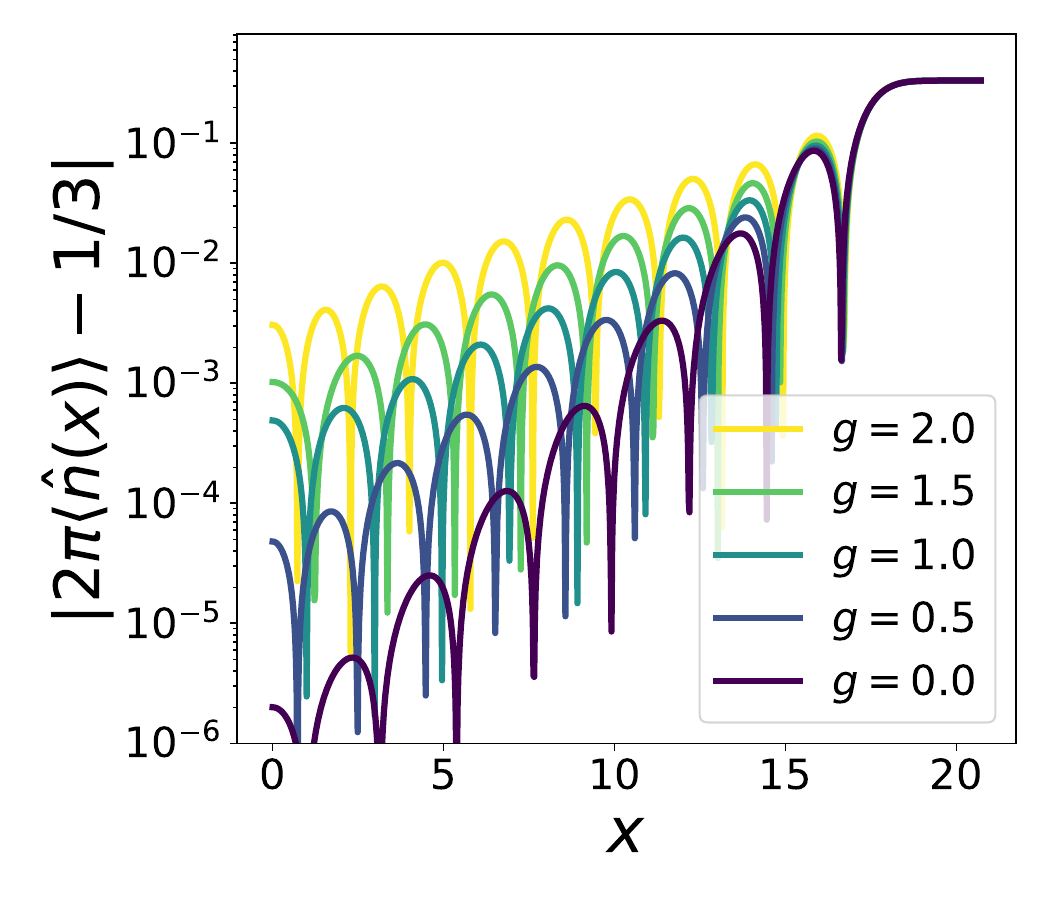}
    \put(25,75){(c)}
    \end{overpic}
    \caption{(a) Sketch of the adiabatic flux insertion; as a unit of time $\tau$ passes a unit of flux $\Phi_0$ is inserted resulting in relative shift in position of the cylinder along $x$. (b) Inverse resistivity in the bulk as measured by the density (Eq.~\eqref{eq:resistivity_density}) with a shaded area representing the values it takes for $x_0\in[-2,2]$ and quantized value $\rho_{xy}^{-1}=1/3$ as a dashed red line. At strong cavity field gradients $g$ the quantization is lost because of strong bulk density oscillations (see Sec. \ref{sec:stripe_num}). (c) Exponential decay from the edge of density modulations $\hat{n}(x)$ around the bulk quantized value. $x=0$ is the center of the cylinder while $x=20$ is already outside of it. Obtained via DMRG at system size $(L_y,N_e)=(14,25)$ in (b) and $(16,30)$ in (c) with $\omega_c=1$ for the Hamiltonian without single particle terms $\hat{H}_\chi$ (Eq. \eqref{eq:Htot_chi})}.
    \label{fig:resistivity}
\end{figure*}
A key feature of the FQH effect is the fractionally quantized transverse resistivity $\rho_{xy}$. An easy way to probe this quantity numerically on the cylinder geometry is via the so-called adiabatic flux insertion \cite{Thouless_1994_fluxinsertion,Zaletel2014_jsm2014_fluxins}, sketched in Fig. \ref{fig:resistivity}(a). In the following we derive independently an expression that links the the transverse bulk resistivity to the density at a certain position. 

Let us now consider the adiabatic insertion of a single magnetic flux quanta $\Phi_0=2\pi\hbar/e$ in the cylinder over a time $\tau$. This process can be described by a uniform time-dependent vector potential $\boldsymbol{A}_p(t)= \frac{\Phi_0}{L_y} \frac{t}{\tau} \boldsymbol{u}_y$, so that $\Phi(t)=\Phi_0 t/\tau$. The vector potential $\boldsymbol{A}_p(t)$, in turn, generates an electric field $\boldsymbol{E}_p=-\partial_t \boldsymbol{A}_p = -\frac{\Phi_0}{\tau L_y}\boldsymbol{u}_y$, which is constant in time, and directed towards the $y$ direction. Hence, we can find the Hall resistivity by calculating the current $\hat{J}_x=\hat{\boldsymbol{J}}\cdot \boldsymbol{u}_x$ flowing transversely to the electric field $\boldsymbol{E}_p$. Since our system is homogeneous in the $y$-direction, but not along the $x$-direction, the result is a spatially dependent resistivity:
\begin{equation}
    \rho_{xy}(x)= \frac{E_y}{e \langle \hat{J}_x(x)\rangle}.
\end{equation}
For the FQHE the bulk value is expected to be quantized as $\rho_{xy}=\frac{1}{\nu}\frac{h}{e^2}$, with $\nu=1/3$ in this work.

In the adiabatic limit $\tau\rightarrow \infty$, and assuming the existence of a many-body gap, we can focus on the ground state of the instantaneous total Hamiltonian $\hat{H}(\Phi)$ which in principle depends on $\Phi$. Now we know that a constant vector potential $\boldsymbol{A}_p$ does not couple to the LLL. Indeed, the effect of a flux $\Phi$ is to change the $y$-momentum quantization of the single particle orbitals $\psi_{n,k}$ (Sec. \ref{sec:LL}) and consequently their position:
\begin{equation} \label{eq:singleparticle_flux}
     k_{\Phi}= \frac{2\pi}{L_y}m_k + \frac{2\pi}{L_y}\frac{\Phi}{\Phi_0},\qquad X_{k_\Phi}= k_\Phi\, l_B^2 ,
\end{equation}
with $m_k\in \mathbb{Z}$. This means that the projection $\mathcal{P}_\Phi$ to the LLL  depends on $\Phi$. This is not an issue as long as we focus on adiabatic processes. Given Eq.~\eqref{eq:singleparticle_flux}, we can now inspect how the many-body light-matter Hamiltonian $\hat{H}(\Phi)$ looks like. The electron-electron interaction clearly remains unchanged as it only depends on difference of momenta. The light-matter coupling, for this purpose, is more conveniently formulated within the Coulomb gauge formulation (Sec. \ref{sec:howtoLL}). Here the interaction is only controlled by difference in momenta $k_{\Phi}-k_{\Phi}'$ and hence the full Hamiltonian $\hat{H}(\Phi)$ remains unchanged. As a key consequence, we have that the ground state wavefunction in second quantization  $\ket{\mathrm{GS}_\Phi}$, hence fixing an orbital basis, will be the same:
\begin{equation}\label{eq:Hpsi_flux}
    \hat{H}(\Phi)\equiv \hat{H}(\Phi=0) \;\;\Rightarrow\; \;\ket{\mathrm{GS}_\Phi}\equiv\ket{\mathrm{GS}_{\Phi=0}}.
\end{equation}
It is important to stress that in Eq.~\eqref{eq:Hpsi_flux}, we have not used the equality sign. The reasons is that the Hilbert spaces in which the two sides of the equations are defined are not, strictly speaking, the same: they refer to different LLL projection $\hat{\Pi}_\Phi$. In order to perform a meaningful comparison, we can consider relevant physical observables such as the charge density $\hat{n}(x)$, which is expressed as:
\begin{align}\label{eq:density_Phi}
    \hat{n}(x) =\frac{1}{L_y}\sum_{k_\Phi} \hat{n}_{k_\Phi} |\phi_0(x-X_{k_\Phi})|^2,
\end{align}
and can be evaluated at different $\Phi$ by just using the $\Phi=0$ solution with a change in the single particle orbitals $\phi_0(x-X_{k_\Phi})$. 

In order to calculate the current $\langle \hat{J}_x\rangle $ we use the continuity equation:
\begin{equation}
     \boldsymbol{\nabla}\cdot \langle \hat{\boldsymbol{J}} \rangle = -\partial_t \langle \hat{n}\rangle,
\end{equation}
where $\hat{\boldsymbol{J}}$ is the current operator. Given our finite cylinder geometry we can integrate both sides of the equation above in a region $\mathcal{V}=\{(x,y)\ \mathrm{s.t.}\ x<x_0\}$ to get:
\begin{equation}
    \langle \hat{J}_x(x_0) \rangle = - \partial_t \int_{-\infty}^{x_0}\dd x \langle \hat{n}(x)\rangle_{\Phi(t)} ,
\end{equation}
with $ \hat{J}_x(x_0)$ the current density along $x$ at position $x_0$. Because of translational invariance on $y$ the current and the density do not depend on $y$ making the integration over $y$ trivial. Using Eq.~\eqref{eq:density_Phi} and the ramp protocol $\Phi(t)\propto t$, we can express the local transverse resistivity in terms of the static density:
\begin{equation}\label{eq:resistivity_density}
    \rho_{xy}(x_0)= \frac{1}{2\pi \langle \hat{n}(x_0) \rangle } \frac{h}{e^2},
\end{equation}
where $h/e^2$ is the von Klitzing constant, and the factor $2\pi$ is the area occupied by a single-particle state (in units of $l_B^2$). The above equation directly links the bulk density to the fractional Hall response of the system. In particular, for a topologically ordered state in the class of the Laughlin $\nu=1/3$, one expects the bulk density to be constant $2\pi \langle \hat{n}(x_0)\rangle=1/3$.

In Fig.~\ref{fig:resistivity}(b) we show the bulk resistivity (shaded area represent all positions $x_0\in [-2,2]$) as a function of cavity field gradients $g$ in the Hamiltonian $\hat{H}_\chi$ (Eq. \eqref{eq:Htot_chi}) where cavity induced inhomogeneous single particle potentials are neglected. In order to compare two different cavity frequencies, we show the results as a function of $g\sqrt{\omega_c}$, as discussed in Sec. \ref{sec:hamiltonian}. The resistivity is quantized up to exponential corrections even at relatively large values of the cavity field gradient $g\sqrt{\omega_c}\simeq 2$. At stronger gradients the bulk density features strong oscillations (see Sec. \ref{sec:stripe_num}) hence a non-quantized Hall response. Here the FQH liquid becomes unstable towards the formation of stripes that, as we will show in Sec. \ref{sec:spectral_num}, is linked to the softening of finite momentum magnetorotons. The fact that changing the cavity frequency by a factor 2 leaves the instability region unchanged indicates that this is not a resonant effect. We remark however that the actual fate of the system in this regime strongly depend on the inclusion of single particle potentials and will be better characterized in Sec \ref{sec:stripe}. In Fig.~\ref{fig:resistivity}(c) we depict the deviations of the density from its bulk value when approaching an edge of the cylinder ($x=0$ is the center of the system while $x=20$ is outside of it). For all depicted couplings in the FQH phase, the corrections decay exponentially in the bulk with a correlation length $l_\mathrm{edge}$ which increase with $g$. 

It has been recently argued \cite{Rokaj_prl2023_losses} that the finite lifetime of the cavity mode can give a correction to the quantized transverse conductivity at temperature $T=0$ in the IQH regime. We note that at the many-body level the flux insertion argument can be adapted to the case of weak cavity losses (see Appendix~\ref{app:cavity_loss}). The key physical insight is that the steady state of the whole system (cavity plus matter) is at thermal equilibrium with the bath \cite{BreuerPetruccione_2007}, hence the ground state results, regarding the Hall resistivity, are expected to hold provided that the photonic bath is at a small enough temperature. It would be interesting to check, in the spirit of Ref.~\cite{Rokaj_prl2023_losses}, what finite temperature corrections are. We leave this to future work.

\subsection{Entanglement spectrum}\label{sec:ent_spectrum}
\begin{figure*}
    \centering
    \begin{overpic}[width=0.295\linewidth]{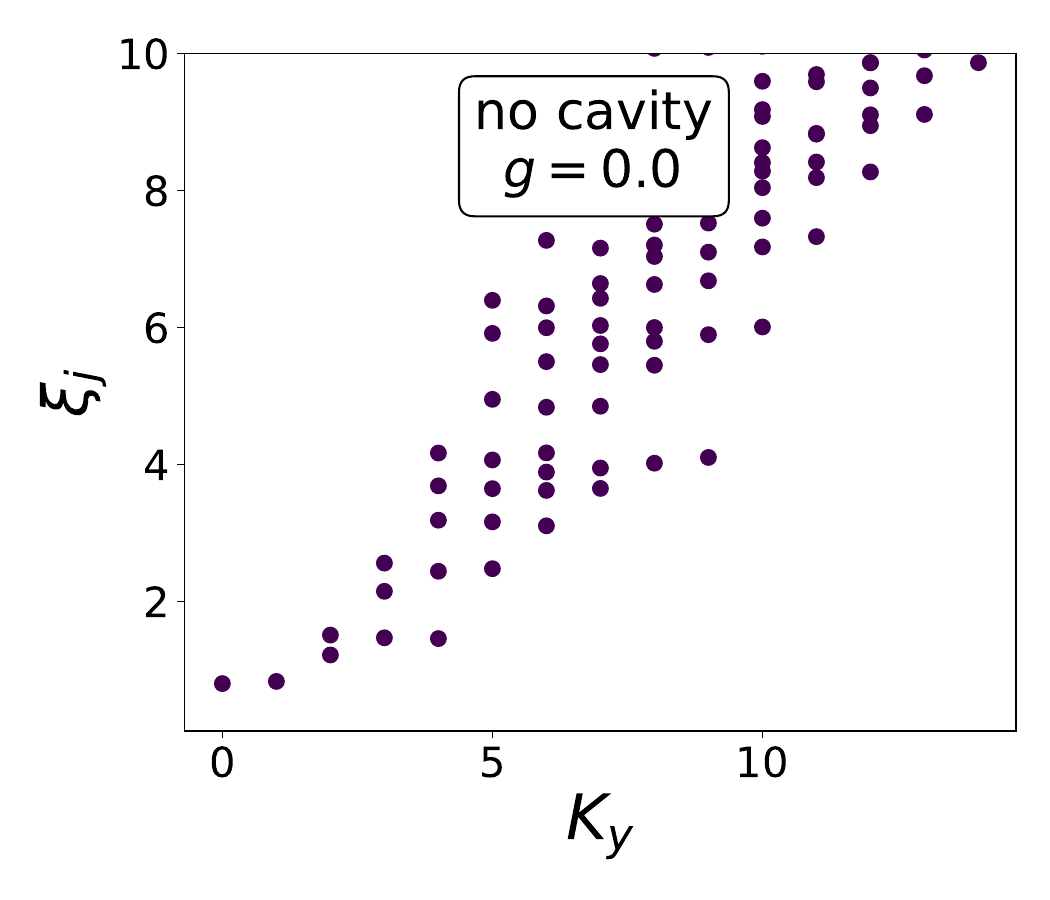}
    \put(20,65){(a)}
    \end{overpic}
    \quad
    \begin{overpic}[width=0.295\linewidth]{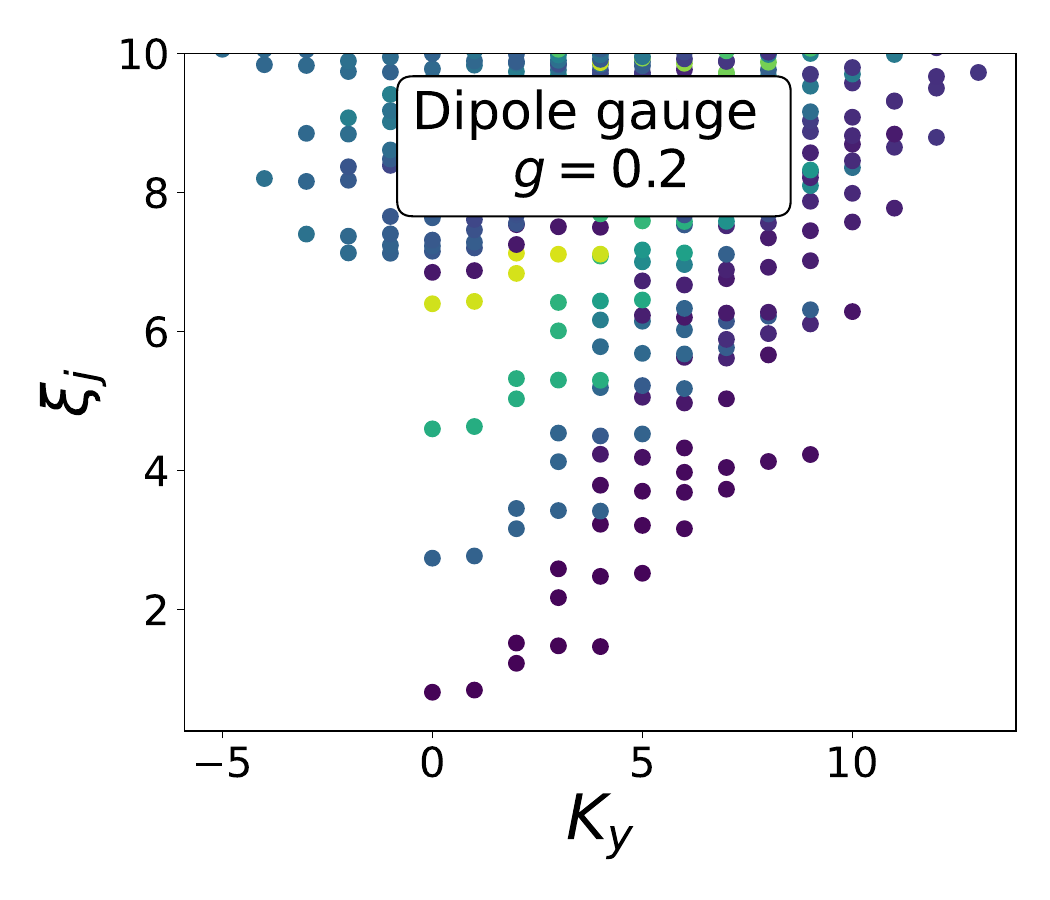}
    \put(20,20){(b)}
    \end{overpic}
    \quad
    \begin{overpic}[width=0.359\linewidth]{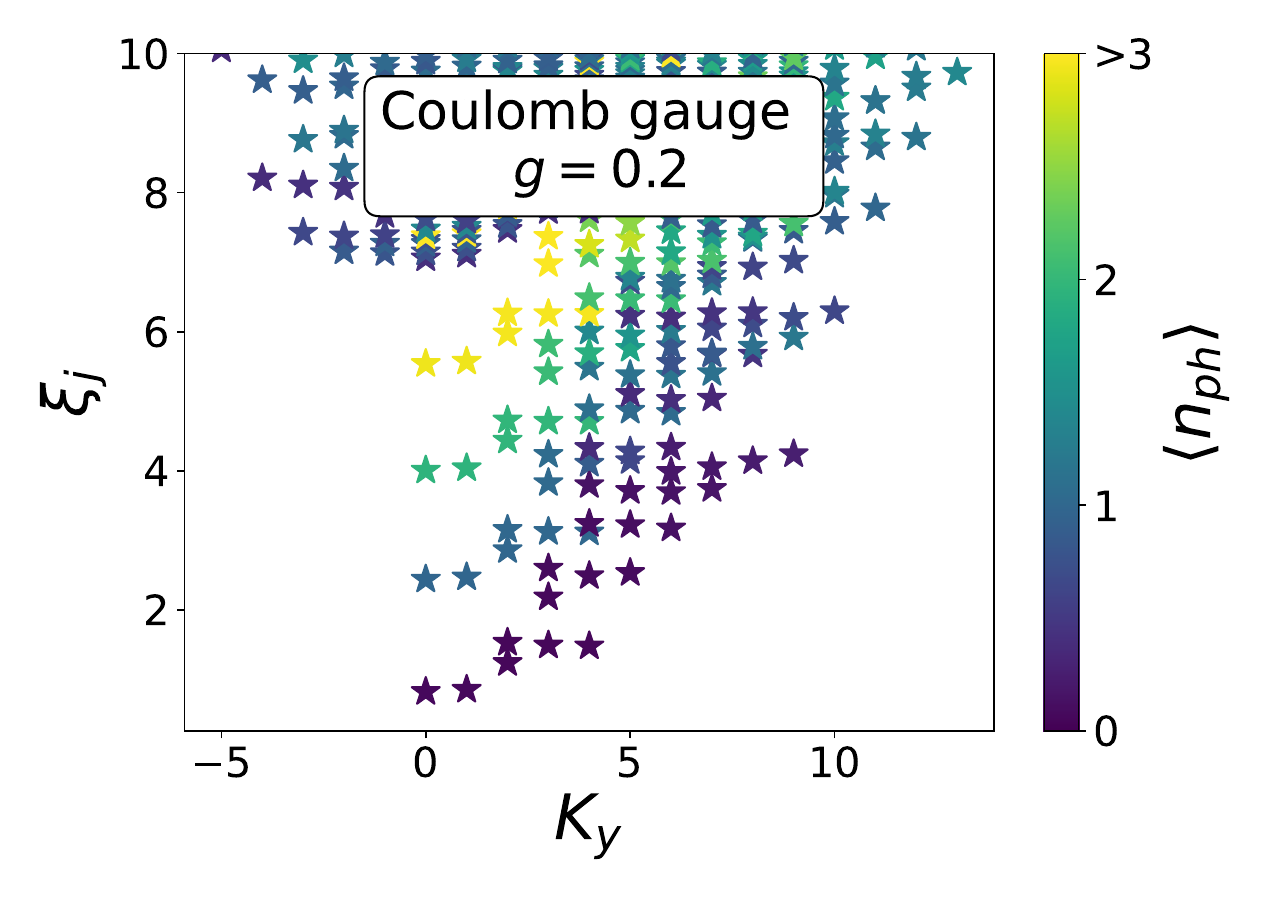}
    \put(17,17){(c)}
    \end{overpic}

    \begin{overpic}[width=0.295\linewidth]{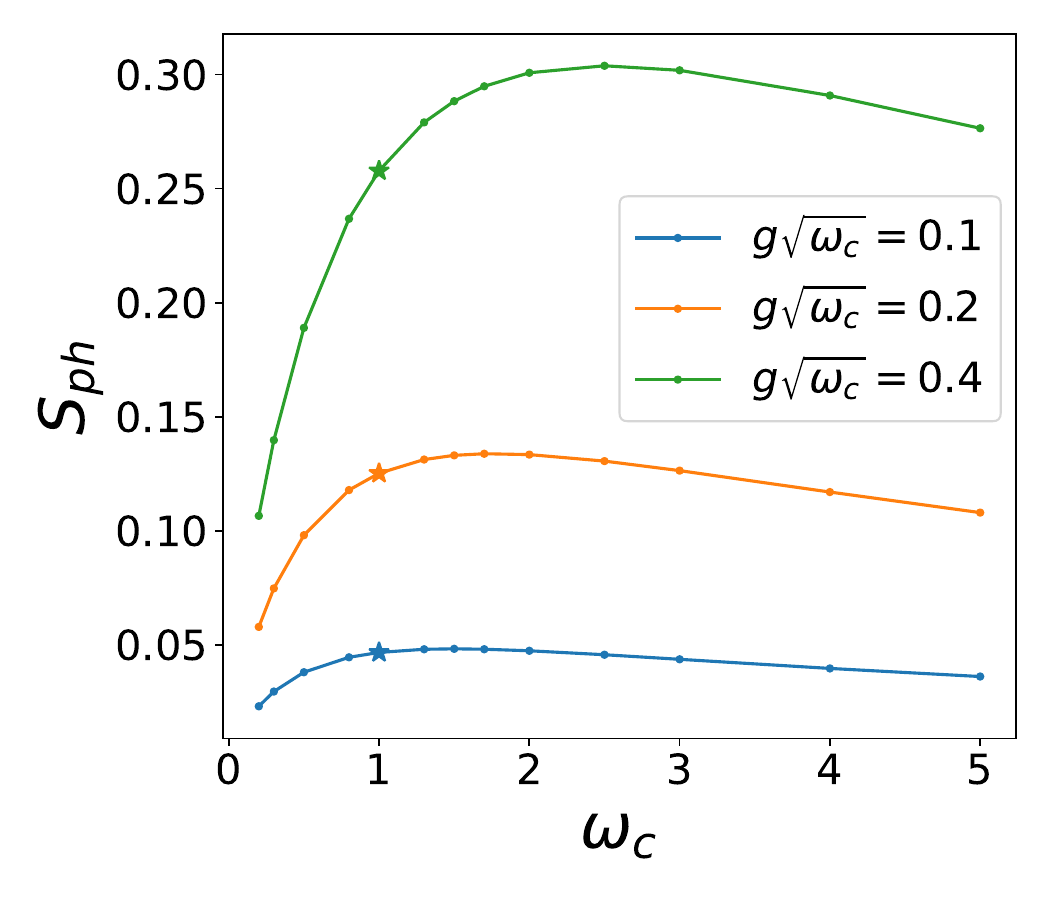}
    \put(23,67){(d)}
    \end{overpic}
    \quad
    \begin{overpic}[width=0.295\linewidth]{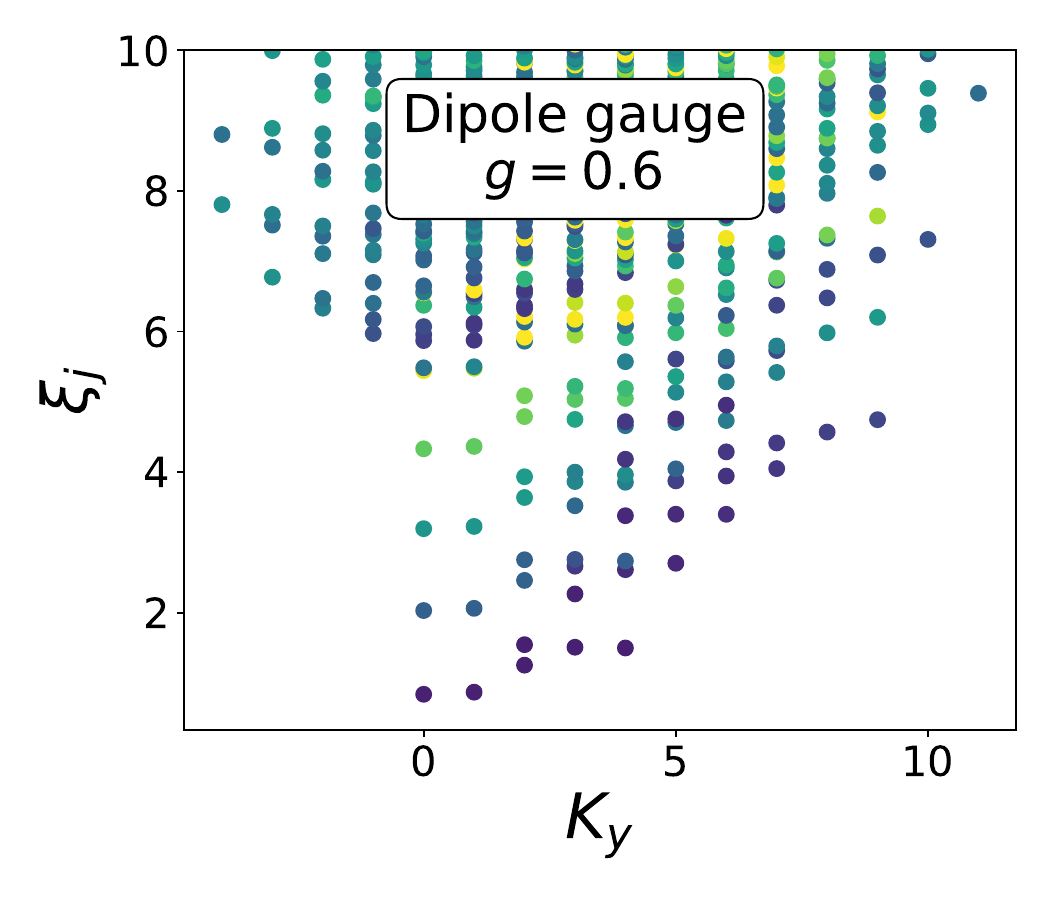}
    \put(20,20){(e)}
    \end{overpic}
    \quad
    \begin{overpic}[width=0.359\linewidth]{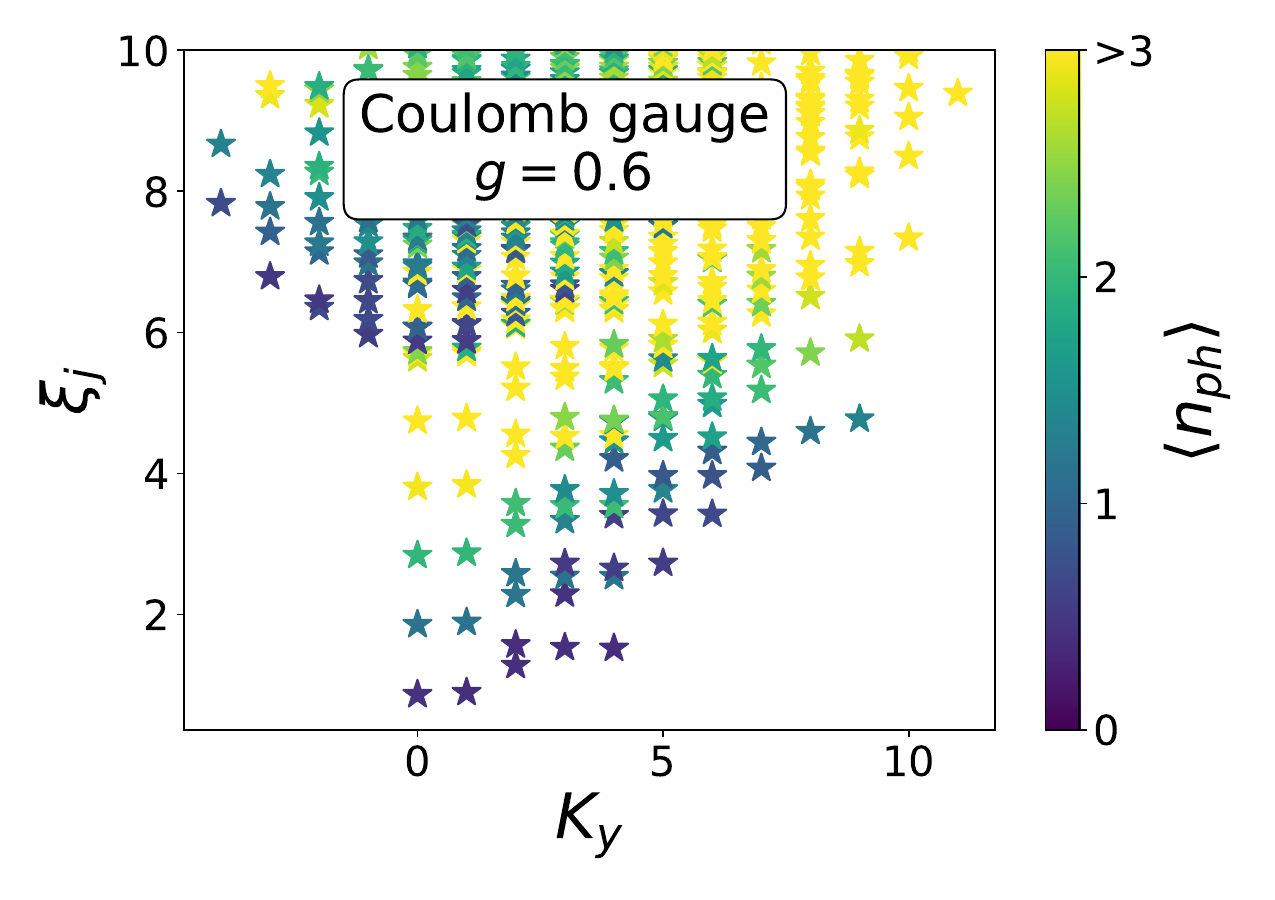}
    \put(17,17){(f)}
    \end{overpic}
    \caption{(a,b,c,e,f) Entanglement spectra obtained via DMRG for the asymmetric bipartition of Eq.~\eqref{eq:bipartition_c} on system size  $L_y=20$ and $N_e=34$ and (d) Entanglement entropy of the cavity in the Dipole gauge as a function of $\omega_c$ and fixed $g\sqrt{\omega_c}$. All results obtained from the Hamiltonian $\hat{H}_\chi$ in Eq. \eqref{eq:Htot_chi}. The charge sector of the shown spectra is $N_e/2=17$ while the momentum along $y$ gives the x-axis. In panel (a) the state is pure Laughlin ($g=0$) while in panels (b,c) and (e,f) we have a finite light-matter coupling $g=0.2$ and $g=0.6$ in Dipole and Coulomb gauge respectively. The color represent the number of photons in each Schmidt state and the colorbar is the same for all (a,b,c,e,f) panels}
    \label{fig:ent_spectrum}
\end{figure*}

The entanglement spectrum at a bipartition of a topologically ordered state exhibits distinct signatures, which can be used to detect topological order \cite{LiHaldane_prl2008,SterdyniakRegnault_IOP2011,Zalatel_prl2013,Zache_2022entspect_proposal}. In particular, one expects to find information about the edge theory of the topological state under consideration. This procedure, dubbed entanglement spectroscopy, is not only widely used as a theoretical tool, but has also been also proposed as an experimental protocol to detect topology in cold atom systems \cite{Zache_2022entspect_proposal}. The deep roots of this topology-entanglement connection lie in a very general results in relativistic quantum field theory, the Bisognano-Wichmann theorem~\cite{bisognano1976duality,dalmonte2022entanglement}, which dictates closed functional form expressions for the entanglement (or modular) Hamiltonian, and explain the Li-Haldane result~\cite{swingle2012geometric}. 
A natural question to ask is, what is left about this  topology footprint on entanglement in the presence of quantum light. 

Before discussing our hybrid cavity-matter setting, we review some general concepts. Let us define the pure state $\ket{\Psi}$ of the full system and the density matrix $\hat{\rho}_A$ of a subsystem $A$ as $\hat{\rho}_A=\Tr_{B} \left[\ket{\Psi}\bra{\Psi}\right]$ with $B$ being the rest of the system. In general, we can write:
\begin{equation}
    \hat{\rho}_A= \exp \left( -\hat{H}_{ee}\right) = \exp \left( -\sum_{{q},i} \xi_{\{q\},i}\ket{\phi_{\{q\},i}}\bra{\phi_{\{q\},i}} \right),
\end{equation}
where $\hat{H}_{ee}$ is the entanglement Hamiltonian, $\xi_{\{q\},i}$ the entanglement energies, $\ket{\phi_{\{q\},i}}$ the Schmidt vectors corresponding to the bipartition, and $(\{q\},i)$ an index tuple labeling the quantum numbers $\{q\}$ and the Schmidt state $i$. Following the Li-Haldane conjecture \cite{LiHaldane_prl2008,Regnault2017_lectnotes}, the entanglement spectrum for a bipartition in the bulk must follow, at low energies, the Hamiltonian of the edge. For Laughlin states, the edge theory is a chiral Luttinger liquid ($\chi$LL) \cite{Wen1990cll,Wen1992cll} which, once the $U(1)$ charge sector is fixed, gives a specific fingerprint in terms of degeneracies at each total momentum:
\begin{equation}
    (d_0,d_1,d_2,d_3,d_4,d_5,\dots)=(1,1,2,3,5,7,\dots),
\end{equation}
with $d_k$ being the degeneracy at momentum quantum number $k$. At finite sizes, the degeneracies are usually broken but a gap still separate a universal low energy part from a non-universal part of the entanglement spectrum \cite{Regnault2017_lectnotes}. This remark that this also happens for energy spectra at physical edges \cite{NardinCarusotto_pra2023_nonlinearedge} and not only for entanglement spectra in bulk bipartitions.

In the case of a cavity embedded systems there is no clear notion of pure state bipartition as the non-local bosonic mode cannot be ``divided" in two. As already done in \cite{bacciconi2023topological,ChiriacòChanda_prb2022}, one needs to define asymmetric bipartitions where the cavity mode resides on one side. A possible choice is the following:
\begin{equation}\label{eq:bipartition_c}
    \hat{\rho}_{X_k>0} = \Tr_{x<0;c}\left[ \ket{\Psi}\bra{\Psi} \right] ,
\end{equation}
where $\ket{\Psi}$ is the many-body ground state, $\hat{\rho}_{X_k>0}$ is the reduced density matrix for electrons in orbitals $k$ with $X_k>0$, and $\Tr_{X_k<0;c}$ is the trace over electrons at $X_k<0$ and the cavity mode $c$. One can, however, always recover the notion of an only matter bipartition by considering the electronic density matrix $\hat{\rho}_{el}=\Tr_c [\ket{\Psi}\bra{\Psi}]$. Taking its  bipartition will give as a result the same density matrix of the asymmetric bipartition with the cavity in Eq.~\eqref{eq:bipartition_c}.

In light of the discussion about different possible representations of the cavity-matter system (Sec. \ref{sec:howtoLL}), we want to stress that (not unexpectedly) the entanglement spectrum is not a gauge invariant quantity. Indeed, it can change under global unitary transformations $\hat{U}$, which implement the change of gauge in the truncated LLL model, since a global change of basis can change the reduced density matrix of subsystems. However, as discussed in Sec.~\ref{sec:num_meth}, we can easily change the gauge in which a state $\ket{\Psi}$ is represented by applying the unitary $\hat{U}$ and just check whether the entanglement spectrum displays features that are stable against the change of gauge. We remark that the change of gauge here implemented (after LLL truncation) leads to results which are in general different from a truncation in the Coulomb gauge.

\subsubsection{Entanglement spectrum bands and polariton entanglement gap}

In Fig.~\ref{fig:ent_spectrum} we show the DMRG results for the entanglement spectrum of the asymmetric bipartition in Eq.~\eqref{eq:bipartition_c} at $g=0$ (a) and at finite $g=0.2$ in the Dipole (b) and Coulomb (c) gauges of the Hamiltonian $\hat{H}_\chi$, described in Eq. \eqref{eq:Htot_chi} in the Dipole gauge. In particular, we fix the number of particles of the bipartition to be $N_e/2$ and look at momentum quantum number $K_y$. It is worth noting that we are interested here solely in regimes where electrons and cavity modes are hybridized, so that their mutual entanglement - portrayed in Fig.~\ref{fig:ent_spectrum}(d) is finite. The limit $\omega_c\rightarrow \infty$ cannot be immediately extrapolated from here, as the LLL truncation made in the model building part is not generically applicable when $\omega_c \sim \omega_B$.

At finite light-matter coupling $g$ the entanglement spectrum still show the $\chi$LL counting, but the higher energy part clearly changes. In order to understand the difference between entanglement eigenvectors, we color the markers based on the number of photons in the respective Schmidt state:
\begin{equation}
    n^{ph}_{\{q\},i }=\bra{\phi_{\{q\},i}}\hat{a}^\dagger \hat{a}\ket{\phi_{\{q\},i}},
\end{equation} 
where the Schmidt states are readily available from the total MPS. This highlights a very informative pattern. The $\chi$LL counting is repeated for number of photons roughly equals to integers. We empirically find for each of these branches:
\begin{equation}\label{eq:polariton_ent}
n^{ph}_{k,j }\simeq j, \qquad \xi_{k,j }\simeq \epsilon_{k}+ j\Delta_{pol},
\end{equation}
with $\epsilon_k$ a size dependent non-universal dispersion, and $j=0,1,2,\dots$ an integer. 

We remark that this phenomenology does not indicate multiple edge modes, but rather the presence of a mixed cavity-matter bulk excitation at zero momentum, akin to what happens for finite momentum magnetorotons in the case of Coulomb FQH ground states \cite{Regnault2017_lectnotes} that also give rise to high entanglement-energy features. In the present case, the repeated $\chi$LL counting of entanglement spectrum without any momentum-shift is a clear signature of the graviton mode getting hybridized with the cavity photons (see Sec. \ref{sec:spectral_eff} for the corresponding spectroscopic analysis).  

The quantity $\Delta_{pol}$, which we call \textit{polariton entanglement gap}, controls the separation between different sectors of the $\chi$LL with different number of photons and needs to be finite to preserve the $\chi$LL structure. Importantly, qualitative features of the entanglement spectrum are observed to persist in different gauges at least up until $g=1$ for the volumes considered here (see. Fig.~\ref{fig:ent_spectrum}b-c and e-f for two direct comparisons).

\begin{figure}
    \centering
    \includegraphics[width=0.9\linewidth]{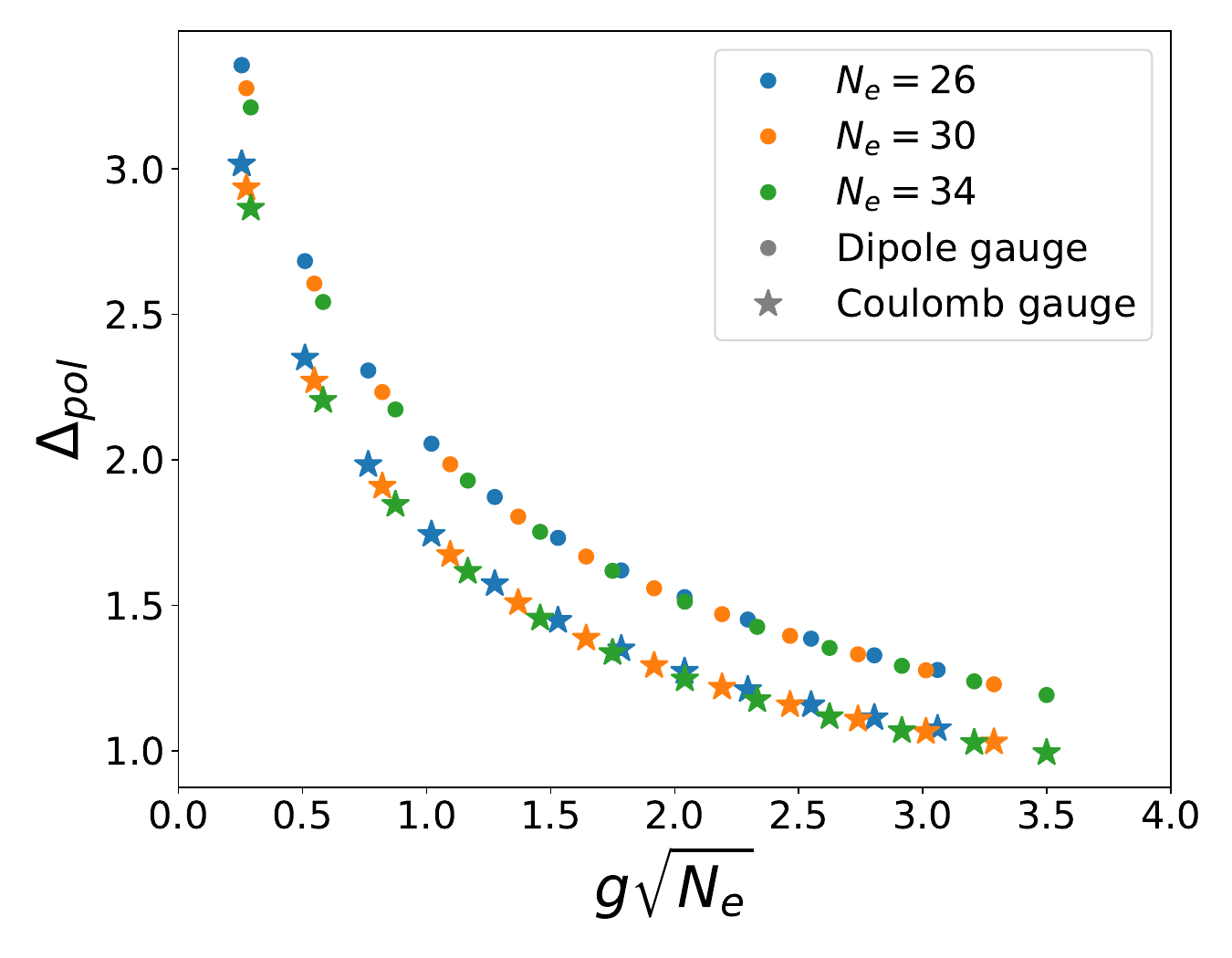}
    \caption{Polariton entanglement gap. Data extracted from DMRG entanglement spectra in the middle of the cylinder for different number of particles $N_e$, and fixed $L_y=20$, in the Dipole gauge (circles) and Coulomb gauge (stars) as a function of the collective coupling $g\sqrt{N_e}$.}
    \label{fig:entgap}
\end{figure}

In Fig.~\ref{fig:entgap} we study the polariton entanglement gap dependence with system size. We show its dependence as a function of a re-scaled collective coupling $g\sqrt{N_e}$ for different number of particles $N_e$. The perfect collapse highlights the collective nature of polariton excitations, i.e., they are controlled by the collective coupling $g\sqrt{N_e}$. By showing the entanglement gap for both gauges it is evident that this is a gauge dependent quantity, nonetheless both gauges reveal the same collective behavior. Moreover, we find that the value of $\Delta_{pol}$ does not depend on $L_y$ (not shown).

The nature of this gapped polaritonic excitation will be clarified in Sec. \ref{sec:spectral}, where we show that a strong hybridization between a collective emergent electronic mode, the momentum $q\simeq0$ part of the magnetoroton spectrum, and the cavity is taking place. An important consequence of the collective coupling is that taking  thermodynamic limit $N_e\rightarrow \infty$ at fixed $g$ breaks the $\chi$LL counting and the topological order of the state. The stability of the FQH phase at finite $g$ then needs to be understood in a mesoscopic sense -- another important scale $g\sqrt{N_e}$ is controlling the many-body gap of the FQH phase and hence its topological order. \\

Overall, our prediction of a stable FQH topology upon the inclusion of the non-local cavity mode should be thought as parallel to recent precise experimental measurements in the IQH regime \cite{enkner_2023_vonklitzing}. The non-local element introduced by the cavity mode can make this prediction a priopri not trivial. We can rationalize this result in two key factors which are still satisfied: (i) there must be a finite bulk excitation gap (see Sec. \ref{sec:spectral}) and (ii) the non-local cavity mode couples to local charge conserving operators. Breaking (ii) would for example immediately generate a cavity-mediated inter-edge scattering thus breaking the topological protection, even in presence of (i). It is indeed a cavity-mediated inter-edge scattering due to small sample size \cite{Ciuti_prb2021}, that likely leads to a loss of good transport properties \cite{AppuglieseFaist_science2022}.

\section{Spectral properties}\label{sec:spectral}

In this section we discuss the effect of the cavity mode on the neutral bulk spectral properties of the FQH liquid. We first give an introduction to the neutral excitations in absence of the cavity mode (Sec. \ref{sec:spectral_mr}), e.g., the magnetoroton spectrum, using the so-called single-mode approximation \cite{GMP1986} for the magnetorotons due to Girvin, MacDonald and Platzman. In Sec. \ref{sec:spectral_num} we present  numerical results which map out the phenomenology of the low-lying excited state spectrum in presence of the cavity mode. Finally, we provide a simple effective polariton model that captures the essential physics of hybridized cavity-matter excitations, the graviton-polaritons, and test its predictions against numerics (Sec. \ref{sec:spectral_eff}).

%%%%%%%%%%%%%%%%%%%%%
\subsection{Magnetoroton spectrum}\label{sec:spectral_mr}
\begin{figure*}
    \centering
    \begin{overpic}[width=0.3\linewidth]{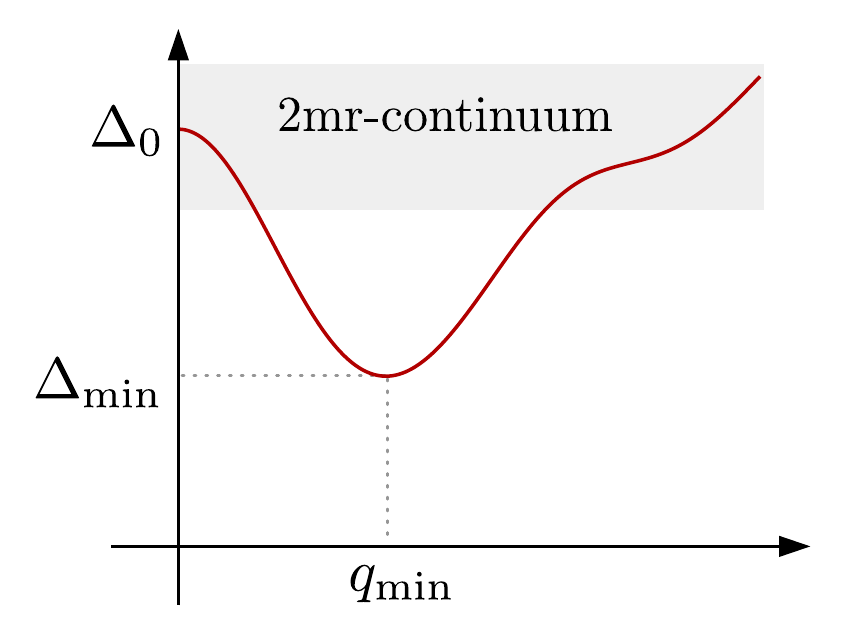}
        \put(10,73){(a)}
    \end{overpic}
        \medskip
    \begin{overpic}[width=0.3\linewidth]{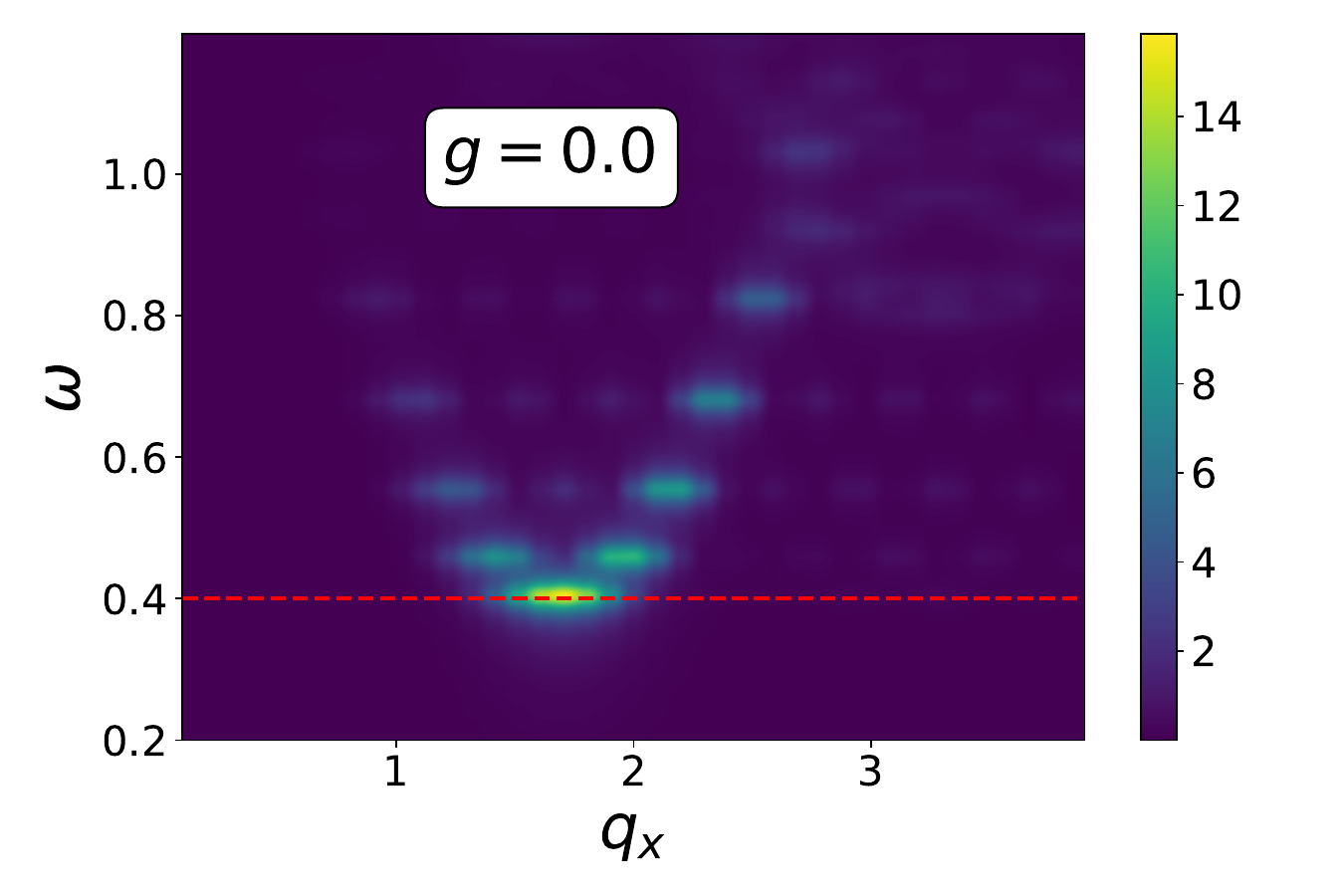} 
    \put(10,73){(b)}
    \end{overpic}
    \begin{overpic}
    [width=0.3\linewidth]{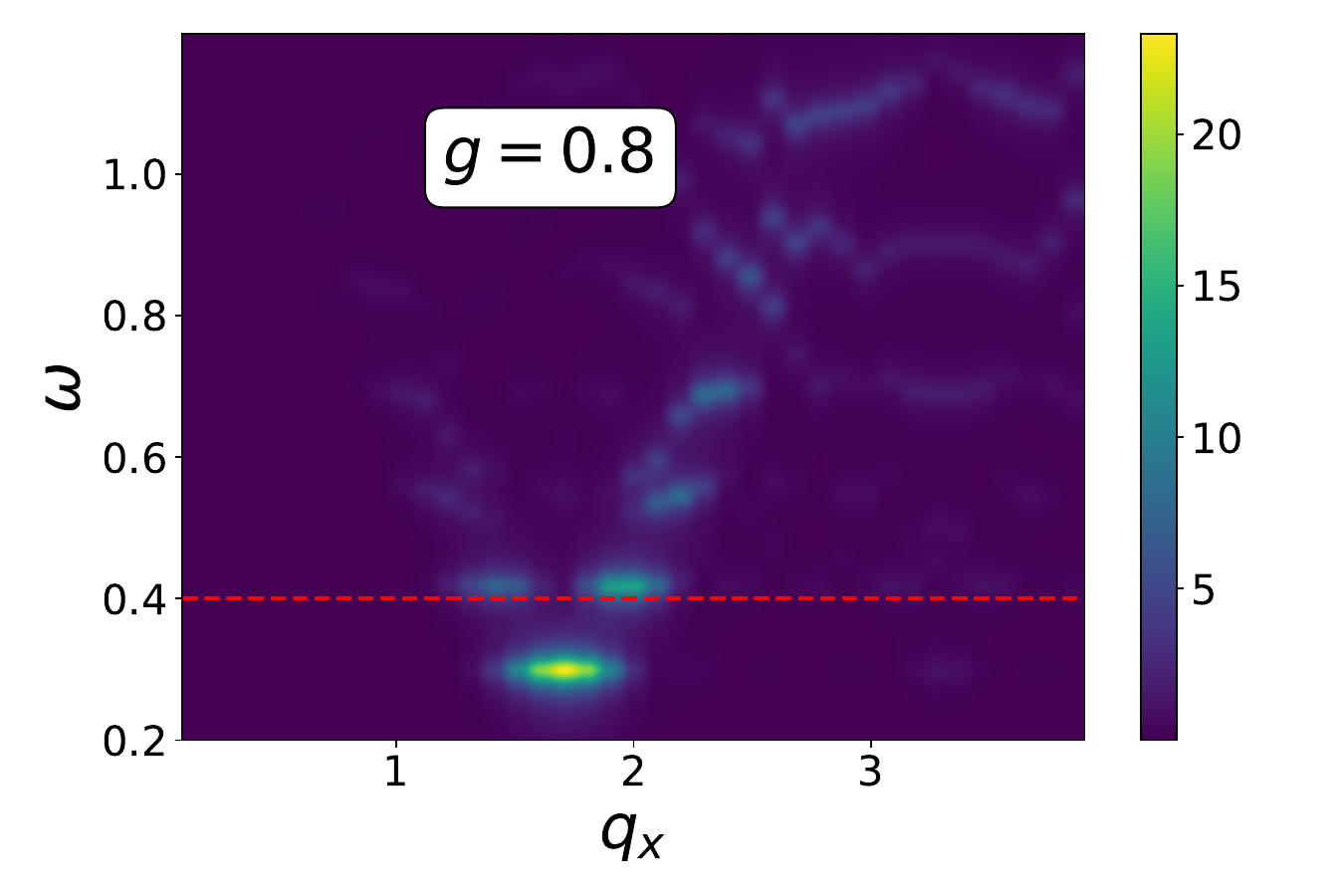}
    \put(10,73){(c)}
    \end{overpic}
    \begin{overpic}[width=0.329\linewidth]{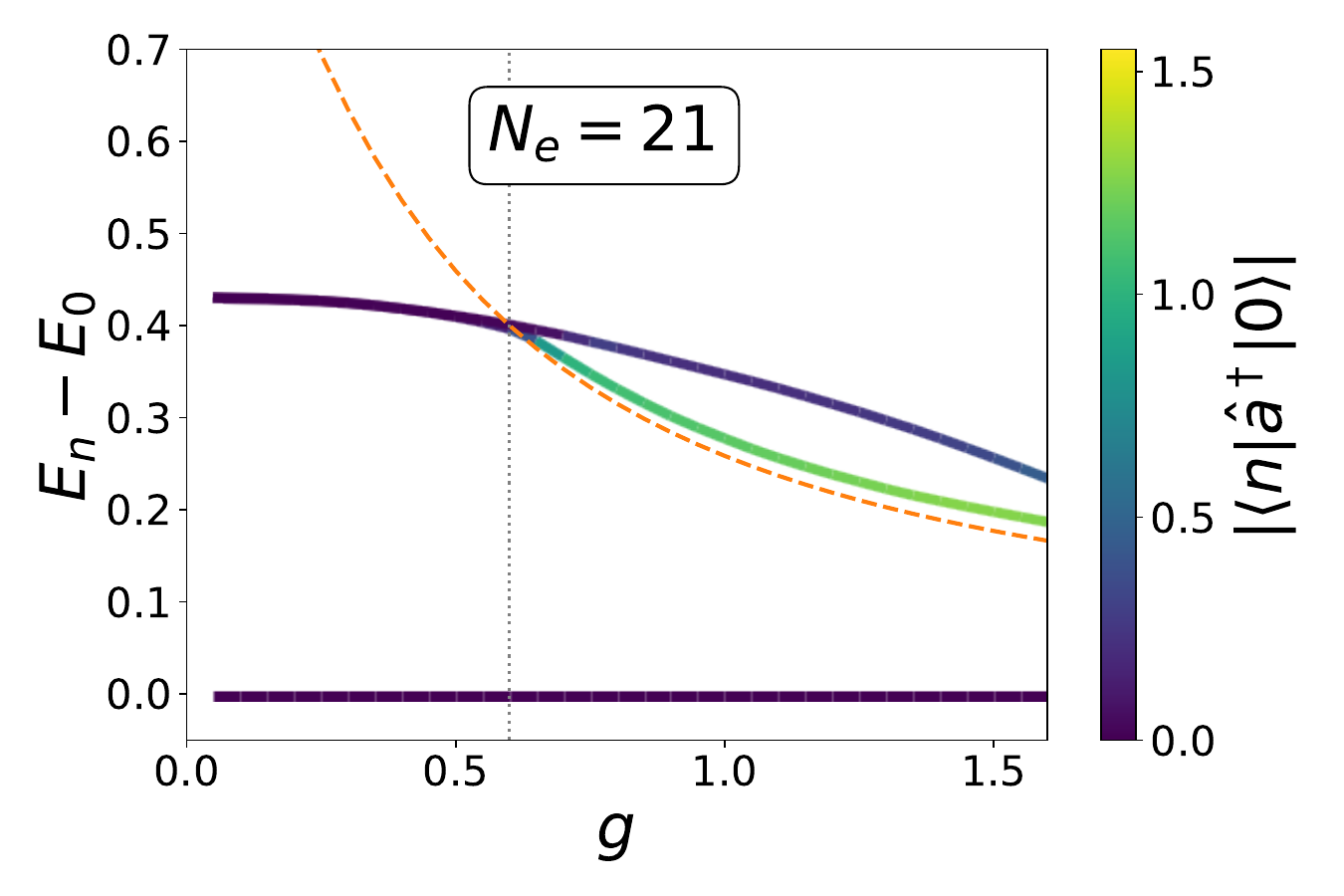}
    \put(15,67){(d)}
    \end{overpic}
    \medskip
    \begin{overpic}[width=0.329\linewidth]{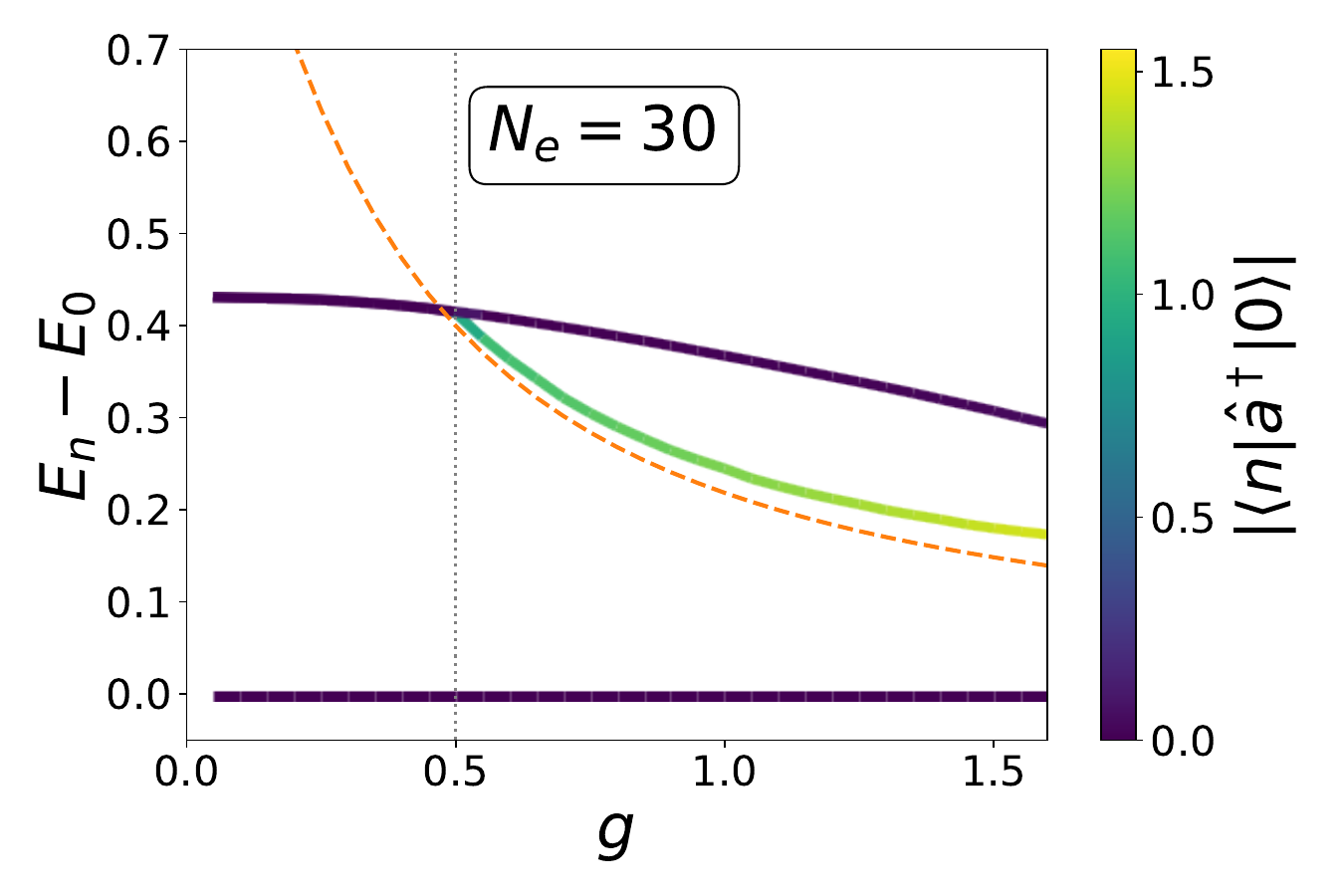}
    \put(15,67){(e)}
    \end{overpic}
    \begin{overpic}
        [width=0.292\linewidth]{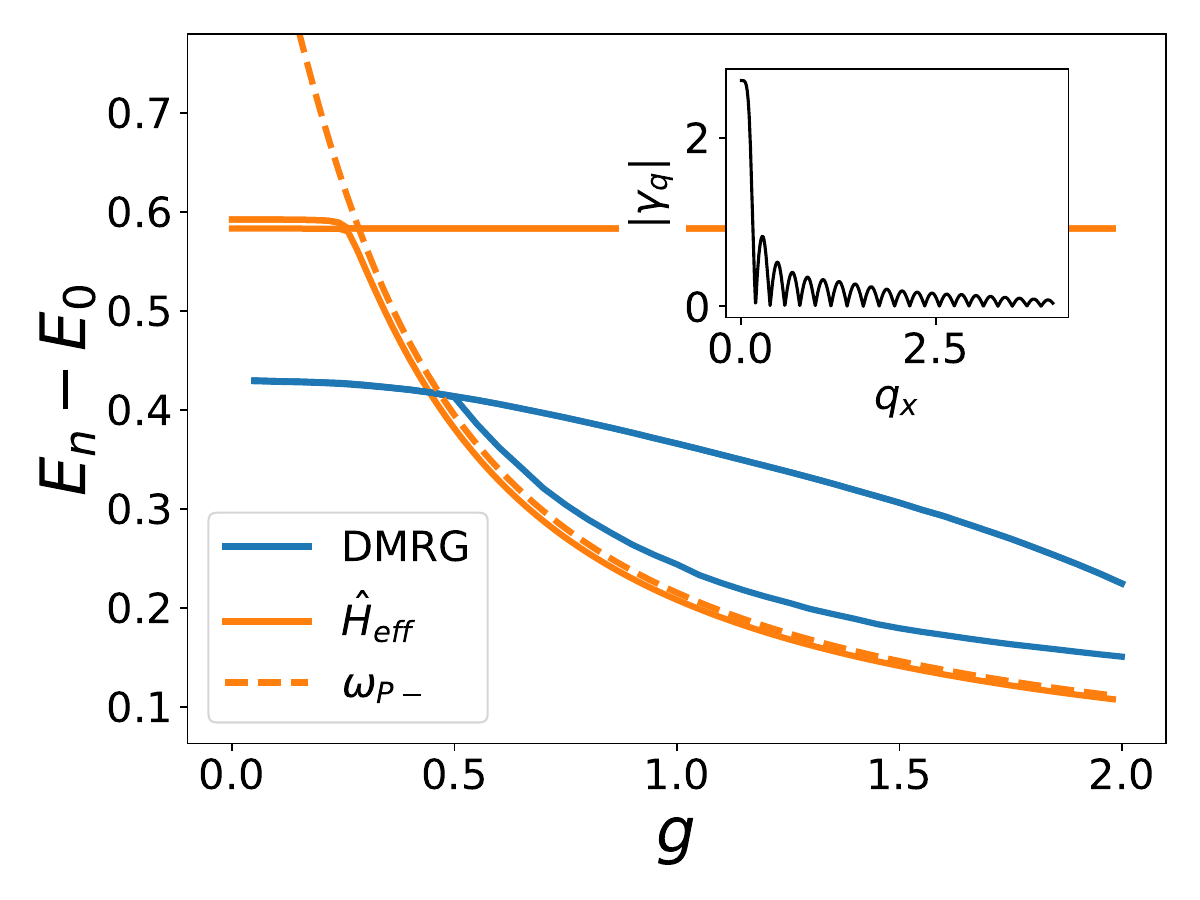}
        \put(15,76){(f)}
    \end{overpic}
    \caption{(a) Magnetoroton dispersion as calculated from the SMA (red) with a sketch of the two-magnetoroton continuum. The SMA magnetoroton gap $\Delta_\mathrm{min}\simeq 0.6$ occurs at a finite wavevector $q_\mathrm{min}\simeq1.7$ and a gap is also present at small wavevectors $\Delta_0\simeq1.5$. (b,c) Dynamical structure factor as a function of momentum along $x$ at $g=0$ (b) and $g=0.8$ (c). The red dashed line is a guide to the eye that signals the energy of the first excited state at $g=0$. $L_y=10$ and $N_e=10$ in (b) and (c). (d),(e) Low energy spectrum in the $K_y=0$ sector as a function of $g$ calculated via DMRG for different number of particles $N_e=21$ (d) and $N_e=30$ (e) at the same $L_y=16$. The colorbar represent the photon creation matrix element from the ground state to each state and the vertical grey dashed line a guide to the eye for the value at which a polariton becomes the first excited state. The orange dashed line represent the lower polariton frequency obtained from the bare graviton-polariton model Eq. \eqref{eq:eff_model_0}. (f) Comparison between DMRG results and the eigenfrequencies of the effective model in Eq. \eqref{eq:effective_H} for the first two excited states. These are always either magnetoroton states or polaritons. The effective model uses the coupling parameters $\gamma_q$ shown in the inset. The orange dashed line represent again the lower polariton frequency from the bare graviton polariton model Eq. \eqref{eq:eff_model_0}. $L_y=16$ in (d),(e),(f). The cavity frequency is always $\omega_c=1$, and cavity vacuum-induced Stark shifts are absent.}
    \label{fig:magnetoroton}
\end{figure*}
%\textcolor{red}{Here we introduce the magnetoroton excitations and the SMA approximation by GMP \cite{GMP1986}. } 

Magnetorotons are the lowest-energy excitations above the gapped FQH ground state. They manifest as charge density modulations within the LLL that arise from the bound state of a quasi-electron and a quasi-hole \cite{Laughlin1984excitons,KamillaJain1996excitonsCF}. Crucial to the stability of the FQH state, the magnetoroton spectrum has been subject of intense study \cite{GMP1986,KamillaJain1996excitonsCF,KumarHaldane_prb2022,Kumar2022prb_stripes,Nguyen2022mganetorotons,Balram_2022prx_fqhgravitons}. Their dispersion relation exhibits a pronounced minimum at finite wavevector $q\sim1/l_B$, which descends below the two-particle continuum, see Fig. \ref{fig:magnetoroton}(a). The minimum is known as the magnetoroton gap, and is the precursor of ordered phases such as the Wigner crystal and stripes \cite{Kumar2022prb_stripes}. 

More recently the $q\rightarrow 0$ magnetorotons have been associated to the fluctuations of the quantum Hall geometry tensor \cite{Haldane_prl2011_quantummetric,haldane2023_quadrupole}. These emergent spin-2 gravitons (gapped and non-relativistic) have a strong quadrupolar moment with a vanishing dipole moment and have a preferred chirality, exact for model wavefunctions. We remark that chiral FQH gravitons have also been recently detected in inelastic scattering experiments with circularly polarized light \cite{Liang_nature2024_graviton}.

We now briefly review a simple physical picture of the magnetoroton mode, provided by Girvin, MacDonald and Platzman 
(GMP) with the single-mode approximation (SMA)~\cite{GMP1986}. Note that here the ``single mode" term does not refer to our cavity model. The SMA is a variational ansatz that describes the magnetoroton excitations as long wavelength charge modulations on top of the liquid ground state of uniform density. With this picture in mind, GMP built a set of excited states as:
\begin{equation}
\ket{\bm{q}}=\hat{\Bar{\rho}}_{\bm{q}}\ket{\Psi_L},
\end{equation}
where $\hat{\Bar{\rho}}_{\bm{q}}$ is the guiding center density operator in the LLL defined in Eq.~\eqref{eq:guidingcenter_def}. Using the static structure factor $S(\bm{q})$ to normalize the variational state, the excitation energy then becomes fixed by the ratio:
\begin{equation}\label{eq:SMA-energy}
    \Delta_\mathrm{SMA}(\bm{q})=\frac{F(\bm{q})}{S(\bm{q})},\qquad
    F(\bm{q})=\bra{\bm{q}}\hat{H}_\mathrm{int}-E_0\ket{\bm{q}},
\end{equation}
where $F(\bm{q})$ is the oscillator strength and $S(\bm{q})$ the guiding center structure factor. The LLL density operator $\hat{\rho}_{\bm{q}}$ obeys the Lie algebra: $\comm{\hat{\rho}_{\bm{q}}}{\hat{\rho}_{\bm{q}}'}=2i\sin(\frac12\bm{q}\times\bm{q}')\hat{\rho}_{\bm{q}+\bm{q}'}$, named after GMP \cite{GMP1986,KumarHaldane_prb2022}, which allows to express the oscillator strength $F(\bm{q})$ as a sole function of $S(\bm{q})$ and the interaction potential \cite{GMP1986}. 

As shown in Fig. \ref{fig:magnetoroton}(a), the GMP ansatz captures the essential features of the magnetoroton mode, reproducing a fully gapped mode with a mininum at finite momenta. It is worth noting that the SMA overestimates the magnetoroton mode energy gap as the wave vector is increased. This is a well-known shortcoming of the ansatz \cite{GMP1986}, given that the density operator also couples to high-energy states containing a greater number of quasi-electron and quasi-hole pairs \cite{KumarHaldane_prb2022}. 

In our notations, the SMA predicts a gap minimum of $\Delta_\mathrm{min}\simeq0.6$ close to the momenta $q_\mathrm{min}\simeq1.4$. This is to be contrasted with the ED calculation of the dynamical structure factor shown in Fig. \ref{fig:magnetoroton}(b). We observe the actual gap is smaller, around $\Delta_\mathrm{min}\simeq0.4$ with a corresponding wave vector $q_\mathrm{min}\simeq1.7$. Long-wavelength magnetorotons hide inside the two-quasi-particle continuum as the SMA predicts $\Delta_0\simeq1.5$. While the SMA fails to predict a quantitatively correct magnetoroton gap, it is believed to capture the graviton energy $\Delta_0$. It is less clear however up to what extent the two particle continuum damps the pristine magnetoroton dispersion found in the SMA. In what follows, we will see that for suitable values of the light-matter coupling, the energy of the lower graviton-polariton gets shifted below the two excitation continuum, making it protected against such an absorption channel.

\subsection{Numerical results}\label{sec:spectral_num}

We now investigate numerically the effect of the cavity mode on the low-lying FQH bulk spectrum. Again, in order to mantain homogeneity in the system, we will work with $\hat{H}_\chi$ (Eq. \eqref{eq:Htot_chi}) in absence of cavity mediated single particle potentials ($\mu_k=0$). Note also that because of the choice of ``hard" boundaries, the edge states are gapped out, and only bulk excitations remain. In order to track the magnetoroton dispersion we look at the dynamical structure factor $S(\boldsymbol{q},\omega)$, Eq.~\eqref{eq:dynamicalSq}. The dynamical structure factor probes the response of the system to density excitation at a certain frequency $\omega$. In the following we will focus on the response to modulation only along $x$ (uniform on $y$) so on $S(q_x,\omega)$. In Fig.~\ref{fig:magnetoroton} we show this quantity for $g=0$ (b) and for $g=0.8$ (c) at a small system size ($L_y=10$ and $N_e=10$) accessible via ED. Note that due to the open boundary conditions on $x$ the momenta is not a good quantum number and excitations are expected to spread over a finite region of $q_x$. Cavity mediated interactions (proportional to $\chi_k$) are clearly lowering the magnetoroton gap with no particular modification of the wavevector at which the minimum is found $q_\mathrm{min}\simeq 1.75$. As we will see later in Sec. \ref{sec:stripe_num}, this is the precursor of a cavity-mediated instability of the FQH liquid towards a stripe state.

In order to access larger system sizes we look directly at the excited state spectrum by targeting excited states via local effective Hamiltonians constructed during DMRG calculations~\cite{Mila_prb2017_excited} as explained in Section \ref{sec:num_meth}. More details about the accuracy of this method can be found in Appendix \ref{app:excited_appendix}. Moreover by looking at the excited states we can gain more information also on the cavity degree of freedom. It is important to remark that this method allows us to probe excitations in the $K_y=0$ sector only, hence exploring only quasi-momenta $q_x$. In Fig. \ref{fig:magnetoroton}(d,e) we show the first two low-lying excited states as a function of $g$ for number of particles: $N_e=21$ (d) and $N_e=30$ (e) at $L_y=16$. The result are obtained from the middle-chain local effective Hamiltonian. The colors of the lines represent the strength of the matrix element $|\bra{n}\hat{a}^\dagger\ket{0}|$, which enters in the cavity density of states $D_c(\omega)$, see Eq. \eqref{eq:Acavity_def}, helping us to spot the polaritonic character of the states. The orange dashed line is an analytical prediction for the lower polariton that will be discussed in Sec. \ref{sec:spectral_eff}. We can distinguish two qualitatively different regimes. These are the following:
\begin{enumerate}
    \item Near $g=0$, the two low lying states are part of the magnetoroton dispersion around the $q_\mathrm{min}$ and they start at around $\Delta_\mathrm{min}\simeq 0.4$ which is the bulk neutral gap of the Laughlin state. All these states show a vanishingly small one-photon matrix element with the ground state. The cavity photon mode is located well above, at his bare energy $\omega_c=1$.
    \item For finite $g$, the cavity mode mixes with matter excitations, in particular with the quadrupole-active graviton mode at $\Delta_0$. This results into a pair of  graviton-polaritons, characterized by a sizable one-photon matrix element with the ground state. In particular, the energy of the lower graviton-polariton starts at the bare cavity frequenxy $\omega_c=1$ and is then shifted down for growing $g$. At a strong enough value of the coupling $g_P$ marked with a vertical dashed line, the lower graviton-polariton becomes the first excited state and gets immune from absorption processes into other matter excitations. This happens sooner for larger systems, for system sizes shown in Figure \ref{fig:magnetoroton} we find $g_P=0.65$ vs $g_P=0.5$ ($N_e=21$ vs $N_e=30$). An effective model (orange dashed line, see Sec. \ref{sec:spectral_eff}) capture this feature. The upper graviton-polariton  (not visible in the figure) starts at $\Delta_0>\omega_c$ and gets blue-shifted for larger $g$. 
\end{enumerate} 
Note that having the polariton state lower in energy with respect to the magnetoroton implies that the gap protecting the topological order from finite temperatures will be the polariton gap, as pointed out in Ref. \cite{Rokaj_prl2023_losses} for the IQH case. We then want to remark that there seems to be a difference between the dependence with system size of the polariton state and of the magnetoroton dispersion. While the latter is empirically controlled by $g$, the polariton state sees the collective enhancement thus it is controlled by $g\sqrt{N_e}$ (we are assuming here the cavity mode volume to be independent of $N_e$). In the next subsection we propose an effective model to explain this feature and the nature of the polariton state.

%%%%%%%%%%%%%%%%%%%%%%%%%%
\subsection{Effective model}\label{sec:spectral_eff}

\begin{figure*}
    \centering
    
    \begin{overpic}[width=0.313\linewidth]{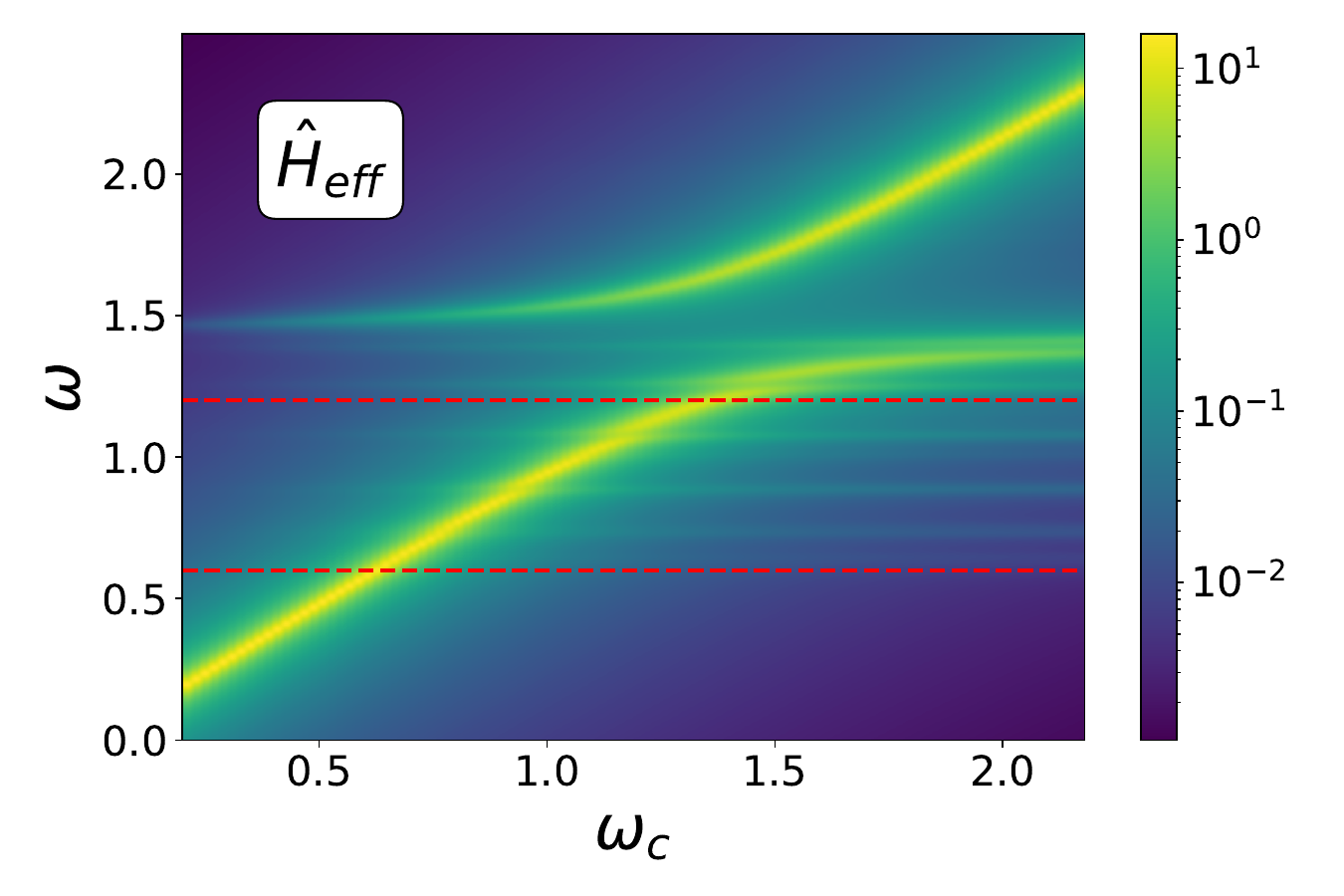}
    \put(0,60){(a)}
    \end{overpic}
    \begin{overpic}[width=0.313\linewidth]{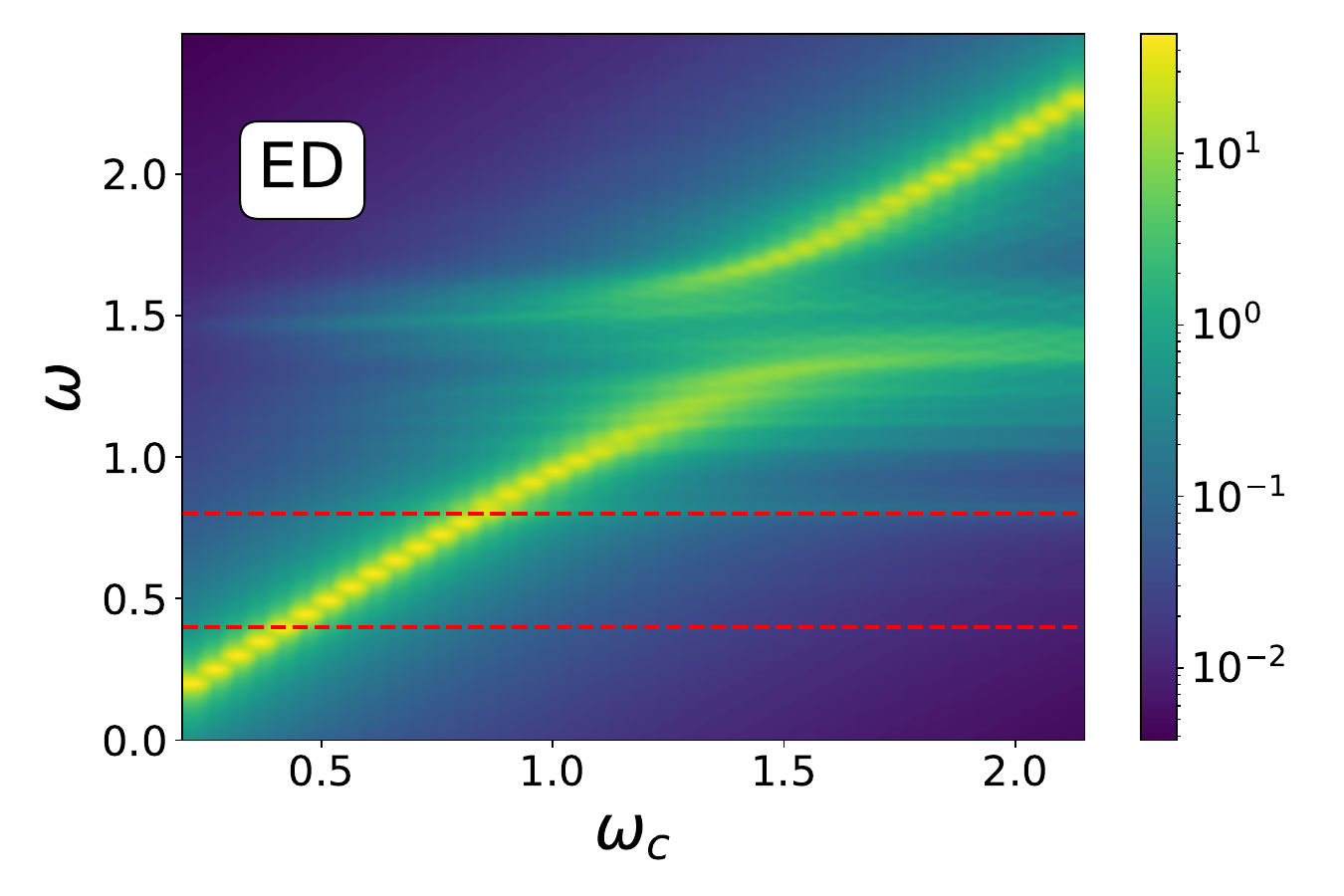}
    \put(0,60){(b)}
    \end{overpic}
    \begin{overpic}[width=0.279\linewidth]{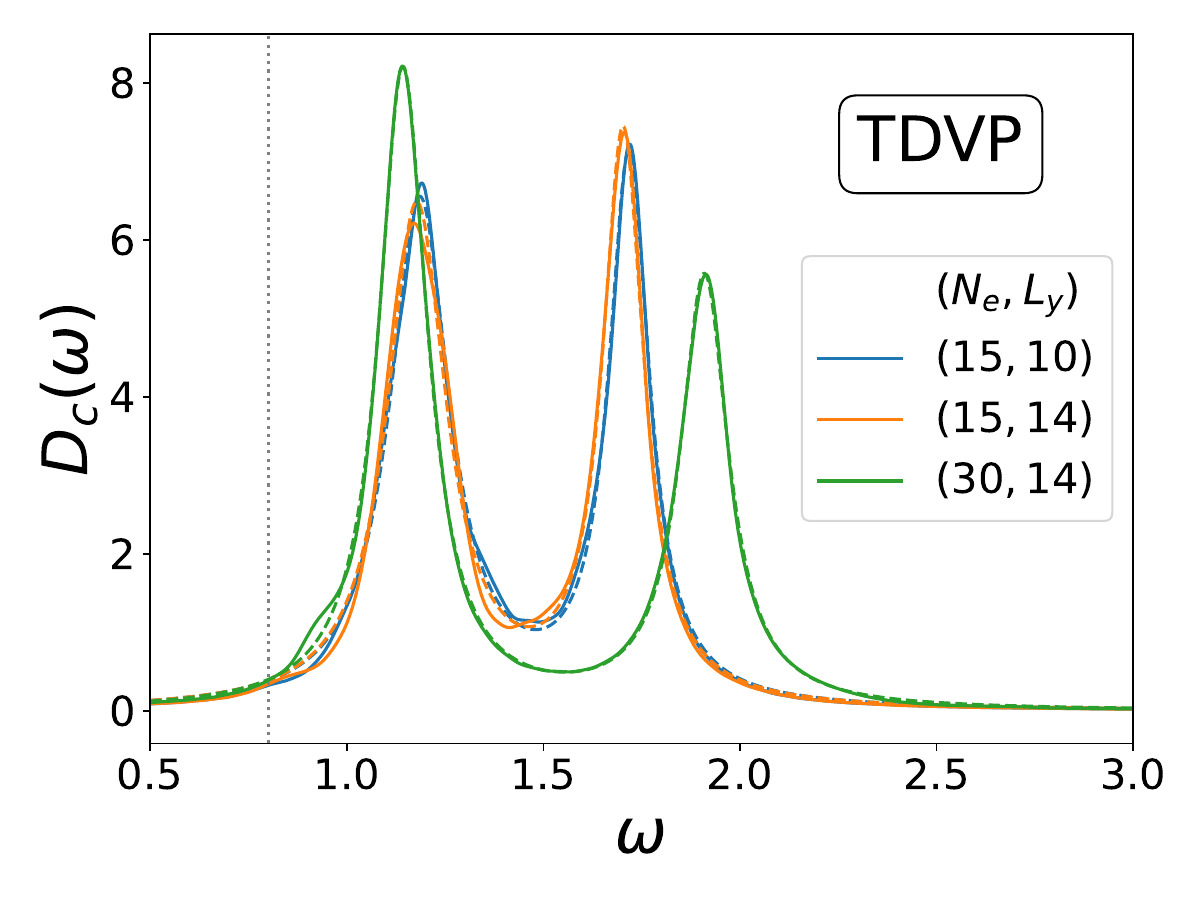}
    \put(-2,65){(c)}
    \end{overpic}
    \caption{Cavity density of states $D_c(\omega)$ in the Dipole gauge with different methods in absence of cavity induced single particle potentials. The ED results (b) and effective model (a) show the avoided crossing when changing the cavity frequency $\omega_c$, signalling the strong coupling between the graviton mode and the cavity mode. Dashed red lines mark the frequency range between the magnetoroton gap $\Delta_{min}$ and $2\Delta_{min}$ where the two-particle continuum begins. In the effective model there is a residual small coupling to the finite $q$ part of the magnetoroton while the ED results only highlight a coupling to the two-particle continuum and not to single magnetorotons states. In panel (c) we show TDVP results with fixed cavity frequency $\omega_c=1.5$, almost resonant with the graviton,  and we change $N_e$ and $L_y$. By fitting with two Lorentzians (dashed lines) we extract the Rabi splittings between the two polariton resonances (Table \ref{tab:rabi_spl}). System size $(N_e,L_y)$ in (a) is $(10,10)$, in (b) is $(30,16)$, in (c) are described in the legend. The couplings for $(a,b)$ are respectively $g=0.1,0.1/\sqrt{30/10}$ to have the same collective coupling. In (c) $g=0.1$. Broadening parameters $\eta=0.02$ in (a,b) and $\eta=0.05$ in (c).}
    \label{fig:Acav}
\end{figure*}

%\textcolor{red}{Here we discuss the effective model for the magnetoroton-polaritons and show the numerical vs effective model result for the cavity spectral function. Then we compare the actual energy states. Hernan's perturbation theory? }

We introduce an effective model to describe the coupling between magnetorotons and the cavity field, neglecting for the moment cavity mediated single particle terms. Treating the magnetorotons as free bosons, we express the matter Hamiltonian as a collection of independent oscillators: $H_\mathrm{MR}\approx\sum_{\bm{q}}\Delta(\bm{q})\hat{b}^\dagger_{\bm{q}} \hat{b}_{\bm{q}}$, where $\hat{b}_{\bm{q}}$ represents the bosonic excitation of the magnetoroton at momentum $\bm{q}$, and $\Delta(\bm{q})$ denotes the energy dispersion. For the sake of simplicity we neglect interactions. In the original model the coupling to the cavity is performed via the polarization operator $\hat{P}=\frac12\sum_k k^2\hat{n}_k$. For the effective model we replace it with $\hat{P}\to\sum_q\gamma_q(\hat{b}_q+\hat{b}_q^\dagger)$, where $\gamma_q$ represents the effective coupling governing the transition, and the momenta $\boldsymbol{q}=(q,0)$ is restricted to the $x$ direction since the electric field does not depend on $y$. The effective model Hamiltonian is then expressed as
\begin{align}\label{eq:effective_H}
    \hat{H}_\mathrm{eff}&=\omega_c \hat{a}^\dagger \hat{a}+ \sum_{q}\Delta(q)\hat{b}^\dagger_q \hat{b}_q+\omega_c g^2 \Big[\sum_q\gamma_q (\hat{b}_q+\hat{b}^\dagger_q ) \Big]^2\nonumber\\
    &+i \omega_c g(\hat{a}-\hat{a}^\dagger) \sum_q\gamma_q (\hat{b}_q+\hat{b}^\dagger_q ),
\end{align}
where we factor out the $g$ coupling in order to facilitate the power counting. The sum on $q$ should be sensitive to boundary conditions: in the present case of a finite cylinder of length $L_x$ in the direction of OBC we will consider modes with momentum $q=\pm(2j_q-1)\pi/L_x$ ($j_q=1,\dots,M/2$), whose symmetric and antisymmetric combination form even and odd standing waves with proper boundary conditions. Then, being Eq. \eqref{eq:effective_H} a quadratic bosonic Hamiltonian, it can be easily solved via Bogoliubov-Hopfield transformations. Note that the $q=0$ mode cannot be constracted from the SMA procedure.

To draw a comparison with the numerical simulations of the actual FQH plus cavity setup, we fix the effective parameters of our model by using the SMA. The energy dispersion follows from the variational ansatz in Eq. (\ref{eq:SMA-energy}) and is sketched in Fig. \ref{fig:magnetoroton}(a). The light-matter interaction parameters $\gamma_q$ are obtained from the matrix elements of the polarization operator $\hat{P}=\frac12\sum_k k^2\hat{n}_k$ with respect to the Laughlin ground state and the SMA excited states:
\begin{align}
    \gamma_q=\frac{1}{\sqrt{MS(q)}}\sum_{k,k'}\frac{1}{2}k^2e^{-i k'q}\bra{\Psi_L}\delta\hat{n}_{k}\delta\hat{n}_{k'}\ket{\Psi_L}.
\end{align}

The matrix element $\gamma_q$ is shown in the inset of Fig.~\ref{fig:magnetoroton}(f). We observe it  displays a prominent peak as $q\to0$ and some smaller oscillations for finite wave vectors. These have a period of roughly $2\pi/L_x$ and are caused by the open cylinder geometry. The behaviour at $q\to 0$ can be obtained from the long-wavelength expansion of the structure factor $S(q)\simeq S_4^{xx}q^4$, which in the thermodynamic limit ($L_y,N_e\to\infty$ and $L_y/L_x\propto L_y/\sqrt{N_e}=const$ for a fixed aspect ratio) yields:
\begin{equation}\label{eq:gamma_en}
    \tilde{\gamma}_0\equiv\lim_{q\to0} \gamma_{q}=-\sqrt{N_eS_4^{xx}/\nu}\propto \sqrt{N_e}.
\end{equation}
Note that, although the $q=0$ momentum state is strictly speaking excluded from the SMA ansatz, by taking the limit we can define a well behaved collective mode. At finite but big enough system sizes, the relevant coupling strength is going to be peaked around $q=0$ and spread over a region of momenta $\delta q \simeq \pi/L_x$ thus allowing a further simplification to single matter mode model that captures this behaviour. The collective enhancement factor $\sqrt{N_e}$ is signalling that the graviton is expected to be a good collective excitation able to couple to the field gradient. 

We remark that in Eq.~\eqref{eq:gamma_en} there is no explicit dependence on $L_y$ nor $N_e$, just on the parameter $S_4^{xx}$ which controls the long-wavelength correlations of the FQH liquid. The behaviour at finite $q$ is a boundary effect and is also the regime where the SMA should be taken with an extra grain of salt. The collective enhancement of the $q\to 0$ mode only suggests that the effective model can be further simplified to a bare graviton-polariton model, with a single collective matter mode $\hat{\tilde{b}}_{0}$ coupled to the cavity mode $\hat{a}$:
\begin{align}\label{eq:eff_model_0}
        \hat{H}_{\text{eff},0}&= \omega_c \hat{a}^\dagger\hat{a}+\Delta_0 \hat{\tilde{b}}_0^\dagger \hat{\tilde{b}}_0 + i\omega_c g\tilde{\gamma}_0(\hat{a}-\hat{a}^\dagger)(\hat{\tilde{b}}_0+\hat{\tilde{b}}_0^\dagger) \nonumber \\
        &+\omega_c g^2\tilde{\gamma}_0^2(\hat{\tilde{b}}_0+\hat{\tilde{b}}_0^\dagger)^2,
\end{align}
where $\Delta_0$ denotes the graviton energy, and $\tilde{\gamma}_0$ is taken from Eq.~\eqref{eq:gamma_en}. By means of a Bogoliubov-Hopfield transformation we can directly get the two polariton energies resulting from Eq.~\eqref{eq:eff_model_0}:
\begin{equation}\label{eq:twopol_en}
    \omega_{P\pm}^2 = \frac{1}{2}\left( \omega_c^2 +\Delta_0^2 +\Omega^2\pm \sqrt{(\omega_c^2+\Delta_0^2+\Omega^2)^2 -4\omega_c^2\Delta_0^2} \right),
\end{equation}
where we have introduced the Rabi frequency:
\begin{equation}\label{eq:rabi_frequency}
    \Omega=2g\tilde{\gamma}_0 \sqrt{\omega_c\Delta_0}.
\end{equation}

In Fig.~\ref{fig:magnetoroton}(f) we compare the predictions of the effective models (orange lines) with the low-lying energy spectrum obtained from DMRG simulations (blue lines). We observe that the effective model $\hat{H}_{\mathrm{eff}}$ successfully captures the emergence of the polariton mode that comes down in energy as a function of $g$. This energy closely follows the lower polariton energy $\omega_{P_-}$ where only an effective $q=0$ mode is taken into account. Small deviations are compatible with the accuracy of the DMRG variational estimate for the polaritonic state (see App. \ref{app:excited_appendix}). In stark contrast, the effective model misses the gap softening of other magnetoroton states which live at finite $q$. The red-shift of the magnetoroton mode under the effect of light-matter interactions seems indeed very important at strong coupling, signaling the emergence of an instability towards a density modulated stripe phase as we describe next section.  

In this respect, we wish to repeat that even at $g=0$ the effective model, being based on the SMA, does not accurately capture the exact value of the gap. Interactions, both matter-matter and cavity-mediated, play a key role in the renormalization of the gap and capturing them requires the treatment of the full many-body problem beyond the effective model.

\subsubsection*{Spectroscopy of the graviton-polariton}

Thanks to the hybrid nature of the polariton excitations~\cite{basov2020polariton}, it is possible to have distinct simple spectroscopic signature of both the upper and the lower modes, which gives direct evidence of the strong-coupling regime. To this end we show (Fig. \ref{fig:Acav}) the cavity density of states $D_c(\omega)$ as defined in Eq.~\eqref{eq:Acavity_def}. Here  we compare the prediction of the effective model (b) with ED (a) and TDVP (c) results. The ED and effective model result indicates the dominant hybridization at small $g$ occurs at the energy scales of the graviton $\Delta_0$. In the ED result we also spot subdominant couplings to other states inside the two-particle continuum which are not captured in the effective model. The latter instead predicts a weight on the finite $q$ part of the magnetoroton that in ED is not present (region between the red dashed lines), consistent with the failure of the effective model in capturing the softening of the magnetoroton at finite momenta. The predicted collective enhancement of the graviton hybridization is confirmed by the TDVP results. By fitting the spectral function with two Lorentzians, we can extract the Rabi splitting between the two polariton resonances, which are reported in Table \ref{tab:rabi_spl}.

\begin{table}[h!]
    \centering
    \begin{tabular}{|c|ccc|c|}
    \hline
       $(N_e,L_y)$  & $(15,10)$ & $(15,14)$ & $(30,14)$ & Effective model\\ \hline

        $\Omega$ & 0.53 & 0.53 & 0.77 & $ 1.0 \times \omega_c g\sqrt{N_e}$ \\
        $\Omega/\omega_c g\sqrt{N_e}$ & 0.91 & 0.91 & 0.94 & 1.0
        \\
    \hline
    \end{tabular}
    \caption{Rabi splittings between graviton-polaritons resonances obtained by fitting TDVP results, Fig. \ref{fig:Acav}(c), compared with the effective model prediction in the thermodynamic limit and assuming $\Delta_0=\omega_c=1.5$. Small corrections due to small detunings are expected, as well as finite-size corrections with both $N_e$ and $L_y$ on $\Delta_0$ and $\Omega$.}
    \label{tab:rabi_spl}
\end{table}

According to the graviton-polariton model, Eq.~\eqref{eq:twopol_en}, and assuming a resonance condition $\Delta_0=\omega_c=1.5$ the Rabi splitting at small enough $g\sqrt{N_e}$ coincide with the Rabi frequency, and hence should increase with $\sqrt{N_e}$. Indeed we find that $\Omega$ increases by a factor $1.45\simeq\sqrt{2}$ when the number of particles $N_e$ is doubled while keeping the mode volume constant, and it does not change with $L_y$, confirming $\tilde{\gamma}_0$ to be independent of it. The precise value is also quite close to the graviton-polariton effective model and the discrepancy gets smaller as the circumference $L_y$ is increased.

We now comment on the role of cavity induced single particle potentials. Since the strength of these is of order $g^2$ (see Eq. \eqref{eq:Htot_chimu}), they will not disturb the resonant, small $g$, splitting of the graviton-polariton doublet. However they will be important for a quantitative prediction of the corrections to the magnetoroton gap (Fig. \ref{fig:magnetoroton}) which only takes off-resonant contributions of order $g^2$. Note also that, while our results with $\hat{H}_\chi$ (withouth cavity vacuum-induced Stark shifts $\mu_k$) indicate no collective enhancement in the renormalization of the magnetoroton gap, the single-particle terms  ($\mu_k$) which appear in Hamiltonian $\hat{H}_{\mu,\chi}$ (Eq. \ref{eq:Htot_chimu}) are \textit{collectively} enhanced by a geometric factor $L_x^2 \sim N_e$ when finite variations of the electric field $g$ are present.

\section{Ultra-strong coupling effects}\label{sec:stripe}

In this section we analyze other important effects which arise in the so-called ultra-strong coupling regime. Although this term is used with specific tresholds for $g$ in the literature, we will be more naive and term ultra-strong coupling the generic regime where $g$ is not infinitesimal. We will however carefully distinguish single-particle ($g$) and collective ($g\sqrt{N_e}$) ultra-strong coupling regimes. 

In Sec. \ref{sec:geometry} we first discuss the impact of the cavity on FQH long-wavelength fluctuation, related to the FQH emergent metric and the finite frequency graviton mode. In Sec. \ref{sec:stripe_num} we then show how, in absence of inhomogeneus cavity-induced vacuum Stark shifts, the FQH liquid becomes unstable towards the formation of stripes. This however only arise at the single-particle strong coupling regime. A much stronger (collective) effect is instead imprinted by the non-uniform cavity vacuum-induced Stark shift (Sec. \ref{sec:qp_ren}) which renormalize the energy cost of charged quasi-particles. 

\subsection{Cavity control of FQH geometry}
\label{sec:geometry}

An important property of FQH states which have received a lot of interest in recent years is their intrinsic geometry or metric \cite{Haldane_prl2011_quantummetric,son2013_newtoncartan_graviton,GolkarSon_2016jhep_graviton,Liu2018quench,Ippoliti_prb2018_geometryfluxattach}. This controls many of its ground state correlations and its long-wavelength gapped excitations quanta that, in analogy with gravitational theories, have been dubbed as ``gravitons" \cite{son2013_newtoncartan_graviton,GolkarSon_2016jhep_graviton,Liou_prl2019_graviton}. Here we focus on ground state referring to Sec. \ref{sec:spectral} for the discussion about excitations.

A key ground state footprint of non-trivial geometry can be found in guiding-center correlations~\cite{haldane2009hallviscosity,Ippoliti_prb2018_geometryfluxattach}, in particular via the guiding-center structure factor $S(\boldsymbol{q})$, Eq.~\eqref{eq:staticSq_def}, at small momenta. For a gapped FQH state it is possible to show \cite{haldane2009hallviscosity,Ippoliti_prb2018_geometryfluxattach} that:
\begin{equation}\label{eq:S4_smallq}
    S(\boldsymbol{q}) \simeq S_4^{xx} q_x^4 + 2S_4^{xy} q_x^2q_y^2 + S_4^{yy} q_y^4. 
\end{equation}
In the case of unbroken rotational invariance, we further have $S_4^{xx}=S_4^{xy}=S_4^{yy}=s_4$, with $s_4$ satisfying the Haldane bound $s_4\geq (1-\nu)/24\nu$ saturated by the Laughlin state \cite{haldane2009hallviscosity,kumarhaldane_scipost2024_bound}. Anisotropic interactions and/or anisotropic LLs \cite{Haldane_prl2011_quantummetric} induce an intrinsic anisotropic geometry on the FQH liquid.

\begin{figure}
    \centering
    \begin{overpic}[width=\linewidth]{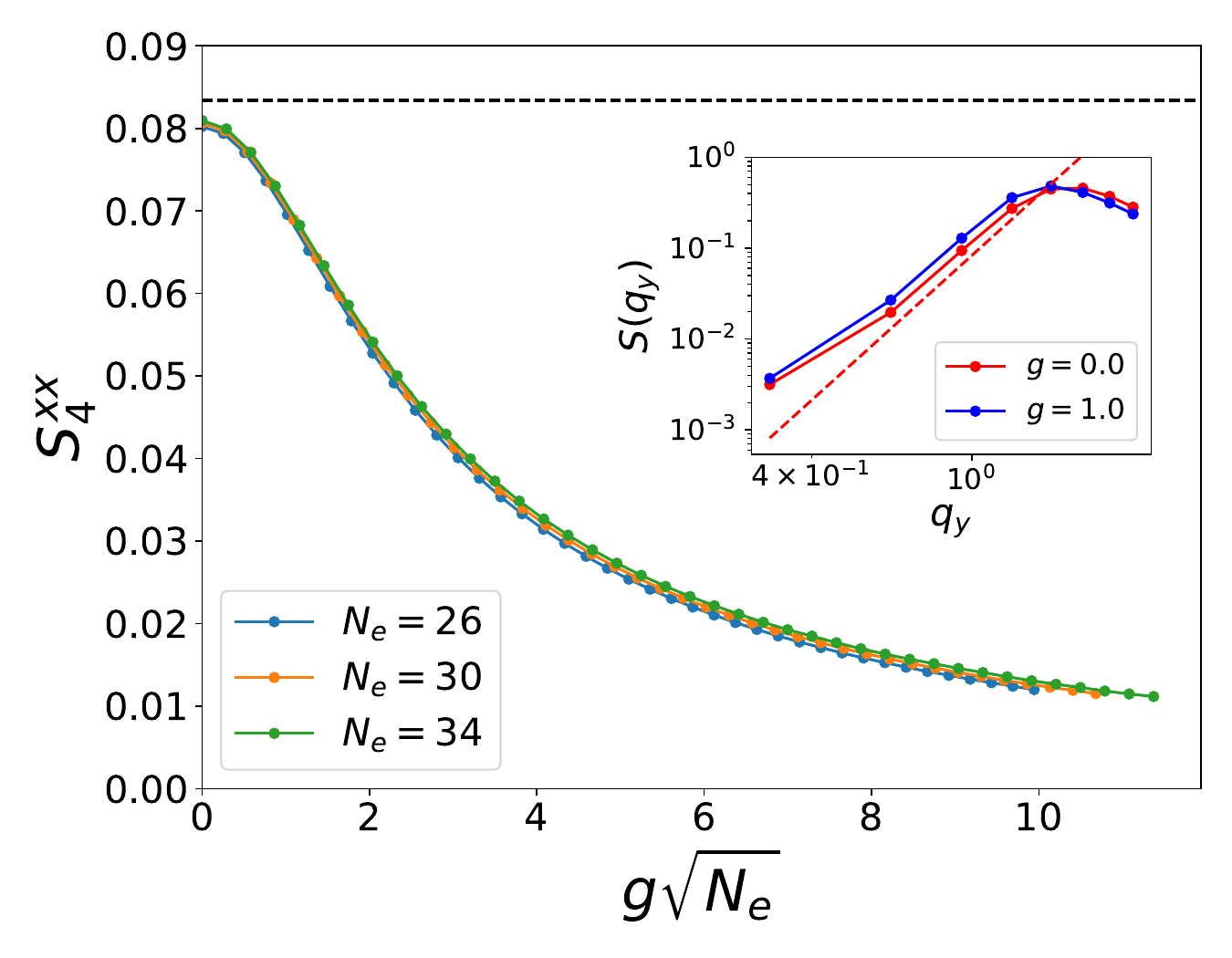}
        \end{overpic}
        \begin{overpic}[width=\linewidth]{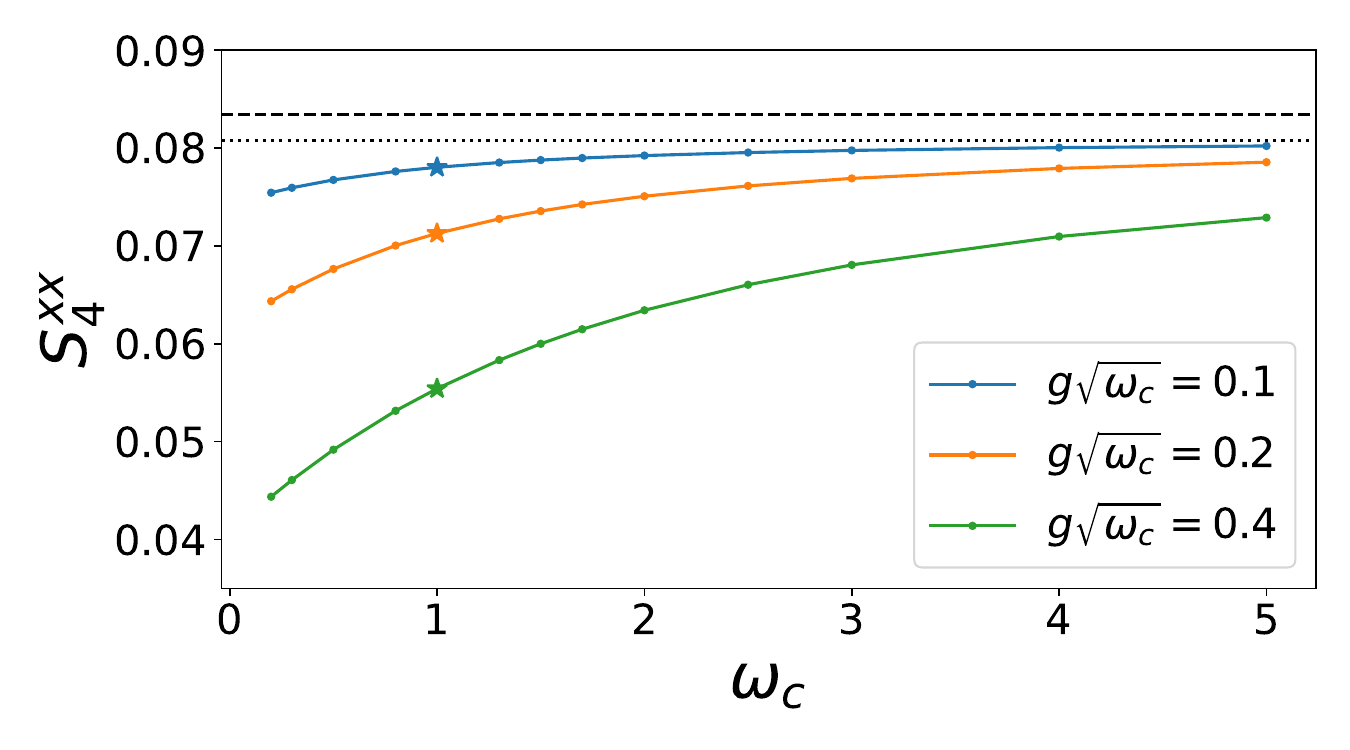}
        \put(20,100){(a)}
        \put(35,15){(b)}

        \end{overpic}
    \caption{DMRG results for the long wavelength static structure factor parameter $S^{xx}_4$ in the model $\hat{H}_\chi$ as a function of (a) the collective coupling $g\sqrt{N_e}$ ($\omega_c=1$) for different number of electrons $N_e$ at fixed $L_y=20$ and (b) as a function of the cavity frequency $\omega_c$ for different fixed coupling $g\sqrt{\omega_c}$ at $N_e=21$ and $L_y=16$. The black dashed line represent in both (a) and (b) the Haldane bound $s_4=(1-\nu)/24\nu$ valid for isotropic interactions and saturated by the Laughlin at $g=0$ in the thermodynamic limit, while the dotted line in (b) represent the Laughlin value ($g=0$) at the finite system size studied there.. Inset: Static structure factor as a function of momenta along $y$ for $g=0$ (red) and $g=1$ (blue) showing enhancement of the small wavevector part. The dashed red line represent the thermodynamic limit behaviour $S(q_y)=s_4 \;q_y^4$ of the Laughlin state. }
    \label{fig:S4x}
\end{figure}

In Fig. \ref{fig:S4x} we show the effect of cavity fluctuations on the long-wavelength properties of $S(\boldsymbol{q})$, in the simplified case of $\hat{H}_\chi$ where single particle potentials are absent. The behaviour for momenta along $x$ is still quartic and the proportionality factor $S_4^{xx}$ is plotted against the collective coupling $g\sqrt{N_e}$ in Fig. \ref{fig:S4x}(a). To extract it we look at the definition of $S(q_x)$ and expand for small $q_x$:
\begin{align}\label{eq:S4-xx}
S_4^{xx}=\lim_{q_x\rightarrow 0} \frac{1}{4!}\partial_{q_x}^4 S(q_x)=\frac{1}{4M}\sum_{k,k'} k^2 k'^2 \langle \delta \hat{n}_k \delta \hat{n}_{k'} \rangle,  
\end{align}
with $\delta\hat{n}_k=\hat{n}_k-\langle \hat{n}_k\rangle$.
Finite-size effects with both $N_e$ and $L_y$ are expected, and signaled by the discrepancy with the known value $s_4=(1-\nu)/24\nu$ at $g=0$ shown as a dashed black line. Nonetheless, the finite $g$ reduction is consistent and likely hold in the thermodynamic limit. The extraction of all three parameters in Eq.~\eqref{eq:S4_smallq} is hindered by strong finite size effects along $y$. This is highlighted in the inset of Fig. \ref{fig:S4x}, where we show the full dependence of $S(q_y)$, which cannot capture the $q_y^4$ dependence. Qualitatively we still see that correlations along $y$ are enhanced at finite $g$, as opposed to those on $x$. There is also an evident dependence on the cavity frequecy as highlighted in Fig. \ref{fig:S4x}(b). As the cavity frequency increases, keeping $g\sqrt{\omega_c}$ fixed, we find a monotonous descrease on the metric squeezing, i.e. the magnitude of the corrections to $S_4^{xx}$ decreases. This hints at the fact that for larger $\omega_c$ the effect of the cavity on the long-wavelength properties of the FQH is reduced, even though in Fig. \ref{fig:resistivity} we found a qualitative similar coupling regime for the loss of transport quantization at two quite different cavity frequencies.

Another key signature of a distorted metric is the shape of the correlation hole of the $G_2$, shown in Fig. \ref{fig:stripe_numerics}(c) for $g=1$ (white line) and compared with the circular one of the $g=0$ Laughlin case (blue line). As already noticed for other anysotropic FQH model wavefunctions \cite{QiuHaldane_prb2012_model}, we also find that the short distance behaviour of the $G_2(r)$ changes from $r^6$ to $r^2$ going from isotropic ($g=0$) to anisotropic cases ($g>0$). We also note that the effect at long wavelength shown in Fig.~\ref{fig:S4x} is greater in magnitude with respect to the reshaping of the correlation hole. As a rule of thumb, we expect the latter to be more sensible to short wavelength properties (electron-electron interactions) while the former to long-wavelength (cavity-mediated interactions). We suggest this to be related to the cavity-induced modifications of the magnetoroton dispersion (Sec. \ref{sec:spectral}), which are collectively enhanced ($g\sqrt{N_e}$) only around the graviton part $q\to 0$, while at finite $q$ seems only to see the single-particle coupling $g$. This is intuitively related to the fact that a uniform field gradient couples collectively to a uniform quadrupolar excitation, the graviton ($q\to 0$), and not to quadrupolar excitations at finite momenta, the rest of the magnetoroton dispersion ($q$ finite). 

We further remark that single-particle potentials, such as the cavity vacuum-induced Stark shift neglected so far, are also expected to shape the emergent geometry of the FQH liquid \cite{Haldane_prl2011_quantummetric}.

\subsection{Stripe instability}\label{sec:stripe_num}
\begin{figure*}
    \centering
    \begin{overpic}[width=0.305\linewidth]{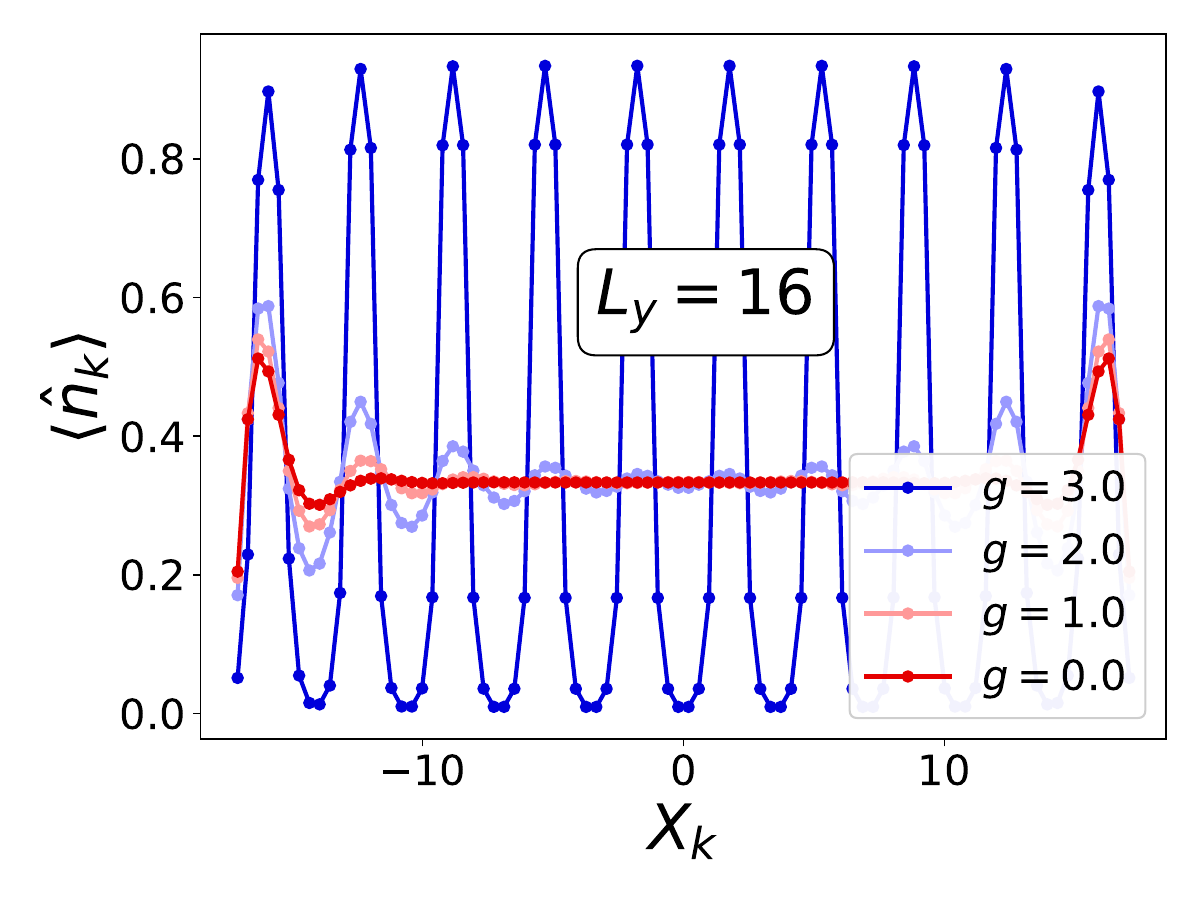}
    \put(20,75){(a)}
    \end{overpic}
    \quad
    \begin{overpic}[width=0.34\linewidth]{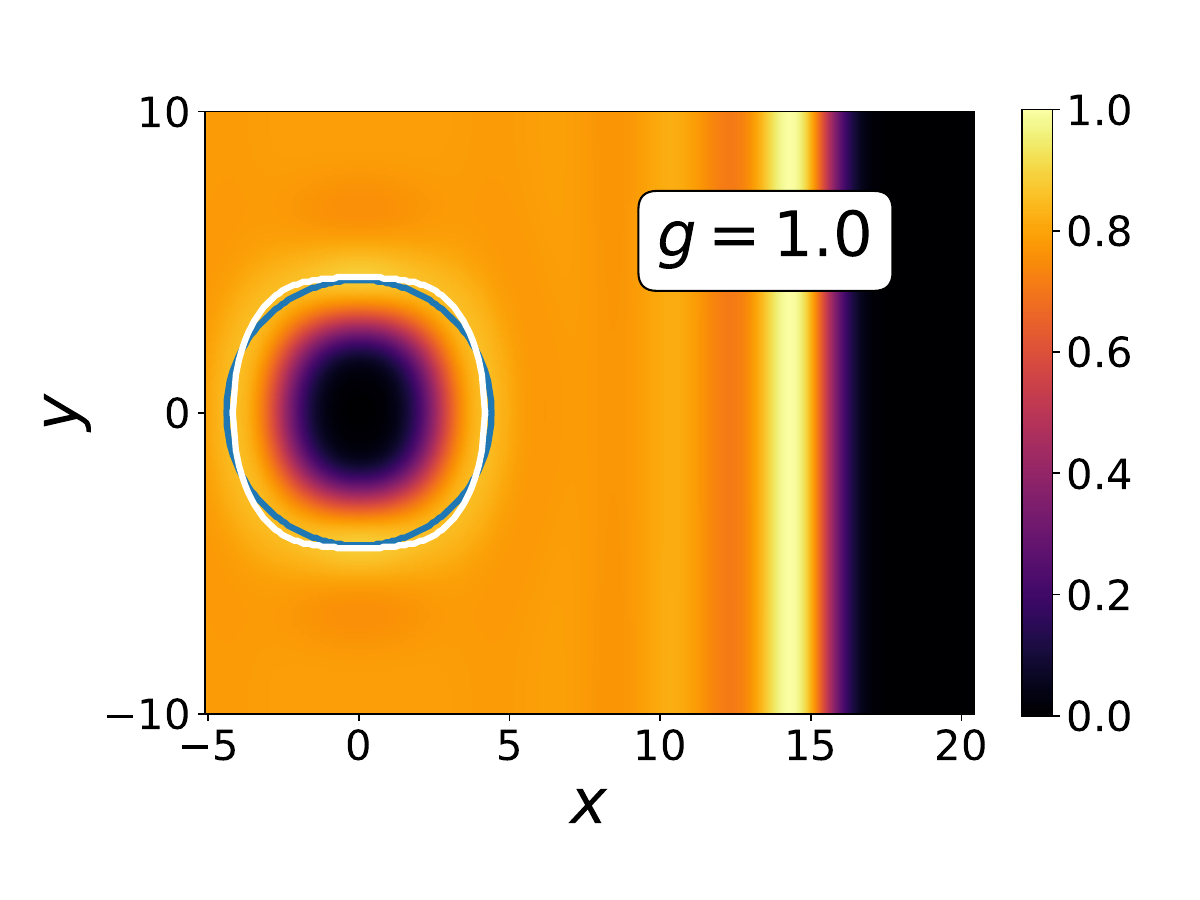}
    \put(20,70){(c)}
    \end{overpic}
    \quad
    \begin{overpic}[width=0.305\linewidth]{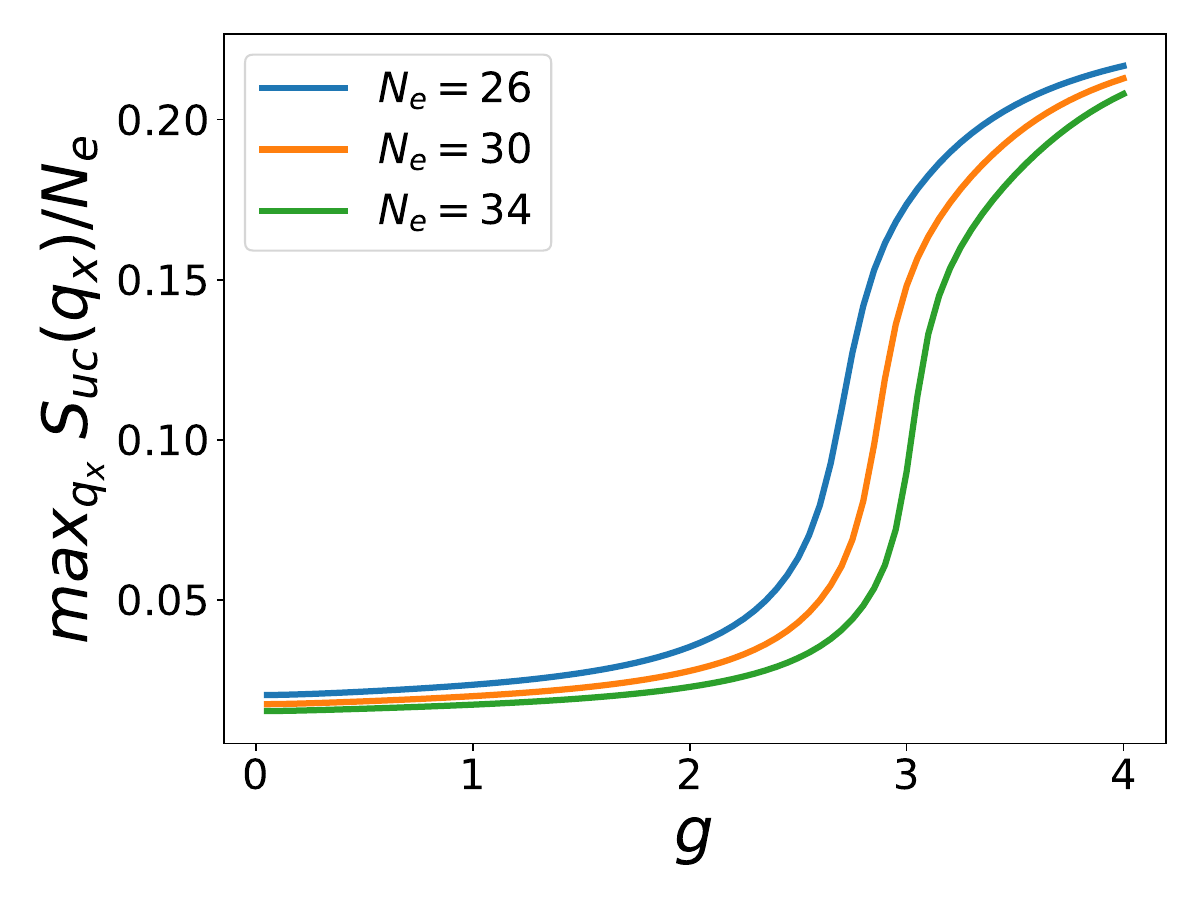}
    \put(20,75){(e)}
    \end{overpic}
    \medskip
    \begin{overpic}[width=0.305\linewidth]{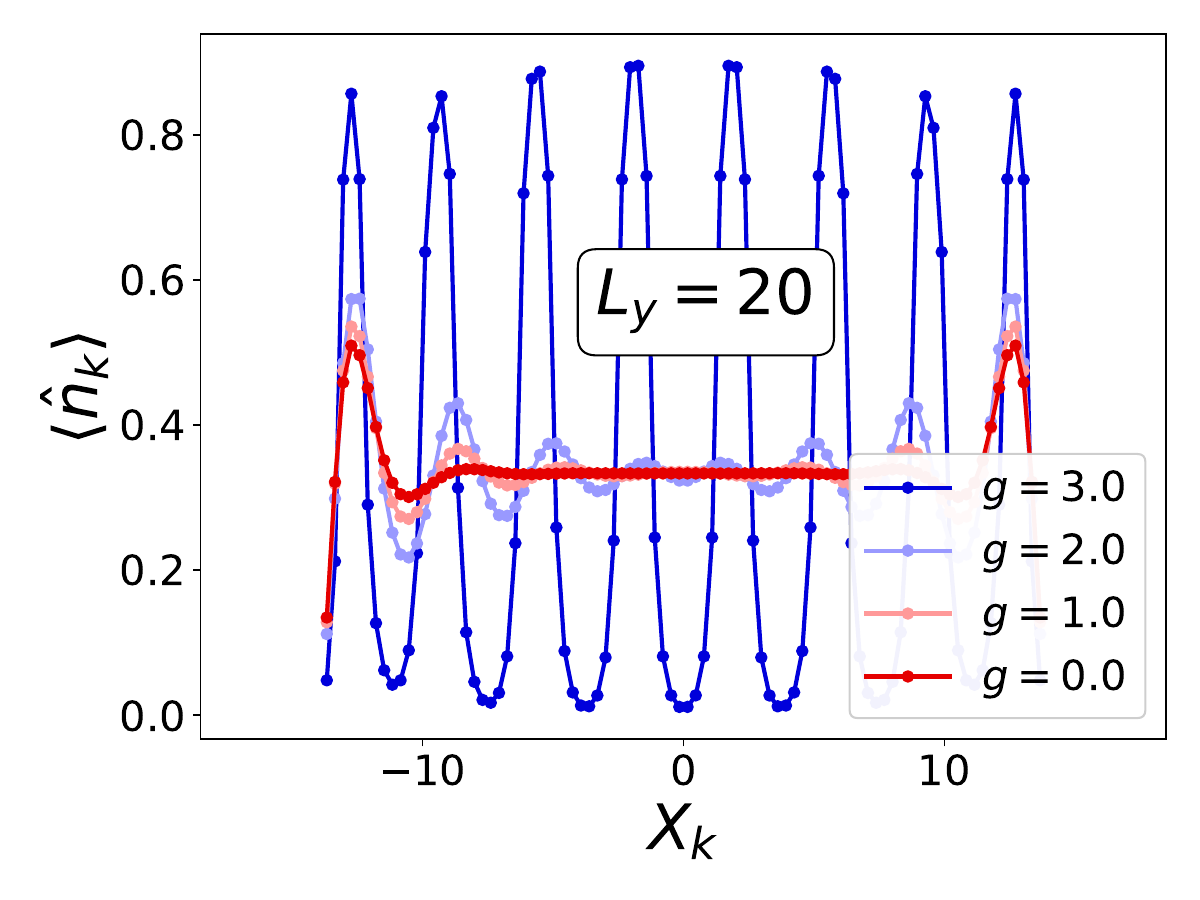}
    \put(20,75){(b)}
    \end{overpic}
    \quad
    \begin{overpic}[width=0.34\linewidth]{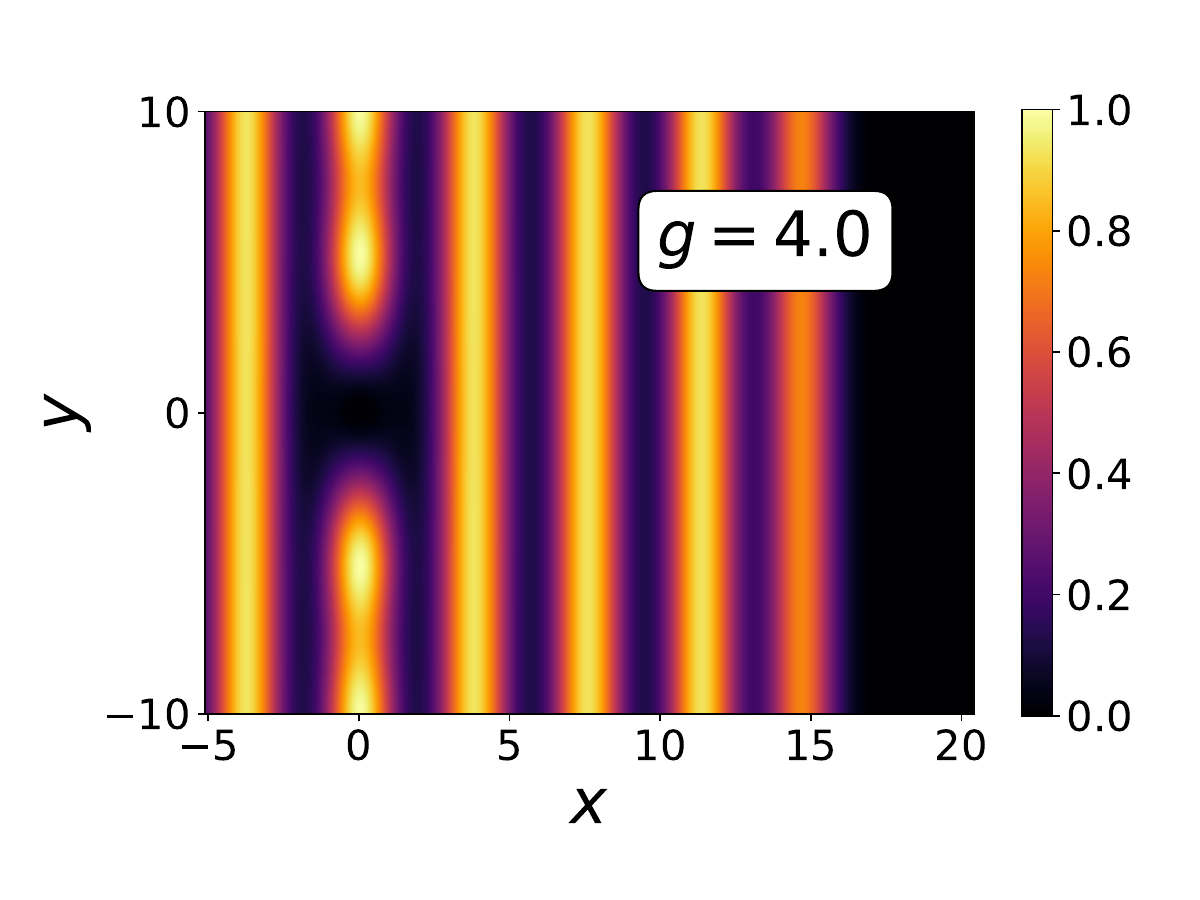}
    \put(20,70){(d)}
    \end{overpic}
    \quad
    \begin{overpic}[width=0.305\linewidth]{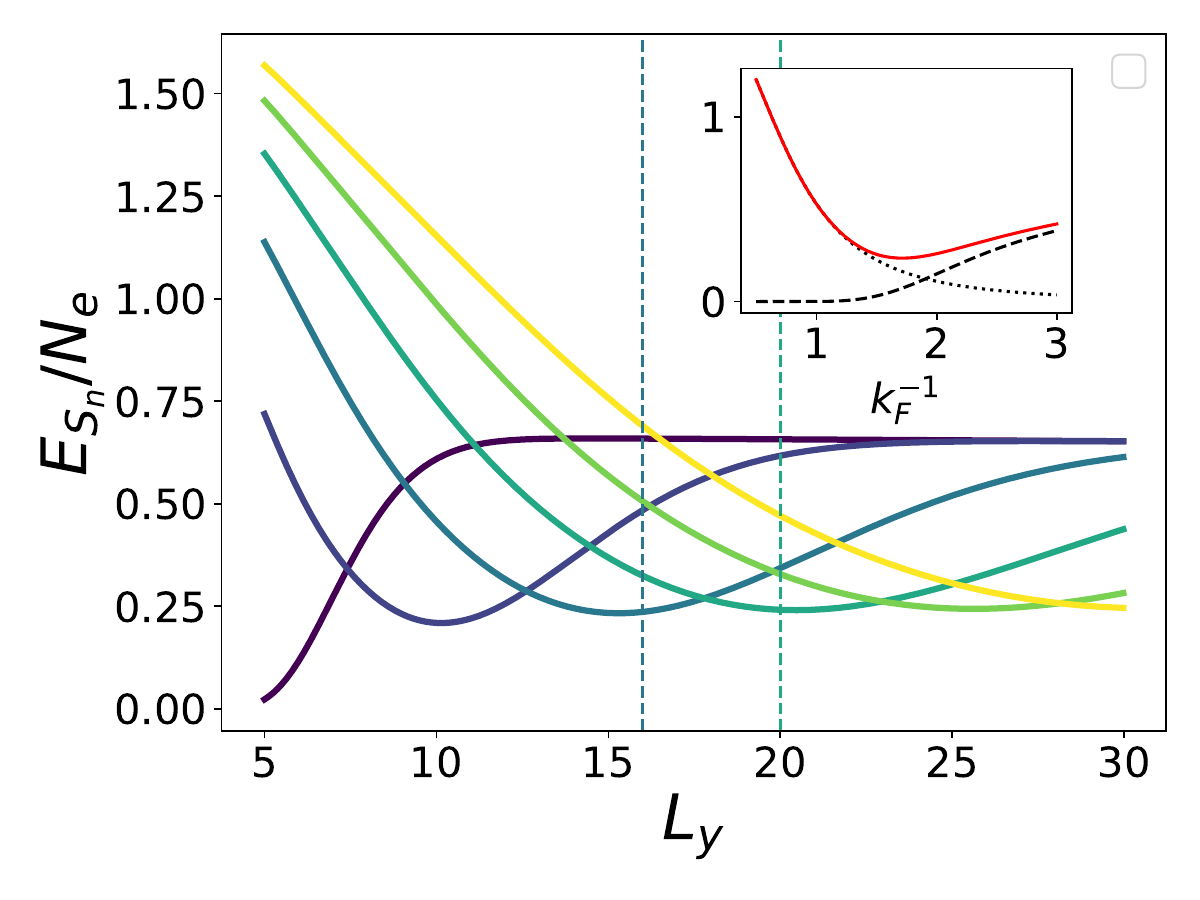}
    \put(20,75){(f)}

    \end{overpic}
    
    \caption{DMRG (a-e) and photon mean-field (f) results  in the ultrastrong coupling regimes in absence of vacuum-induced Stark shifts $\hat{H}_\chi$. Left panels (a,b) show the orbital occupancies from Laughlin state at $g=0$ (red) to stripe phase at $g=3$ (blue) for two circumferences of the cylinder $L_y=16$ (a) and $L_y=20$ (b) at fixed $N_e=30$. Note that at fixed number of particles the size of the system along $x$ reduces. Center panels (c,d) show the densisty-density correlations $G^{(2)}(0,\boldsymbol{r})$ in the two phases, at $g=1$ for the FQH phase in panel (c) and at $g=4$ for the stripe phase in panel (d). In the FQH phase (panel (c)) we also compare the shape of the correlation hole between $g=1$ (white line) and the Laughlin $g=0$ (blue line) by tracking the local maximum of the $G^{(2)}$. For both panels (c,d) $L_y=20$ and $N_e=34$. Panel (e) show the peak height of the static structure factor as defined in the main text and serve as an order parameter, $L_y=20$. Panel (f) show the mean-field energy density of a stripe state $\ket{S_n}$ (see main text) for differnt number of electrons $n$ per stripe (from $n=1$ dark blue to $n=6$ yellow) as a function $L_y$. The inset shows the thermodinamic limit ($L_y,n\to\infty$) value of the energy density as a function of $k_F^{-1}=2\pi n/L_y$. Vertical dashed lines mark the $L_y=16$ and $L_y=20$ used in the other panels.}
    \label{fig:stripe_numerics}
\end{figure*}

We now explore even stronger values of the coupling $g$ within the toy model $\hat{H}_\chi$ (Eq. \eqref{eq:Htot_chi}) with a uniform gradient and no single particle terms.

In Fig. \ref{fig:stripe_numerics} we present a sample of the numerical results obtained via DMRG simulations as discussed in Sec. \ref{sec:num_meth}. In order to get immediate insight on the state of the system we can look at the orbital density $\hat{n}_k$ in the left panels (a) and (b). The Laughlin state ($g=0$ red line) shows a quantized bulk density $\hat{n}_k$ with modulations only close to the edge due to the open boundaries. A strong bulk modulation instead appears around $g\simeq 2-3$ for both system sizes $L_y=16$ and $L_y=20$. The modulation in orbital space follows a different pattern for the two system sizes. In particular, we have that the number of electron per peak are $n_S=3$ and $n_S=4$ for system sizes $L_y=16$ and $L_y=20$ respectively. In order to understand this we recall that the distance between two neighboring orbitals decreases as $\Delta X_k=2\pi/L_y$. Hence to keep the distance between the peaks $\lambda_S$ independent of $L_y$, the number of particle per peak $n_S$ must increase. This constrains the generic behavior of $\lambda_S$ to be
\begin{equation}\label{eq:lambda-S}
    \lambda_S= \frac{2\pi}{L_y}3n_s(L_y),
\end{equation}
with the factor $3$ coming from the filling and where we expect  $n_s(L_y)\propto L_y$ in the thermodynamic limit.

In order to check the 2D stripe nature of the phase we show the density-density correlator $G^{(2)}(\boldsymbol{r}_1,\boldsymbol{r}_2)$ in the center panels (c) and (d) as defined in Eq.~\eqref{eq:g2_def}. In the FQH phase, $g=1$ panel (c), we find the characteristic correlation hole at short distances and a constant value at long distances as one should expect from a liquid state. Note that, as discussed in Sec. \ref{sec:geometry}, the anisotropic nature of the cavity mode reflects into an anisotropic correlation hole. To highlight this, two lines are shown which tracks the maximum of the $G^{(2)}$, one for the Laughlin at $g=0$ (blue line) and one for the $g=1$ case (white line) whose $G^{(2)}$ is actually plotted. In the strong coupling regime instead, $g=4$ panel (d), there is clear ordering on $x$ and absence of ordering on $y$, confirming the interpretation of the phase as a stripe phase. Weak density modulations on $y$ are present within the stripe, likely due to finite size effects and not to a true crystalline order in 2D.

In the right top panel (e) we show an electronic order parameter which tracks the instability. We consider the maximum of the disconnected static structure factor $S_{uc}(\boldsymbol{q})$ at finite momenta, expected to scale with $N_e$ in a charge density wave phase and not in a liquid phase. $S_{uc}(\boldsymbol{q})$ is defined as in $S(\boldsymbol{q})$, Eq.~\eqref{eq:staticSq_def}, but with the full guiding center density operator $\hat{\Bar{\rho}}(\boldsymbol{q})$ replacing fluctuations $\delta\hat{\Bar{\rho}}(\boldsymbol{q})$ in the definition. Panel (e) shows the normalized finite-momenta peak height of the disconnected static structure factor $\max S_{uc}(\boldsymbol{q})/N_e$ which is found around $\boldsymbol{q}\simeq(1.75 ,0)$. At fixed $L_y$ the transition point seems to shift towards higher values of $g$ with increasing $N_e$. However, a more careful investigation is hindered by commensurability effects on $y$ which likely renormalize the energy cost of the stripes for finite $L_y$. We note that considering the thermodynamic limit in the $x$ direction with an infinite MPS is not possible with the present cavity-matter structure. 

In order to better understand this instability, we propose a photon mean-field argument, detailed in Appendix \ref{app:PMF}. While we have also investigated a different cavity elimination scheme based on a large cavity frequency, i.e. a Schrieffer-Wolff transformation, this results in vanishing cavity corrections to the only-matter model up to second order in $g$ (see App. \ref{app:large-omegac-limit}) and hence cannot capture the instability. The photon-mean field instead can be seen as the limit case for $g\omega_c\gg \omega_c$ (or $\omega_c \to 0$ with $g\sqrt{\omega_c}\sim const$ as shown in Fig. \ref{fig:ent_spectrum}(d) ). In this limit, the state of the cavity has to be \textit{classical} to leading order (vanishing light-matter entanglement) as $\hat{H}_c \ll \hat{H}_{DP}$ and only one of the two non-commuting terms for the cavity survives.

Neglecting light-matter correlations we can write the following photon mean-field (PMF) Hamiltonian:
\begin{equation}
    \hat{H}_\mathrm{PMF} = \hat{H}_\mathrm{int} +  \omega_c g^2 \left[ \sum_k \frac{k^2}{2} \big(\hat{n}_k -  \bra{\psi} \hat{n}_k \ket{\psi}\big) \right]^2,
\end{equation}
where $\hat{H}_{int}$ is the electron-electron interaction (Haldane pseudopotential), $\ket{\psi}$ a generic electronic many-body state to be found self-consistently. The second term encodes the effect of cavity mediated interactions which in this form becomes quite simple to interpret. It exactly correspond to the variance of the quadrupole moment along $x$ in the LLL:
\begin{align}
    \hat{Q}^{xx}= \frac{1}{2}\sum_k X_k^2 \,\hat{n}_k\;
\end{align}
whose fluctuations define the $S_4^{xx}$ component of the FQH geometry (Eq. \eqref{eq:S4_smallq}). The squeezing of the geometry observed in Sec. \ref{sec:geometry}, i.e. the reduction of $S_4^{xx}$, is a consequence of penalizing quadrupole fluctuations in one direction. The energy of the FQH liquid, at the photon mean-field level, is indeed renormalized as:
\begin{equation}
    E_{FQH}/N_e\simeq \nu^{-1} S_4^{xx} \omega_c g^2
\end{equation}
The extreme limit where no fluctuations are present is represented by stripe states of the form:
\begin{equation}\label{eq:definition_stripe}
    \ket{S_n}=|{{\underbrace{0...0}_{n}\underbrace{1...1}_n\underbrace{0\dots0}_{n}}}\rangle\;
\end{equation}
which exactly minimize the cavity-mediated interaction energy. The energy cost of forming stripes is rather contained in the interaction energy $\hat{H}_{int}$, which in the case of Haldane pseudopotentials is shown in Fig. \ref{fig:stripe_numerics}(f) as a function of cylinder circumference for different number $n$ of electrons per stripe. In the thermodynamic limit ($L_y,n\to \infty$ with $L_y/n\to const$) the energy per particle is fixed $E_{S_n}/N_e\simeq 0.25$ and should be compared with the cavity mediated energy cost of the FQH state. Using the Laughlin state value for the quadrupole fluctuations $S_4^{xx}$, we get a crossing between $E_{FQH}$ and $E_{S_n}$ as a function of $g$ around a transition point:
\begin{equation}
    g_{PMF}^* \sqrt{\omega_c}\simeq 1\;.
\end{equation}
This value does not account for the reduction of $S_4^{xx}$ which indeed push the instability towards higher value of $g$, as observed in numerics, but correctly captures the fact that once we look at rescaled $g\sqrt{\omega_c}$ the instability arises at roughly the same value ($\sim 2$ for the volume studied in Fig. \ref{fig:resistivity}). Another ingredient which is missing and instead should favor the stripe state is the inclusion of fluctuations on top of the classical state depicted in Eq. \eqref{eq:definition_stripe}. Already the numerical results on the orbital occupation (Fig. \ref{fig:stripe_numerics}(a,b) ) show that at finite $g$ the state is not exactly $|S_n\rangle$. The nature of the phase at $g$ large but finite should be understood as an array of Tomonaga-Luttinger liquids, one for each stripe, with a sliding symmetry as elaborated in Appendix \ref{app:stripe_ft}. We also remark that all this discussion assumes a cancellation of single particle potentials, external potentials $W$ and cavity vacuum-induced Stark shifts $\mu_k$.

Still, within the assumption of cancellation of single-particle potentials,
we emphasize our results regarding
the stripe transition are not conclusive. In fact, such
a conclusive statement about the strong-coupling states
is very hard to formulate within computational methods
available (see Appendix \ref{app:excited_appendix}). The answer to this question likely depends on the inclusion of longer range Coulomb interactions and is left for future works. 

\subsection{Vacuum-induced Stark shift of quasi-particle energy}\label{sec:qp_ren}
\begin{figure}
    \centering
    \begin{overpic}[width=0.95\linewidth]{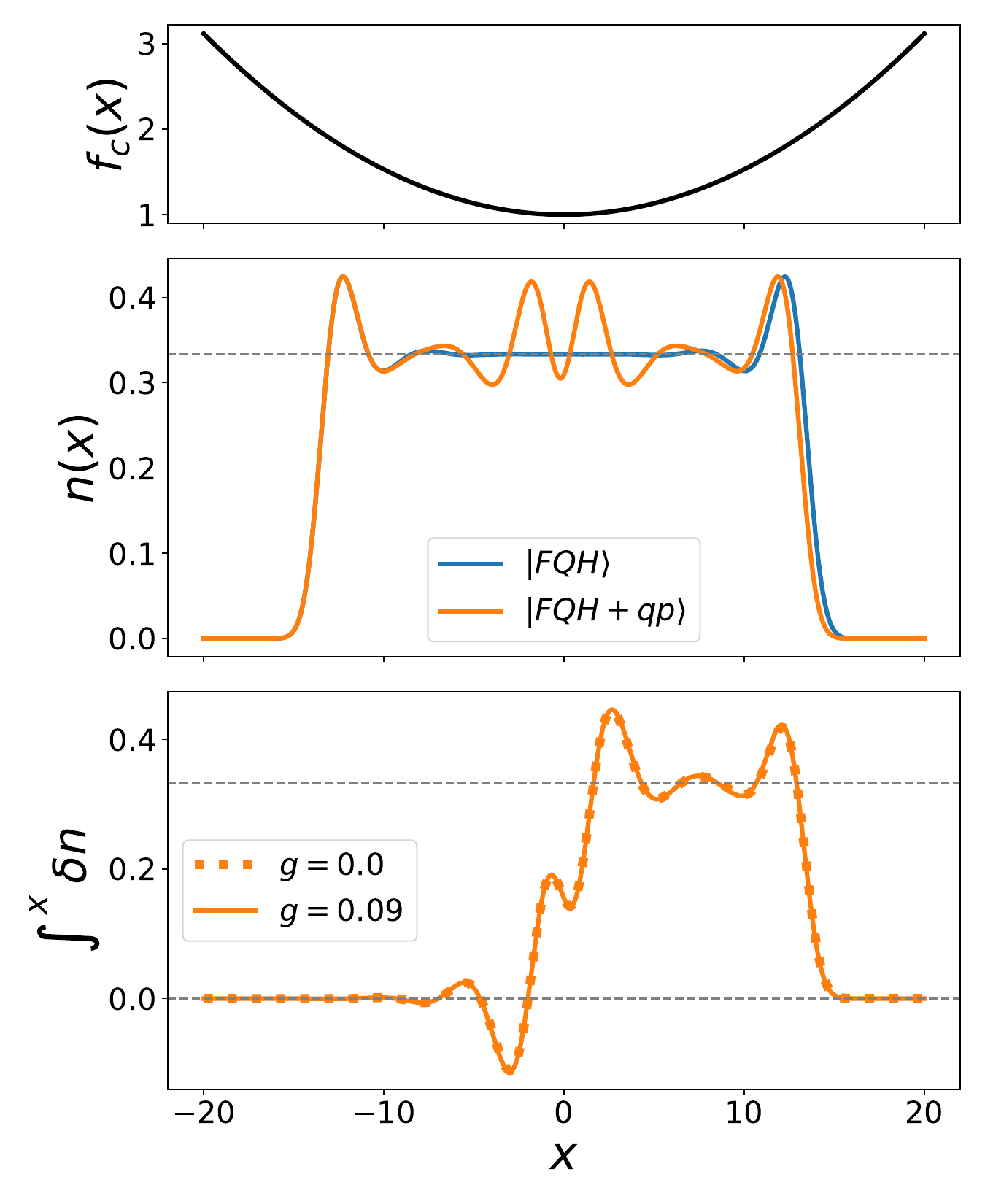}
        \put(15,85){(a)}
        \put(15,52){(b)}
        \put(15,38){(c)}
    \end{overpic}
    \caption{(a) Cavity mode profile under consideration. Note that the spatial dependence of the vacuum-induced Stark shift is governed by $f_c^2(x)$. (b,c) DMRG results for the density profile quasi-particle states in the full Hamiltonian $\hat{H}_{\chi,\mu}$ of Eq. \eqref{eq:Htot_chimu}. (b) Real space density of the FQH ground state (blu) and of the quasi-particle state (orange). (c) Integrated density difference between ground state and quasi-particle state at $g=0$ (dotted) and $g=0.09$ (full). Two jumps of $\pm1/3$ occur at $x\simeq0$ and $x\simeq L_x/2$ corresponding to a bulk quasi-particle and a quasi-hole pushed to the edge of the system. The cavity does not affect the quasi-particle state density. The system size is $L_y=16$ and $N_e=24$. Dashed lines mark the $1/3$ }
    \label{fig:qp_density}
\end{figure}

We now reintroduce the cavity vacuum-induced single particle Stark shift and study the full LLL Hamiltonian $\hat{H}_{\chi,\mu}$ of Eq. \eqref{eq:Htot_chimu}. Moreover, we change the spatial profile of the cavity mode, which is now a parabola with a minimum at the center, thus a gradient varying linearly from $-2g$ to $2g$, as shown in Fig. \ref{fig:qp_density}(a) and explained in Sec. \ref{sec:hamiltonian}. This mimics the experimental configurations of Ref. \cite{AppuglieseFaist_science2022} where the cavity field intensity is minimum in the middle and higher towards the edges, producing a potential which tries to confine charge.

In Fig. \ref{fig:qp_density} we show DMRG simulations for the ground state and charged quasi-particle states of $\hat{H}_{\chi,\mu}$. Quasi-particle states have been already studied via DMRG on both infinite \cite{Zalatel_prl2013} and finite \cite{Misguich_iop2021} cylinders. Converging such excited states is relatively convenient as they live in a different total momentum sector $K_y$. Note that in finite momentum sectors, states are not inversion symmetric with respect to $x=0$. In order to check the charged quasi-particle nature, we show the density of such states compared to the \textit{vacuum} FQH liquid (Fig. \ref{fig:qp_density}(b) ) and the integrated charge difference (Fig. \ref{fig:qp_density}(c) ) between the vacuum case and FQH state:
\begin{equation}
    \int^x \delta n= \int^x\dd x'\, \langle\hat{n}(x')\rangle_{FQH+qp}-\langle\hat{n}(x')\rangle_{FQH}
\end{equation}
This quantity shows a jump of $q=+1/3$ near the center of the cylinder where the quasi-particle is created and a corresponding jump of $q=-1/3$ towards the right edge where a quasi-hole is also created in order to ensure charge neutrality. The $g=0$ and $g=0.09$ case overlap, giving (inderect) evidence of a stable quasi-particle charge.

\begin{figure}
    \centering
    \begin{overpic}[width=0.95\linewidth]{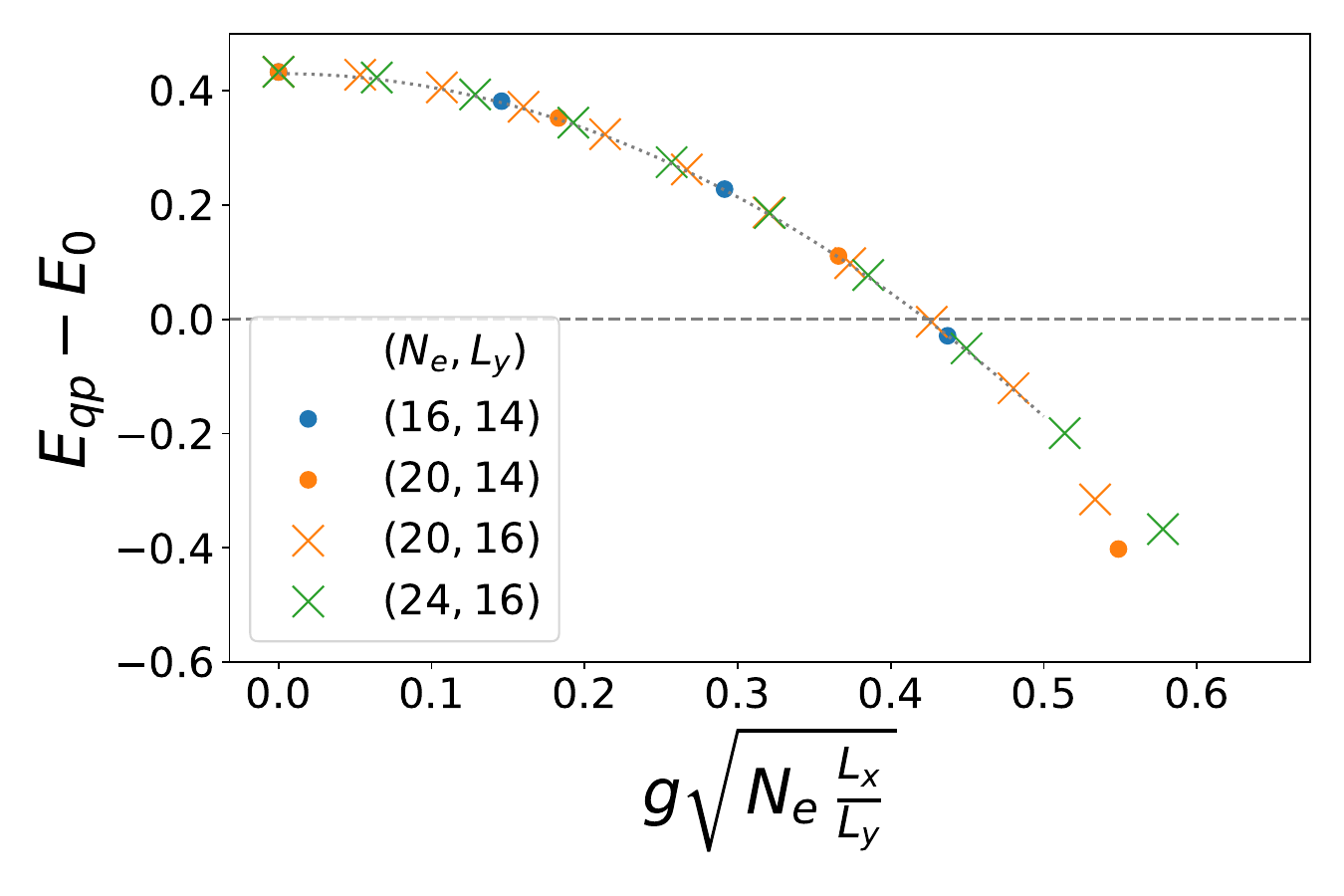}
    \end{overpic}
    \caption{DMRG results for the quasi-particle energy gap over the ground state as a function of a rescaled coupling for various system sizes $(N_e,L_y)$ in presence of the vacuum Stark shift (using $\hat{H}_{\chi,\mu}$). The data collapse indicate that the energy scale governing the renormalization is a collective one $\sim g\sqrt{N_e}$ and is the one of the vacuum Stark shift (see Eq. \ref{eq:Htot_chimu}). Dashed horizontal line marks zero energy, while the dotted line is a parabolic fit $y=0.43-2.4x^2$. } 
    \label{fig:qp_energy}
\end{figure}

The main effect of the cavity is in the energy of such quasi-particles, as shown in Fig \ref{fig:qp_energy}. Here we measure the energy of the quasi-particle state $E_{qp}$ described above relative to the ground state energy $E_0$, hence the gap $\Delta E_{qp}=E_{qp}-E_0$. The data collapsed performed with different system sizes shows how the energy scale controlling the quasi-particle energy renormalization is:
\begin{align}
    \Delta E_{qp}(g) -&\Delta E_{qp}(0)\simeq -2.4\; r^2\omega_c  g^2 N_e \simeq \nonumber\\
   & \simeq-2.4\;\frac{r^2\Omega^2}{\Delta_0} 
\end{align}
where $\Omega$ is the same collective Rabi frequency controlling the graviton-polaritons (Eq. \eqref{eq:rabi_frequency}) and $r=L_x/L_y$ is the aspect ratio. The prefactor is fitted from the numerical data.  The physical origin of the lower cost for creating quasi-particles is the vacuum-induced Stark shift introduced by the spatially dependent cavity mode. Indeed this favors having a charged quasi-particle created in the center where the field fluctuations are lower. At strong enough couplings, the pristine FQH liquid alone is not the total ground state anymore ($\Delta E_{qp}<0$) and quasi-particles can be created in the bulk. This can also be understood in terms of a renormalized area and hence filling $\nu$ of the system, which is now subject to an extra trapping potential. At even stronger couplings this single particle potential likely give rise to phase separation, as we find for small systems in Appendix \ref{app:LLL}. This scenario bears a lot of similarities with cold atom set-ups, where the role of trapping on many-body system has been extensively studied. Note that in our case, thanks to the incompressibility of the FQH liquid which is mantained in presence of the non-local cavity mode, creating a quasi-particle always cost a finite amount of energy.

\section{Experimental discussion}\label{sec:experimental_disc}
%\textcolor{red}{We can discuss the cases of non-uniform gradients, with density modulations only at the edges, and the the cavity spectral function in this case (still we see the polariton splitting). Then also the discussion that now is in the conclusions in red.}
In this section, we discuss in more detail the implications of our theoretical and numerical findings for relevant experimental conditions in solid-state systems.

\begin{figure}
    \centering
    \begin{overpic}[width=\linewidth]{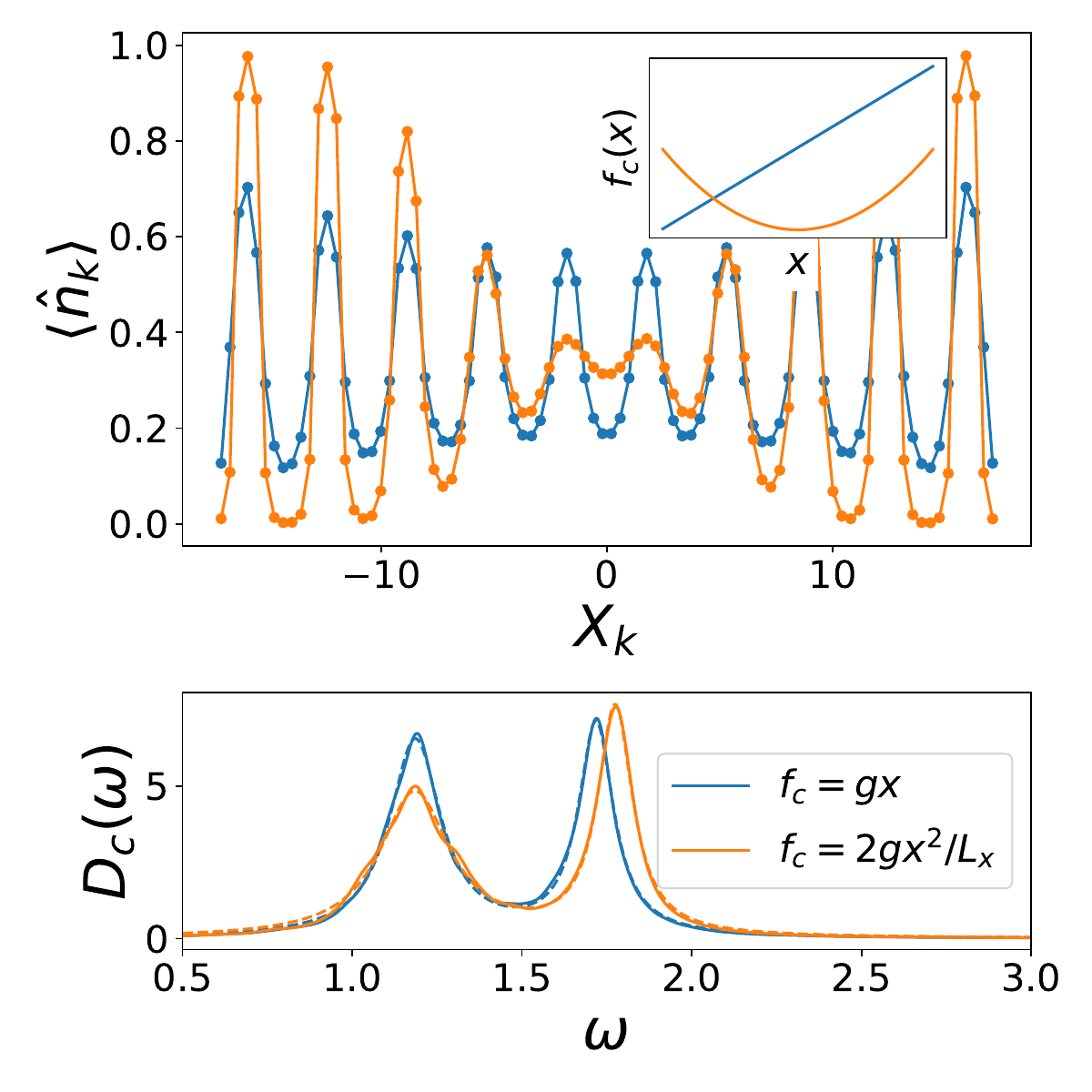}
        \put(20,100){(a)}
        \put(20,40){(b)}
    \end{overpic}
    \caption{Sample of numerical results comparing uniform and non-uniform cavity electric field gradients, using $\hat{H}_\chi$ without vacuum Stark shifts. In panel (a) we show DMRG results for the orbital occupation in the two cases of a uniform gradient (blue) and of a non-uniform gradient (orange). The corresponding mode profiles $f_c(x)$ are shown in the inset which correspond to a straight line $f_c=g x$ (uniform) and a parabola $f_c=2g x^2/L_x$ (non-uniform). Note that with this choice the non-uniform case has field gradients going form zero (bulk) to twice the value of the uniform case (edge). In (a) $N_e=30$,$L_y=16$ and $g=2.5$ so that the uniform case is unstable towards a uniform modulation but the non-uniform case clearly show two different behaviours in the bulk and at the edge. In panel (b) we show the cavity density of states in the two cases obtained via TDVP which still show the formation of graviton-polaritons with fitted rabi splittings $\Omega=0.53$ and $\Omega=0.59$ for uniform and non-uniform case respecively. In (b) $N_e=15$, $L_y=10$ and a smaller $g=0.1$ so that the ground state is still a FQH liquid. }
    \label{fig:nonuniform}
\end{figure}

\paragraph{Non-uniform gradients.---}
The choice of studying uniform gradients has simplified our analysis so far as it enabled to characterize a simpler uniform effect of the cavity mode. However in realistic scenarios \cite{AppuglieseFaist_science2022} this is not in general the case. In particular the vacuum Stark shift effect discussed in Sec. \ref{sec:qp_ren} will depend on the structure of the cavity mode. On the other hand the formation of graviton-polariton is not hindered by the inclusion of smoothly varying field gradients (inset of Fig. \ref{fig:nonuniform}(a)), as we explicitly verify in Fig. \ref{fig:nonuniform} (b). The blue line represent the reference uniform linear gradient (i) while the orange one is a parabolic profile (ii). Note that we choose to compare two cases such that the ``average" gradient is the same, namely in the case (ii) the gradient is $0$ in the bulk and twice the uniform value of case (i) at the edge. The bigger Rabi splitting observed in the non-uniform case can be rationalized by noting that the effective coupling constant is likely averaged as $g_\mathrm{eff}\simeq \sqrt{\int \dd x\, g^2(x)}$, thus giving two slightly different Rabi splittings in the two cases. For completeness we also show that the stripe instability appearing in absence of vacuum Stark shifts ($\hat{H}_\chi$) depends on the local value of the electric field gradient (Fig. \ref{fig:nonuniform}(a)), possibly leading to an inhomogeneous phase.

We further note that the assumption of a fixed mode function $f_c$ is also an approximation, which implicitly derive from the single-mode restriction of QED. Setting a precise limit of validity for this approximations is an open problem for most of cavity QED set-ups which try to achieve non-perturbative regimes, and is beyond the scope of this work. 

\paragraph{Energy scales.---} We now verify that a resonance condition between the typical energy scales of the electronic component (interaction $V$) and of the cavity (frequency $\omega_c$) can be obtained in realistic systems. The strength of interactions are controlled by the Coulomb energy $E_C=e^2/4\pi\epsilon_0\epsilon_r l_B\simeq 56 \, \mathrm{meV} \sqrt{B[\mathrm{T}]}/\epsilon_r $, which for a typical GaAs quantum well (with $n_e= 10^{11}\,\mathrm{cm}^{-2}$ and $\epsilon_r= 13$) at filling $\nu=1/3$ (with $l_B=\sqrt{\nu/2\pi n_e}\simeq 7\,\mathrm{nm}$ and $B=\hbar c/e l_B^2\simeq 12\,\mathrm{T}$) give $E_C\simeq 14 \,\mathrm{meV}$. The energy of the collective modes is then a fraction of the Coulomb energy. In Ref. \cite{Liang_nature2024_graviton} the graviton energy has been measured as $\Delta_0 \simeq0.05 E_C\simeq 0.65\,\mathrm{meV}$ for $\nu=1/3$ but the precise value of the prefactor likely depends on the experimental details such as the well thickness and the strength of disorder ~\cite{Rosales_prl2021_energygaps}. 
On the cavity side, split-ring resonator devices with similar frequencies $\omega_c\simeq 2\pi \cdot 0.1-1\mathrm{Thz}\simeq 0.4-4\,\mathrm{meV}$ have been realized. In order to observe the polariton anticrossing (Fig. \ref{fig:Acav}), direct tunability of the cavity frequency~\cite{paulillo2014circuit} or a full series of cavity devices covering a range of frequencies $\omega_c$ are likely to be needed.

In standard semiconductor systems there is in fact very little tunability of the frequency of the matter excitations. Changing $E_C$ via a change in the magnetic field $B$ imply also a corresponding change in the filling $\nu$, which is expected to destabilize the targeted FQH state \cite{Pinczuk_prl1993collectiveobs,Liang_nature2024_graviton}. We suggest that this issue may be resolved  using graphene samples which allow for tuning the electron density $n_e$ via a suitable gate \cite{Bolotin_nat2009_fqhgraphene}, so to achieve a tunable interaction energy $E_C$ at a constant filling $\nu$.

\paragraph{Coupling strength.---} The other important discussion is on the magnitude of the dimensionless coupling constant $g$ used in this work. To do so we remind the definition of the coupling:
\begin{equation}
    g=e \frac{E_c l_B^2}{\omega_c}  \partial_x f_c(x)\; ,
\end{equation}
where $f_c$ is the mode function, and $E_c=\sqrt{\hbar\omega_c/(2\epsilon_0 V_\mathrm{mode})}$ is the strength of vacuum fluctuations which account for the mode confinement.

As a first example, we consider the micron-sized split-ring resonator device used in Ref.~\cite{AppuglieseFaist_science2022}. We extract an average electric field gradient of roughly $\partial_x E^x_c= E_c \partial_x f_c(x) \simeq (1 \mathrm
{V/m} )/ 10 \mathrm{\mu m} =10^5 \, \mathrm{V/ m^2}$ in relevant sample area $S\simeq 40\mathrm{\mu m} \times 200 \mathrm{\mu m}$ and a cavity frequency $\omega_c=2\pi \times 140\,\mathrm{GHz}$.\footnote{Higher field gradients by three orders of magnitude have been recently reported in Ref. \cite{enknerfaist_2024_fqhenhanced}. } Considering again $n_e=10^{11}\,\mathrm{cm}^{-2}$ and $\nu=1/3$ which give $l_B=7\,\mathrm{nm}$, we get:
\begin{equation}
    g \simeq \frac{10^5 \,\mathrm{eV/m^2}}{2\pi \times 140 \,\mathrm{GHz}} (7\, \mathrm{nm})^2 \simeq 10^{-9}.
\end{equation}
The number of particle in the estimated sample area $S$ is $N_e=n_e S \simeq 8\times 10^6$, giving a rather small collective coupling $g\sqrt{N_e}\simeq 3\times 10^{-6}$. Using Eq.~\eqref{eq:rabi_frequency} for the Rabi frequency and reminding that the adimensional coefficient $\tilde{\gamma}_0$ is of order unity, we would get:
\begin{equation}\label{eq:rabifrequ_estimate}
    \frac{\Omega}{\omega_c} = 2g\tilde{\gamma}_0\sqrt{N_e}\simeq 6\times 10^{-6}.
\end{equation}
A much stronger value of the coupling can be obtained using the nano-cavities of Ref.~\cite{Keller_nanolett2017_strongcoupling}: here, the tighter in-plane confinement of the electric field over distances on the order of a few $100$~nm allows for a dramatic enhancement to the quadrupolar coupling to the graviton mode. The distance between capacitor plates $L^c_x$ roughly controls the volume of the mode as $V_{mode}\sim (L_x^c)^3$, assuming a similar size of the resonator on $y$. While the reduced effective surface of the mode $(L_x^c)^2$ is compensated by an analogous reduction of the maximum number of electrons that can be fit in the cavity region $N_e \sim n_e (L_x^c)^2 $, the reduction in the thickness of the cavity mode orthogonally to the plane $L_z\sim L_x$ strongly enhances field gradients. In particular we can expect the collective Rabi frequency of Eq. \eqref{eq:rabifrequ_estimate}, controlled by field gradients, to be proportional to $L_x^{-3/2}$; contrary to the case of Landau polaritons and cyclotron transitions where the collective Rabi frequency is controlled by the electric field intensity \cite{HagenmullerCiuti_prb2010_landaupolaritons,Ciuti_prb2021} hence scaling as $L_x^{-1/2}$. The larger value of the in-plane gradients are then expected to bring the graviton-polariton Rabi frequency up to values around $\Omega/\omega_c\sim 0.005$.

Given the rather small value of the quality factor of state-of-the-art micro- and nano-cavity devices $Q=\omega_c / \Gamma \simeq 10$~\cite{AppuglieseFaist_science2022,Keller_nanolett2017_strongcoupling}, these designs still fall in the weak coupling regime $\Omega/\omega_c<Q^{-1}$ but the active on-going progress in the fabrication technology of high-Q cavities~\cite{messelot2020tamm}
 make us optimistic about entering the strong-coupling regime with well-resolved graviton-polaritons in a next generation of samples. Considering other type of resonators could be beneficial as well as a more ad-hoc design for the resonators which enhances field gradients. In this respect we remark that a strictly uniform gradient is not necessary and also slowly varying gradients couple to the $q\simeq 0$ collective mode. 

Then we also mention that slightly greater couplings can be also be achieved by looking at more dilute FQH liquids where, at a given filling $\nu$, one has a larger $\ell_B$. For example reducing the electronic density of the quantum well gives $g\sqrt{N_e}\propto l_B^2 \sqrt{n_e} \propto 1/\sqrt{n_e} $ or focusing on higher fractions like $\nu=n+\Bar{\nu}$ (with $n$ integer and $\Bar{\nu}<1$) gives $g\sqrt{N_e^{\text{eff}}}\propto \nu \sqrt{N_e\Bar{\nu}/\nu}$ with $N_e^{\text{eff}}$ the electrons available in the last partially filled LL. 

\paragraph{Beyond single cavity mode. ---}
As a last remark we discuss the possible effect of a more complex electromagnetic enviroment, expecially related to physics discussed in Sec. \ref{sec:stripe}. Without entering too much into the details, we point out two different contributions that are neglected in our modelling. The first is the presence of higher frequency LC resonances or other plasmonic modes which and are expected to give contributions similar to what has been discussed so far. These will certainly contribute to vacuum Stark shift effects as well as enter in the determination of the FQH emergent geometry. The second important one is the modification of Coulomb forces, both at the single particle and many-body level \cite{BlazquezRabl_prl2023}. The former will have a similar role as the vacuum Stark shift of the LLL. The latter has actually been automatically neglected by considering an electron-electron interaction, the first Haldane pseudopotential, independent from the cavity setting. Changing the precise form of the only-matter interaction potential could provide a nice handle on the isotropy of the quantum metric in the underlying FQH liquid \cite{Haldane_prl2011_quantummetric} as well on the magnetoroton dispersion or even drive itself instabilities in higher LL \cite{Kumar2022prb_stripes}.

\section{Conclusions and outlook}\label{sec:conclusions}
In this work we have developed a theoretical framework, readily generalizable to experiments, in which it is possible to understand key effects of confined electromagnetic modes on lowest Landau level physics. Importantly, together with the proposed QED model, we have discussed a tensor network architecture which allows for an in depth study of ground state and spectral properties for large number of particles.

This allowed us to uncover many aspects of the interplay between quantum light and fractional quantum Hall physics. Here we summarize and discuss the main points:
\begin{itemize}
    \item Constant cavity fields are decoupled from intra-Landau level correlations \cite{Kohntheorem} in absence of an external potential. The coupling strength of the quantum light-matter interaction is directly proportional to field gradients and quadrupole moment of the electron fluid. Ref. \cite{AppuglieseFaist_science2022} reported a higher resilience of FQH physics to cavity vacuum fluctuations with respect to the IQH effect. This can be traced back to the smaller value of the effective Rabi frequency, Eq. \eqref{eq:rabifrequ_estimate}, coming from relatively weak field gradients.
    \item The quantized Hall response of the FQH liquid is stable against non-local cavity fluctuations. This theoretical results for FQH states is parallel to recent precise measurements of quantized Hall resistivity in the integer quantum Hall regime \cite{enkner_2023_vonklitzing}. 
    \item A new entanglement structure in hybrid quantum Hall states is found. The role of quantum light is to introduce a ``band" of chiral Luttinger liquids multiplets, each with an approximately quantized photon number and separated by a finite entanglement polariton gap.
    \item We predict the formation of  graviton-polariton modes, describing the hybridization of the long-wavelength magnetorotons with cavity photons. The Rabi frequency is collectively enhanced and is directly proportional to the quadrupole matrix element of the gravitons. We confirm numerically the presence of a polariton doublet in the cavity density of states, smoking gun of the strong coupling regime. 
    \item Cavity vacuum fluctuations can squeeze the FQH geometry \cite{Haldane_prl2011_quantummetric}. The latter can be thought as a hidden variational parameter for the FQH liquid that is thus able to adapt to the long-range anisotropic interactions induced by the cavity. 
    \item Off-resonant cyclotron transitions induce a vacuum Stark shift of the LLL independent of the cyclotron frequency $\omega_B$. This can be used to locally renormalize the energy cost of quasi-particles, provided that the cavity mode is non-uniform. 
    \item At electric field gradients that are strong at the single-particle level and assuming a cancellation of single-particle potentials, we find the FQH phase to be unstable towards sliding Luttinger liquids with strong density modulations in the cavity field gradient direction. The instability depend on the local value of the electric field gradient, thus strong non-uniform gradients give rise to non-uniform states.
\end{itemize}
The simple but rich scenario we discussed here gives a proof of principle on how cavity vacuum fields can be used to probe and control intra-Landau level correlations. This hopefully lays the ground for more detailed and complex many-body investigations, as well as serves as a motivation for a more careful analysis on which experimental cavity set-up best adapts to the need of strong field gradients. Below we give few possible interesting future directions.

As a first example, we remark that the peculiar chiral nature of the graviton excitation \cite{Liang_nature2024_graviton} has not entered our discussion, mainly because of the choice of the cavity mode. However, a set-up with a pair of chiral cavity modes have been recently demonstrated to be in a chiral-selective strong coupling regime with cyclotron transition \cite{andberger_arxiv2023_chiralterahertz}. We expect this set-up to be sensible to the chirality of the gravitons. This will also likely play a role in the FQH regime of a recently demonstrated optical pumping experiment \cite{Hafezi2023opticalpumping}, paving the way for a coherent control of FQH excitations. Tuning the cavity-modes can also be interesting to explore periodic spatial modulation with periodicity on the scale of few magnetic lengths, ranging from approximately $10$nm to $100$nm, achievable with subwavelength set-ups. This will results in coupling to finite momentum magnetoroton excitations. Another effect, likely important away from perfect fillings, is the interplay between the cavity electric field and disorder which allows for a non-trivial coupling to the constant part of the field.

From a more theoretical perspective, it would be interesting to extend our treatment to more complex FQH phases and the effect of a non-local mode on non-Abelian topological ordered states, such as the Moore-Read state. We can anticipate for example that the presence of multiple magnetoroton branches will enrich the entanglement structure on one side and the landscape of polaritons on the other, possibly giving distinct signatures to the nature of the phases. Moving away from gapped liquids; it is also tempting to imagine scenarios where, starting from gapless states such as the $\nu=1/2$ composite Fermi sea, the cavity can be used as a handle to induce instabilities towards desired states. It would also be intriguing to connect some of our findings -- which are inherently related to bulk properties -- to recent works focusing on edge dynamics~\cite{winterZilberberg_arxiv2024}.

Finally, from an experimental viewpoint, one immediate follow up is the extension of our treatment to different setups, including atomic systems in cavities and other solid state platforms. In those contexts, it will be interesting to exploit the strong matter-light correlations we report as a mean to probe so far unexplored features of quantum Hall states, such as their entanglement structure, and eventually understand whether such correlations can be utilized to access and control the functioning of topological quantum memories.

\textit{Note added:} soon after posting this manuscript, an experiment/theory preprint \cite{enknerfaist_2024_fqhenhanced} was also posted on arXiv, investigating fractional quantum Hall samples coupled to cavity modes with strong field gradients, in the off-resonant cavity regime. 

\section*{Acknowledgments} 
We thank G.M. Andolina, A. Asenjo-Garcia, D. Chang, G. Chiriac\'o, D. De Bernardis, O. Dmytruck, D. Fausti, R. Fazio, C. Mora, A. Nardin, R. Colombelli, M. Schirò, and O. Zilberberg for discussions. 
M.D. was partly supported by the QUANTERA DYNAMITE PCI2022-132919, by the EU-Flagship programme Pasquans2, by the PNRR MUR project PE0000023-NQSTI, and by the PRIN programme (project CoQuS). H.B.X. was supported by MIUR Programme FARE (MEPH). I.C. acknowledges financial support by the PE0000023-NQSTI project by the Italian Ministry of University and Research, co-funded by the European Union - NextGeneration EU, and from Provincia Autonoma di Trento (PAT), partly via the Q@TN initiative. 
Z.B., I.C., and M.D. would like to thank the Institut Henri Poincaré (UAR 839 CNRS-Sorbonne Université) for hospitality, and the group of C. Ciuti for inspiring discussions during the stay. 
T.C. was partly supported by the Young Faculty Initiation Grant (NFIG) at IIT Madras (Project. No. RF24250775PHNFIG009162).
T.C. acknowledges the support of PL-GRID infrastructure for the computational resource for a part of the numerical simulations.
Z.B. thanks IIT Madras for kind hospitality through the Centers of Excellence, QuCenDiEM (Project No. SP22231244CPETWOQCDHOC) and CQuICC (Project No. SP22231228CPETWOCQIHOC).

MPS-based numerical calculations have been performed with the help of ITensors.jl~\cite{itensor22} and TenNetLib.jl~\cite{tennetlib} libraries.

\appendix
\section{Test of the LLL truncation}\label{app:LLL}
In this Appendix we study more in detail the LLL truncation by directly including a finite number of Landau Levels $n_{LL}>1$.
We are going to compare results obtained with the two Hamiltonians used in the main text for a single LL $n_{LL}=1$, namely $\hat{H}_{\chi}$ (Eq. \eqref{eq:Htot_chi}) and $\hat{H}_{\chi,\mu}$ (Eq. \eqref{eq:Htot_chimu}), with the difference between the two being the single-particle cavity-induced vacuum Stark shift. For the sake of simplicity we are going to consider a mode function of the form $f_c(x)=C_1 x$ without an homogenues part. Whenever we indicate a number of $n_{LL}>1$ we imply an Hamiltonian of the form:
\begin{align}
    \hat{H}^{n_{LL}}=& \hat{H}_{int}^{n_{LL}} + \hat{H}_0^{n_{LL}} + \omega_c \hat{a}^\dagger \hat{a} +\nonumber \\
    &+i\omega_c(\hat{a}^\dagger-\hat{a}) \hat{P}^{n_{LL}} + \omega_c\left(\hat{P}^{n_{LL}} \right)^2
\end{align}
where the supscript $n_{LL}$ implies a projection over the Hilbert space of $n_{LL}$ LLs. Note that this form  already for $n_{LL}=2$ incorporate all relevant light-matter matrix elements of the LLL in the $n_{LL}\to \infty$ limit, but not all interaction matrix elements.

\begin{figure}
    \centering    
    \begin{overpic}[width=0.45\linewidth]{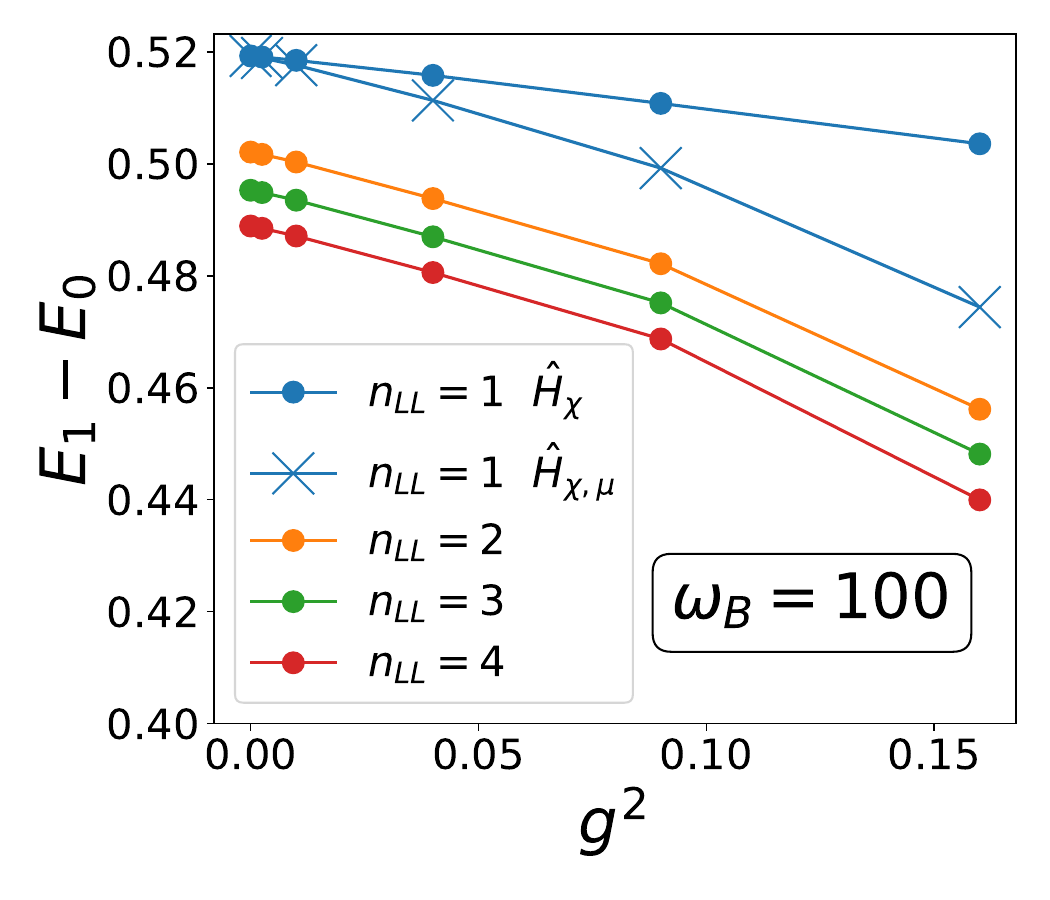}
    \put(10,90){(a)}
    \end{overpic}
    \begin{overpic}[width=0.45\linewidth]{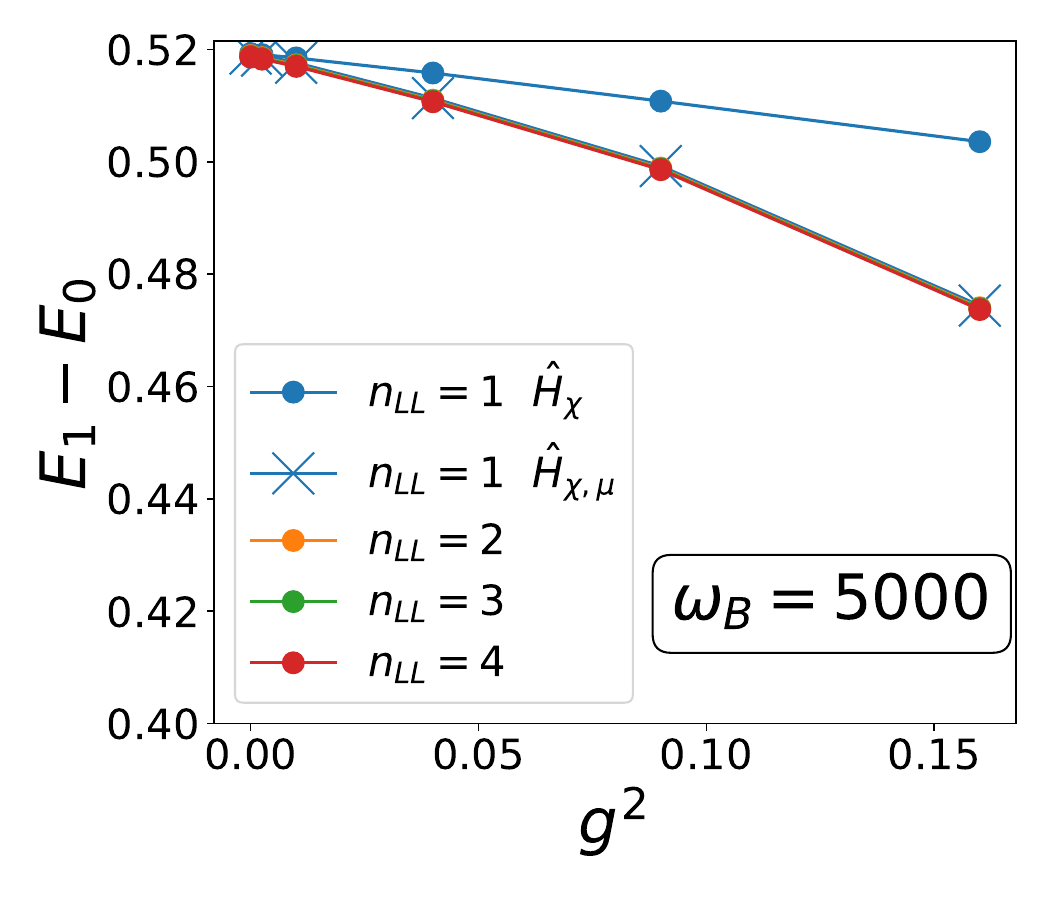}
    \put(10,90){(b)}
    \end{overpic}
    \caption{Energy gap from ED as a function of $g^2$ for different Landau Level truncations $n_{LL}$. The case of $n_{LL}=1$ shows both scenarios with (cross) and without (dots) cavity-induced vacuum Stark shifts. The cyclotron frequency is $\omega_B=100$ and $\omega_B=5000$ in panel (a) and (b). Note that the Landau Level truncation affect the energy gap also at $g=0$, with corrections of order $1/\omega_B$. At $g>0$ instead the discrepancy between different $N_{LL}=1$ and $N_{LL}$ remains finite (order $g^2$ at small $g$) if the cavity-induced vacuum Stark shifts ($\mu_k$) are ignored. Other parameters $N_e=4$, $L_y=10$ and $\omega_c=1$ . }
    \label{fig:gap_LLconv}
\end{figure}

In Fig. \ref{fig:gap_LLconv} we show the energy gap to the first excited state as a function of the coupling $g$ (linear gradient) for two cyclotron frequencies $\omega_B=100$ (a) and $\omega_B=5000$ (b). Already at $g=0.$ we have that higher LL create a discrepancy between different $n_{LL}$. It is indeed known that truncating electron-electron interaction inter-LL matrix elements has a slow convergence with $n_{LL}$, sometimes even giving physically different results by passing from $n_{LL}=2$ to $n_{LL}=3$ \cite{Zalatel_prb2015_LLmixing}. This discrepancy at $g=0.$ is in any case well controlled by $1/\omega_B$ and indeed is heavily reduced in panel (b) where $\omega_B=5000$. At finite $g>0$ we instead clearly see that the model without vacuum-induced Stark shifts $\hat{H}_\chi$ does not recover higher $n_{LL}$ models, while $\hat{H}_{\chi,\mu}$ falls on top. Note that the discrepancy between $\hat{H}_\chi$ and $\hat{H}_{\chi,\mu}$ is of order $g^2$ as expected.   

\begin{figure}
    \centering    
    \begin{overpic}[width=0.45\linewidth]{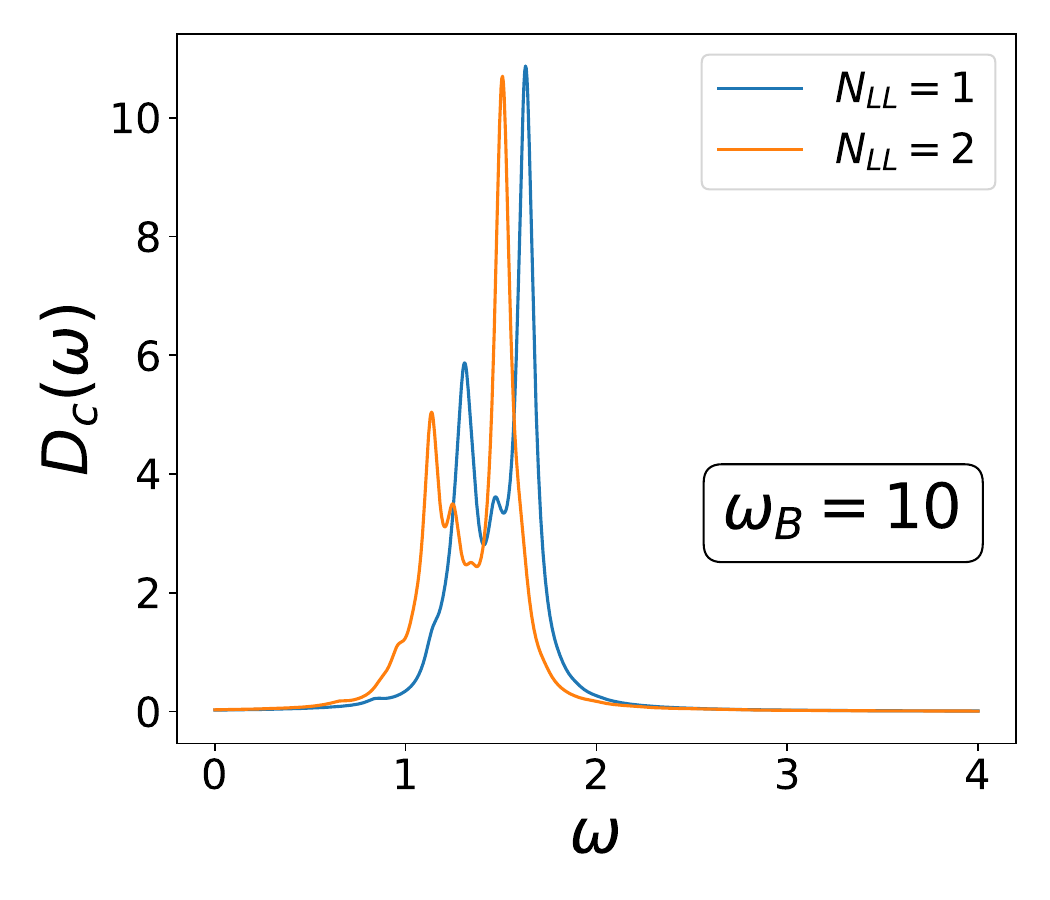}
    \put(20,60){(a)}
    \end{overpic}
    \begin{overpic}[width=0.45\linewidth]{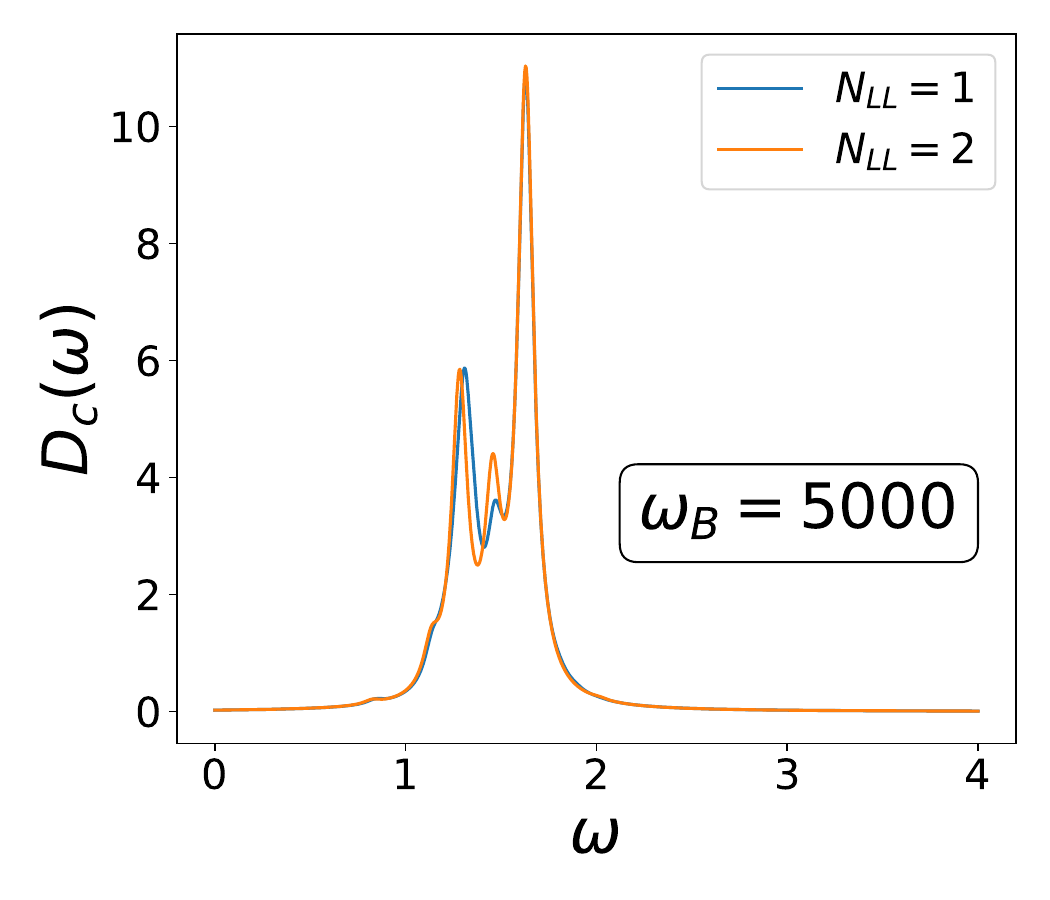}
    \put(20,60){(b)}
    \end{overpic}
    \caption{Cavity density of states $D_c(\omega)$ obtained with ED for different number of Landau Levels $N_{LL}=1,2$ both showing the graviton-polariton splitting with a resonant cavity $\omega_c=1.5$. The cyclotron frequency is $\omega_B=10$ and $\omega_B=5000$ in panel (a) and (b). Other parameters $g=0.1$, $N_e=6$, $L_y=10$ and $\eta=0.05$. }
    \label{fig:Acav_LLconv}
\end{figure}

We now also check that the prediction of graviton-polaritons persist. To this end we show in Fig. \ref{fig:Acav_LLconv} the cavity density of states for two cyclotron frequencies, now $\omega_B=10$ (a) and $\omega_B=5000$ (b). In blue we show the result obtained with the LLL model $\hat{H}_\chi$ while in orange the model at $n_{LL}=2$. Even at realistic values of the cyclotron frequency in terms of interaction energy scales $\omega_B=10$, the polariton doublet is quite clear, featuring a shift in the graviton energy due to higher Landau Level mixing ($\sim 1/\omega_B$). This is confirmed by the fact that the shift of the doublet vanishes when $\omega_B$ is much larger (panel (b) ).

\begin{figure}
    \centering    
    \begin{overpic}[width=0.95\linewidth]{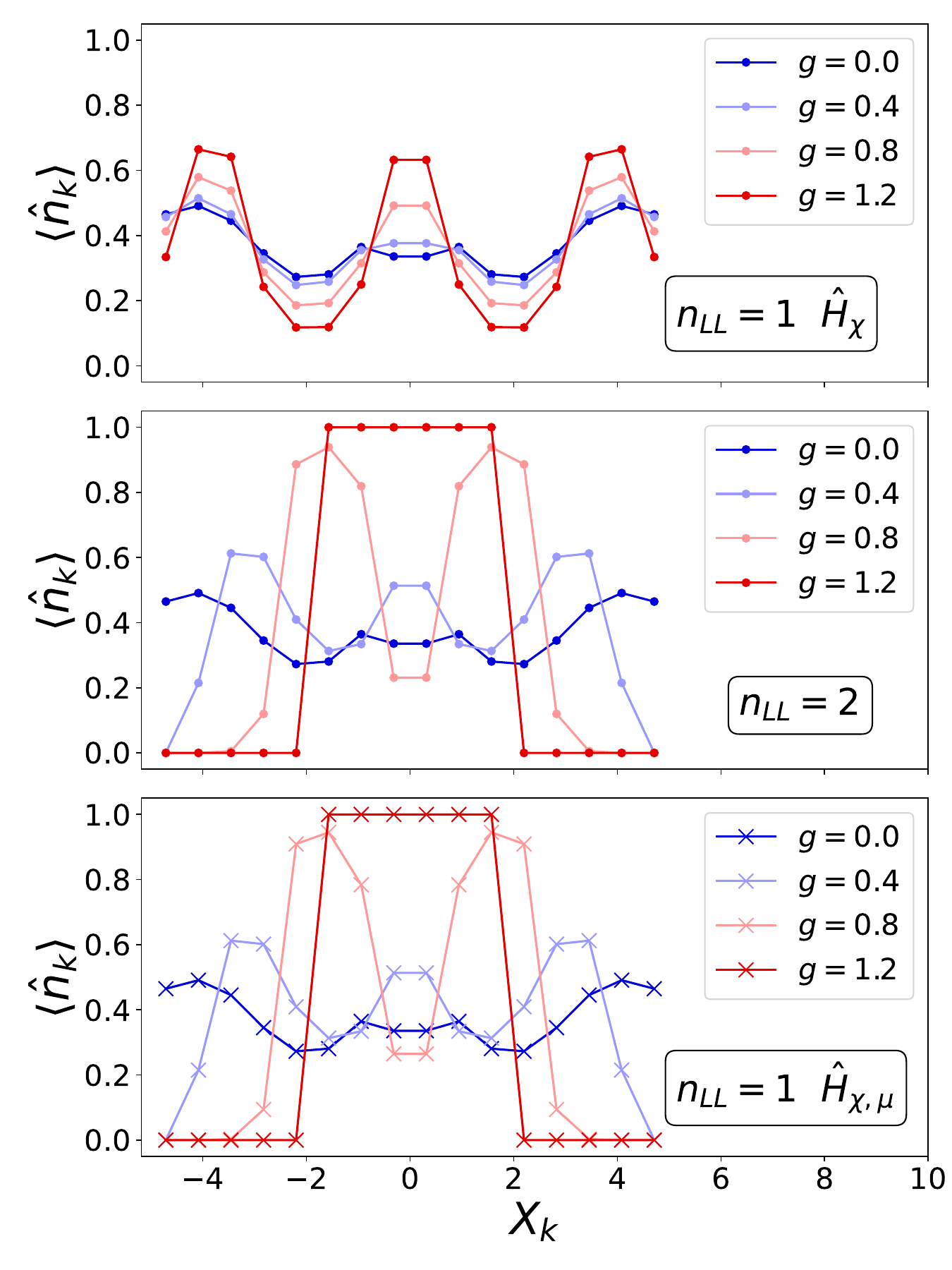}
    \put(13,92){(a)}
    \put(13,60){(b)}
    \put(13,30){(c)}

    \end{overpic}
    \caption{Electronic orbital occupation $\hat{n}_k=\sum_n \hat{c}^\dagger_{k,n}\hat{c}_{k,n}$ for different values of $g$ as a function of orbitals position $X_k=k l_B^2$. Large cyclotron frequency $\omega_B=5000$ but different Landau Level truncation to $N_{LL}=1$ and $N_{LL}=2$ respectively in panel (a) and (b). Other parameters $\omega_c=1.$, $N_e=6$, $L_y=10$ and $\omega_B=5000$. }
    \label{fig:stripe_LLconv}
\end{figure}

As a last check we show how the LLL model $\hat{H}_{\chi,\mu}$ correctly captures also the ultrastrong-coupling behaviour when $\omega_B\to \infty$. This is done in Fig. \ref{fig:stripe_LLconv} where we compare orbital densities for both LLL models and the $n_{LL}=2$ model up to high values of $g\simeq1$. While in the LLL model $\hat{H}_\chi$ we see formation of stripes with $n=2$ electrons, in both cases of $n_{LL}=2$ and $\hat{H}_{\chi,\mu}$ we see how the cavity-induced vacuum Stark shift attracts electrons in the center forming phase separation between a fully filled LLL in the middle and nothing away from it. This is the ultimate fate of the system in the strong coupling limit of what has been explored in Sec. \ref{sec:qp_ren}. Intermediate values of the coupling are difficult to analyze for such small system and might also show intermediate phases. We also remark that actually reaching phase separation will strongly depend on the presence of a neutralizing charge background absent in our case.

\section{Cavity losses}\label{app:cavity_loss}

We consider the effect of losses by introducing a weakly coupled photonic bath with Hamiltonian $\hat{H}_B$ and temperature $T_B$. In the weak system-bath coupling regime it is possible to derive a Lindblad master equation for the evolution of the system density matrix \cite{BreuerPetruccione_2007}, where by system here we mean both cavity and electrons together. Under standard approximations in this derivation (Born, Markov and secular) it is possible to show that the steady state of the system is going to be a thermal state \cite{BreuerPetruccione_2007}:
\begin{equation}\label{eq:thermal_ss}
    \hat{\Theta}= \frac{1}{\Tr\left[ \exp \left(-\frac{1}{T_B} \hat{H}\right)\right]} \exp \left(-\frac{1}{T_B} \hat{H}\right)
\end{equation}
where $\hat{\Theta}$ is the steady state density matrix of the electron-cavity system and $\hat{H}$ its Hamiltonian. The relevant information of equation \ref{eq:thermal_ss} is that for a weakly coupled bath, i.e. weak cavity losses, it is sufficient to look at the cavity plus electron system, independently on the strength of the cavity coupling. Moreover we remark that there is no conceptual difference with the coupling to an external electronic bath, the steady state for weakly coupled thermal environment is indeed a thermal state. In the $T_B\rightarrow 0$ limit we then have that the state of the system is up to exponential corrections the ground state of the electron-cavity Hamiltonian $\hat{H}$. 

The deep consequence then is that the topological properties of the system will not be affected by cavity losses when the temperature of the photon bath is low. For example we can adapt the adiabatic flux insertion procedure of Section \ref{sec:flux_ins} to the case of an environment, assuming adiabaticity can still be reached in the open system scenario. The reason behind this robustness it is that the quantized resistivity is a ground state property and no photon losses can occur from the ground state when $T_B=0$. At finite photon bath temperature $T_B>0$ corrections to the quantization of transport are expected by the fact that we are giving a finite weight to excited states. Non-trivial effects may arise when the bath temperature for the electrons $T_B^{el}$ is different from the bath temperature of the cavity $T_B^{c}$ \cite{JarcFausti_2023,chiriacò2023thermal}, giving rise to a competition whose study is beyond the scope of this work.
\section{Photon mean-field argument}\label{app:PMF}
In this Appendix we provide a more in depth discussion of the photon mean-field theory used in Sec. \ref{sec:stripe_num}.

Due to the linear nature of the coupling term, the mean-field state generally reads:
\begin{equation}
    \ket{\Psi}= \ket{\alpha}\ket{\psi},
\end{equation}
with $\ket{\psi}$ being an electronic wavefunction defined on the LLL, and $\ket{\alpha}$ being a coherent state. Minimizing the mean-field state energy over the coherent state value gives:
\begin{equation}
    \alpha = i\sum_k \chi_k\bra{\psi}  \hat{n}_k \ket{\psi}.
\end{equation}
In order to find the best mean-field ansatz, one should then take the photon mean-field Hamiltonian for the electrons:
\begin{equation}
    \hat{H}_\mathrm{PMF} = \hat{H}_\mathrm{int} +  \omega_c \left[ \sum_k \chi_k \big(\hat{n}_k -  \bra{\psi} \hat{n}_k \ket{\psi}\big) \right]^2,
\end{equation}
and find the ground state $\ket{\psi}$ with a self-consistent procedure \cite{Bacciconi_2023scipost}. 

This minimization still requires the solution of a complicated many-body Hamiltonian. However we can now clearly see two different terms competing, the first is the electron-electron interaction and the second is due to effective cavity mediated interactions. The latter has a very specific form, e.g., when calculated on the ground state $\ket{\psi}$, it is the variance of the polarization operator $\hat{P}$. This has two important consequences. First it is a positive number and second it is, in general, an extensive quantity, being proportional to $N_e$. These are not specific to the FQH set-up, but a more general property of Dipole gauge Hamiltonians where the so-called self polarization term is present \cite{SchulerDebernardis_scipost2020}. For other models where only the linear coupling is present, e.g., the Dicke model, the all-to-all nature of the mode leads to collective enhanced cavity-mediated interactions which contribute superextensively, and grow as $g^2 N_e^2$. For this reason one can usually work with a collective coupling $g\sqrt{N_e}$. Here instead we clearly see that the cavity mediated interactions are controlled by the so-called single particle coupling constant $g$.

We can now discuss two limiting cases: $g=0$ and $g\rightarrow \infty$. In the first case, we recover a pure Laughlin state, eigenstate of the first part of the Hamiltonian. In the second case instead, there will be a massive degeneracy of states with no fluctuations of the polarization operator $\hat{P}$ lifted only by the interactions $\hat{H}_\mathrm{int}$. All product states in the orbital basis will, for example, have no orbital density fluctuations and hence give identically zero contributions to the second term. Among these states we can write down stripe states with index $n$:
\begin{equation}
    \ket{S_n}=\ket{\{0_n 1_n 0_n\}},
\end{equation}
where $0_n$ ($1_n$) indicates the repetition of zero (one) electron in $n$ consecutive orbitals and the curly bracket indicate the repetition of such unit cell. Note that these states satisfy the bulk filling factor $1/3$ by construction. The electron-electron interaction energy contribution of these states is also relatively simple to calculate. The mean-field energy of the stripe state $\ket{S_n}$ is:
\begin{align}\label{eq:stripe-energy-MF}
    E_{S_n}=\bra{S_n}\hat{H}_\mathrm{int}\ket{S_n}= \sum_{k_1, k_2\in K_\mathrm{occ}}2V_{k_1,k_2,k_2,k_1},
\end{align}
where $K_\mathrm{occ}$ is the set of orbitals which are occupied in $\ket{S_n}$, and $V_{k_1,k_2,k_2,k_1}$ are matrix elements given by Eq. (\ref{eq:matrix-elements-Hint}). By changing the circumference of the cylinder $L_y$ the energy per particle will also change, as shown in Fig.~\ref{fig:stripe_numerics}(f) of the main text. For each number of particles per stripe $n$, there is an $L_y$ that minimizes the mean-field energy, which increases with $n$.

This behavior can be well understood by working in the thermodynamic limit $L_y\to\infty$. In this setting, the stripe state can be essentially viewed as an array of filled Fermi seas, each with Fermi momentum $k_\mathrm{F}=\pi n/L_y$, and whose centers are separated in momentum by $\Delta k=6k_\mathrm{F}$. The circumference $L_y$ controls the range of the electron-electron interactions $V_{k_1,k_2,k_2,k_1}$ on the momentum basis via an exponential $\propto e^{-(k_1-k_2)^2/2}$. Thus, if we start from small $L_y$, i.e., large $k_\mathrm{F}$ for some $n$ fixed, the most prominent contribution to $E_{S_n}$ must come from interactions between electrons living on the same stripe. Then, by increasing $L_y$, interactions among electrons on different stripes are expected to become stronger. We thus expand the energy $E_{S_n}$ as the sum
\begin{equation}
    E_{S_n}=E_{S_n}^{(0)}+E_{S_n}^{(1)}+\cdots,
\end{equation}
where $E_{S_n}^{(0)}$ and $E_{S_n}^{(1)}$ respectively represent the energy contributions of electrons on the same and nearest neighbor stripes. Taking the continuum limit $\sum_k\to\frac{L_y}{2\pi}\int\dd k$,  we estimate these two energy contributions (details on Appendix \ref{app:stripe-energy}) to be 
\begin{equation}
    E_{S_n}^{(0)}=2N_e\Big[\erf(\sqrt{2}k_\mathrm{F})+\frac{2e^{-2k_\mathrm{F}^2}-2}{\sqrt{2\pi}k_\mathrm{F}} \Big],
\end{equation}
and
\begin{align}
    E_{S_n}^{(1)}&=4N_e\Big[\erf(\sqrt{8}k_\mathrm{F})-3\erf(\sqrt{18}k_\mathrm{F})+2\erf(\sqrt{32}k_\mathrm{F})\nonumber\\
    &+\frac{e^{-8k_\mathrm{F}^2}-2e^{-18k_\mathrm{F}^2}+e^{-32k_\mathrm{F}^2}}{\sqrt{2\pi}k_\mathrm{F}}\Big],
\end{align}
where $\erf(x)$ is the Gauss error function. In Fig.~\ref{fig:stripe_numerics}(f) we plot the approximate form of $E_{S_n}$ as a function of $1/k_\mathrm{F}$. Within this approximation, we find the minimum lies at $k_\mathrm{F,{min}}\simeq0.62$, with a corresponding energy gap $E_{S_n}/N_e\simeq0.25$. This means the number of electrons per stripe $n_S\approx L_y/5$ is only dictated by $L_y$ in the thermodynamic limit as discussed above, see Eq. (\ref{eq:lambda-S}).

On the other hand we have that the Laughlin state $\ket{\Psi_L}$ is an eigenstate of $\hat{H}_\mathrm{int}$ with zero energy. This means the variational energy of the Laughlin state depends only on the light-induced interaction:
\begin{equation}
    E_{L} = \omega_c \sum_{kk'} \chi_k\chi_{k'}\bra{\Psi_L} \delta\hat{n}_k \delta\hat{n}_{k'} \ket{\Psi_L},
\end{equation}
where we have defined $\delta \hat{n}_k=\hat{n}_k-\bra{\Psi_L}\hat{n}_k\ket{\Psi_L} $. Given that $\chi_k\propto k^2$, it is possible to compute this energy contribution via the long-wavelength behaviour of the static structure factor $S(\boldsymbol{q})$. From Eq. (\ref{eq:S4-xx}), we observe that $E_L$ is directly tied to the $S_4^{xx}$ component, so when we send $N_e\to\infty$, the mean-field energy scales as 
\begin{equation}\label{eq:E-L}
    E_L/N_e\approx S_4^{xx}\omega_cg^2/\nu,
\end{equation}
where $\nu=1/3$. 

The outcome is that a transition to the stripe phase must take place once the energy of the Laughlin state $E_L$ becomes equal or larger than the energy $E_{S_n}$. Replacing the value for the unperturbed Laughlin state, $S_4^{xx}=(1-\nu)/24\nu$, we find the magnitude of the critical point $g=g_\star$ goes as $\omega_c g_\star^2 \approx 1$. This is far below the observed values in the numerics, which happen close to $g\simeq3$. One key reason behind this difference is the renormalization of $S_4^{xx}$ as a function of the light-matter interaction as shown in Fig. \ref{fig:S4x}. Given $S_4^{xx}$ becomes smaller, the energy of the real ground state scales slower than $g^2$, and the transition is pushed towards a higher value of $g$. In particular, by using the numerical data for $S_4^{xx}$ as a function of $g$, we have checked the true ground state energy agrees well with the formula in Eq. (\ref{eq:E-L}).
The strong resilience of the FQH liquid to cavity fluctuations is then to be attributed to the change in its geometry, interpreted as a hidden variational parameter \cite{Haldane_prl2011_quantummetric}. 

%%%%%%%%%%%%%%%%%%%%%%%%%%
\section{Stripe energy estimation}\label{app:stripe-energy}

In this appendix we explicit our derivation for the approximate form of the mean-field energy $E_{S_n}$ associated to the stripe state $\ket{S_n}=\ket{\{0_n1_n0_n\}}$. From the photon mean-field argument, the energy $E_{S_n}$ is given by the expectation value $E_{S_n}=\bra{S_n}\hat{H}_\mathrm{int}\ket{S_n}$ as shown in Eq. (\ref{eq:stripe-energy-MF}). Writing out explicitly the matrix elements of $\hat{H}_\mathrm{int}$, we have
\begin{equation}
E_{S_n}=\frac{\sqrt{8\pi}}{L_y}\sum_{k_1,k_2\in K_\mathrm{occ}}(k_1-k_2)^2 e^{-(k_1-k_2)^2/2},
\end{equation}
where we recall $K_\mathrm{occ}$ denotes the set of orbitals that are occupied in the stripe state. Note the sums over $k_1$ and $k_2$ run independently. Let us then proceed to the thermodynamic limit by sending $L_y,n\to\infty$, while keeping their ratio $n/L_y$ fixed. From the ratio $n/L_y$ we define the Fermi momentum $k_\mathrm{F}=\pi n/L_y$ of the stripes. With this, we may replace the original restricted sum over $K_\mathrm{occ}$ by $\sum_{k\in K_\mathrm{occ}}\to\sum_\sigma\frac{L_y}{2\pi}\int_{-k_\mathrm{F}}^{k_\mathrm{F}} \dd q$, where we sum over stripes labeled by the integer index $\sigma$, and integrate over occupied states within each stripe $|q|<k_\mathrm{F}$. The electronic filling $\nu=1/m$, with $m=3$, fixes the relative distance among neighboring stripes, so we further replace the original momenta $k$ in the summand by $k\to q+2mk_\mathrm{F}\sigma$ within the integral, where $q$ is the momenta measured from the stripe center. We are then led to
\begin{align}\label{eq:energy-Sn-integral}
E_{S_n}&=\frac{L_y/\pi}{\sqrt{2\pi}}\sum_{\sigma_1\sigma_2}\int_{-k_\mathrm{F}}^{k_\mathrm{F}}\dd q_1\dd q_2(q_{12}+6k_\mathrm{F}\sigma_{12})^2\nonumber\\
&\times e^{-\frac12(q_{12}+6k_\mathrm{F}\sigma_{12})^2},
\end{align}
where the integrand only depends on the relative variables $q_{12}=q_1-q_2$ and $\sigma_{12}=\sigma_1-\sigma_2$. 

To extract a series, which only depends on the relative stripe coordinate $|\sigma_{12}|$, we assume translation invariance along $x$, so that we may simplify the formula in Eq. (\ref{eq:energy-Sn-integral}) to
\begin{align}
E_{S_n}&=\frac{N_e}{\sqrt{2\pi}k_\mathrm{F}}\sum_{\sigma_{12}}\int_{-k_\mathrm{F}}^{k_\mathrm{F}}\dd q_1\dd q_2(q_{12}+6k_\mathrm{F}\sigma_{12})^2\nonumber\\
&\times e^{-\frac12(q_{12}+6k_\mathrm{F}\sigma_{12})^2},
\end{align}
where we use that the total number of stripes must be fixed so that $N_\mathrm{stripes}=N_e/n$. In this form we see the energy can be naturally expanded as $E_{S_n}=\sum_{r}E_{S_n}^{(r)}$ for $r=|\sigma_{12}|$, where thanks to the exponential factor, these contributions decrease quite quickly as a function of $r$.

Let us then evaluate the first two contributions, corresponding to $\sigma_{12}=0$ and $\sigma_{12}=\pm1$. When $\sigma_{12}=0$, both $q_1$ and $q_2$ belong to the same stripe. The integral is elementary and gives the energy contribution
\begin{equation}
E_{S_n}^{(0)}=2N_e\big[\erf(\sqrt{2}k_\mathrm{F})+\frac{2e^{-2k_\mathrm{F}^2}-2}{\sqrt{2\pi}k_\mathrm{F}}\big],
\end{equation}
where $\erf(x)$ is the error function, defined as $\erf(x)=\frac{2}{\sqrt{\pi}}\int_0^x\dd t\,e^{-t^2}$. Likewise, we evaluate the contribution for $\sigma_{12}=\pm1$, which comes from the interaction of neighboring stripes. The integral yields the same result for both $\sigma_{12}=+1$ and $\sigma_{12}=-1$, of course, and the energy contribution $E_{S_n}^{(1)}$ is found to be
\begin{align}
E_{S_n}^{(1)}&=4N_e\Big[\erf(\sqrt{8}k_\mathrm{F})-3\erf(\sqrt{18}k_\mathrm{F})+2\erf(\sqrt{32}k_\mathrm{F})\nonumber\\
&+\frac{e^{-8k_\mathrm{F}^2}-2e^{-8k_\mathrm{F}^2}+e^{-32k_\mathrm{F}^2}}{\sqrt{2\pi}k_\mathrm{F}}\Big].
\end{align}
These two functions are plotted in the inset of Fig. \ref{fig:stripe_numerics}(f). When added together they provide an estimate for the minimum of $E_{S_n}$, which sets the minimum energy gap to the stripe state as $E_{S_n}/N_e\simeq0.25$ and fixes the Fermi momentum of that stripe state to $1/k_\mathrm{F,min}\simeq1.62$.

%%%%%%%%%%%%%%%%%%%%%%%
\section{Stripes as an array of Tomonaga-Luttinger liquids}\label{app:stripe_ft}

In this appendix we discuss the potential low-energy theory for the light-induced stripe phase found numerically. We start from the photon mean-field ansatz $\ket{S_n}$, incorporating fluctuations phenomenologically. First, we consider the (infinitely) large light-matter coupling limit, where we assume the energy penalty to introduce $\delta n_k$ fluctuations is so large we can only spare to build excitation with definite density occupation $n_k=0,1$. This is a great simplification that essentially allow us to treat the problem as an array of 1D free electron systems. This state is naturally gapped along $x$, but remain gapless along the $y$ direction, and can be viewed as an archetype of a smectic metal state \cite{Emery2000smectic,Vishwanath&David2001sll,Mukhopadhyay2001sll}, a state with zero shear modulus \cite{Lawler&Fradkin2004smectics}. The gapless modes take the form of particle-hole excitations near the Fermi points of each stripe $k_\sigma=\pm k_\mathrm{F}+6k_\mathrm{F}\sigma$, where $\sigma=0,\pm1,\pm2,\dots$ denotes the stripe index. To verify this, we first consider a particle excitation with momentum $k=k_\mathrm{F}+q$, assuming $q\ll k_\mathrm{F}$ so the excitation lies close to the right Fermi point of the stripe $\sigma=0$. In this limit, similar to the analysis carried out in Appendix \ref{app:stripe-energy}, we approximate the energy cost of adding one electron by the integral:
\begin{align}
    \varepsilon(k_\mathrm{F}+q)&\simeq\sqrt{\frac2\pi}\int_{-k_\mathrm{F}}^{k_\mathrm{F}}\dd k(k_\mathrm{F}-k+q)^2e^{-\tfrac12(k_\mathrm{F}-k+q)^2},
\end{align}
where we integrate over the nearest stripe only, throwing out energy contributions from distant stripes that are suppressed by the exponential factor. Evaluating the integral in the limit $q\to0$, we find the linear dispersion relation $\varepsilon(k_\mathrm{F}+q)\approx\varepsilon(k_\mathrm{F})+v_\mathrm{F}q+\cdots$, where $v_\mathrm{F}=\sqrt{\frac2\pi}4k_\mathrm{F}^2e^{-2k_\mathrm{F}^2}$ plays the role of the Fermi velocity. It thus follows that the particle-hole excitation $c_{k+q}^\dagger c_{k}\ket{S_n}$ in the vicinity of the Fermi level has a dispersion relation of the form $\omega(q)\approx v_\mathrm{F}|q|+\cdots$. 
Taking into account the pertubative addition of interactions, the free fermion
behavior is then compatible with a sliding Tomonaga-Luttinger liquid (sTLL) model description of the stripe phase. 

In this setting, the reintroduction of the light-matter coupling with a linear gradient profile couples to the number of excitations in each stripe separately, penalizing certain processes that change this number. In this way the gradient of the electric field fights against correlated hopping operators, such as the one used in coupled wire constructions to drive a transition to the FQH state \cite{Kane2002cwc}. If the gradient is strong enough even processes leading to stripe crystallization \cite{Mukhopadhyay2001sll} (i.e., phase locking with $2k_\mathrm{F}$ density modulations along the stripes) are disfavored. Note that these couplings are controlled by the field variations across the intra- and inter-stripe distances $\Delta x\sim k_\mathrm{F}l_B^2$. The detailed interplay among these competing operators lies beyond our current mean-field ansatz, and we leave these refinements for future work.

\begin{figure*}
$\vcenter{\hbox{\begin{overpic}[width=0.3\linewidth]{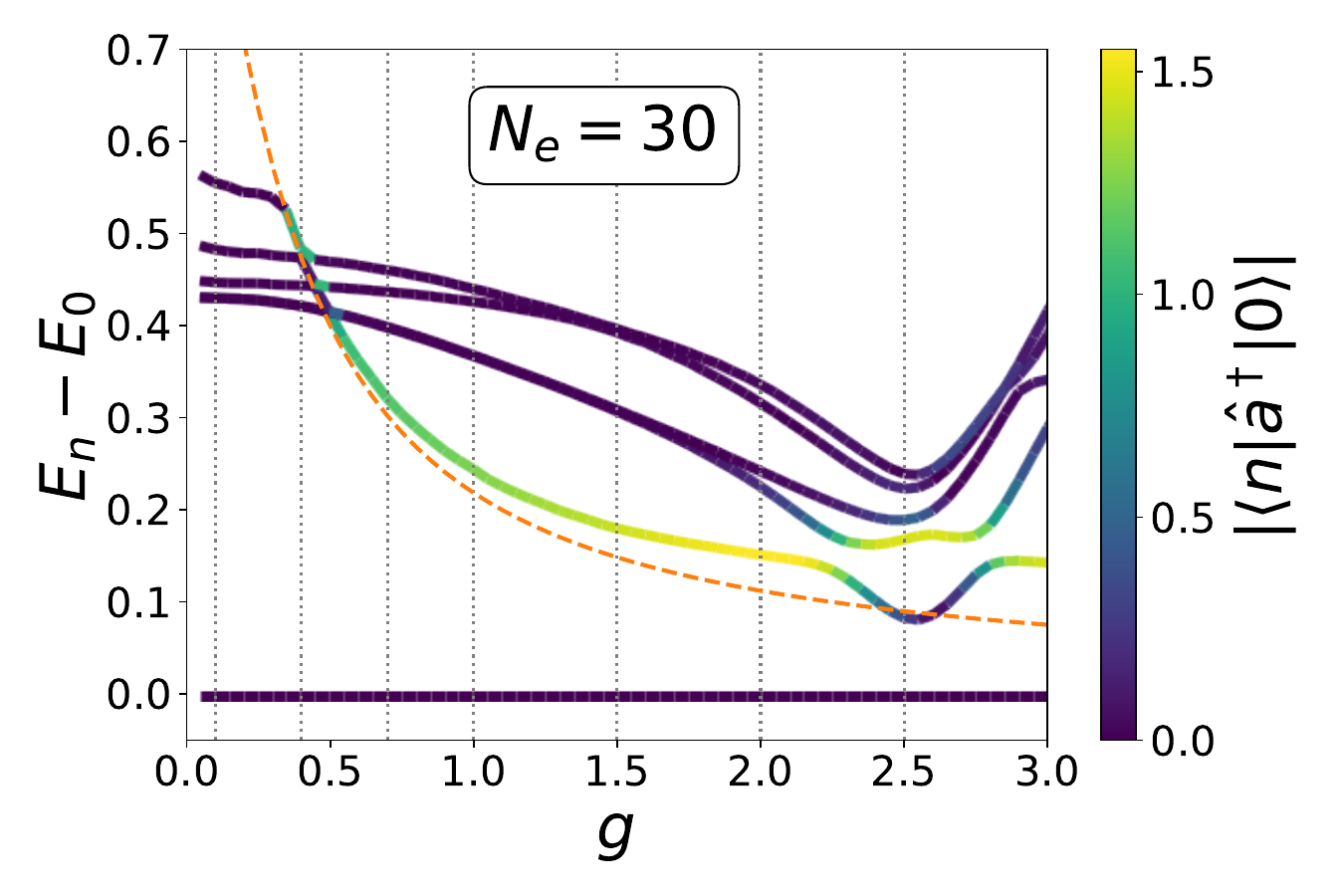}\put(15,70){(a)}
     \end{overpic}}}$
     $\vcenter{\hbox{
     \begin{overpic}[width=0.65\linewidth]{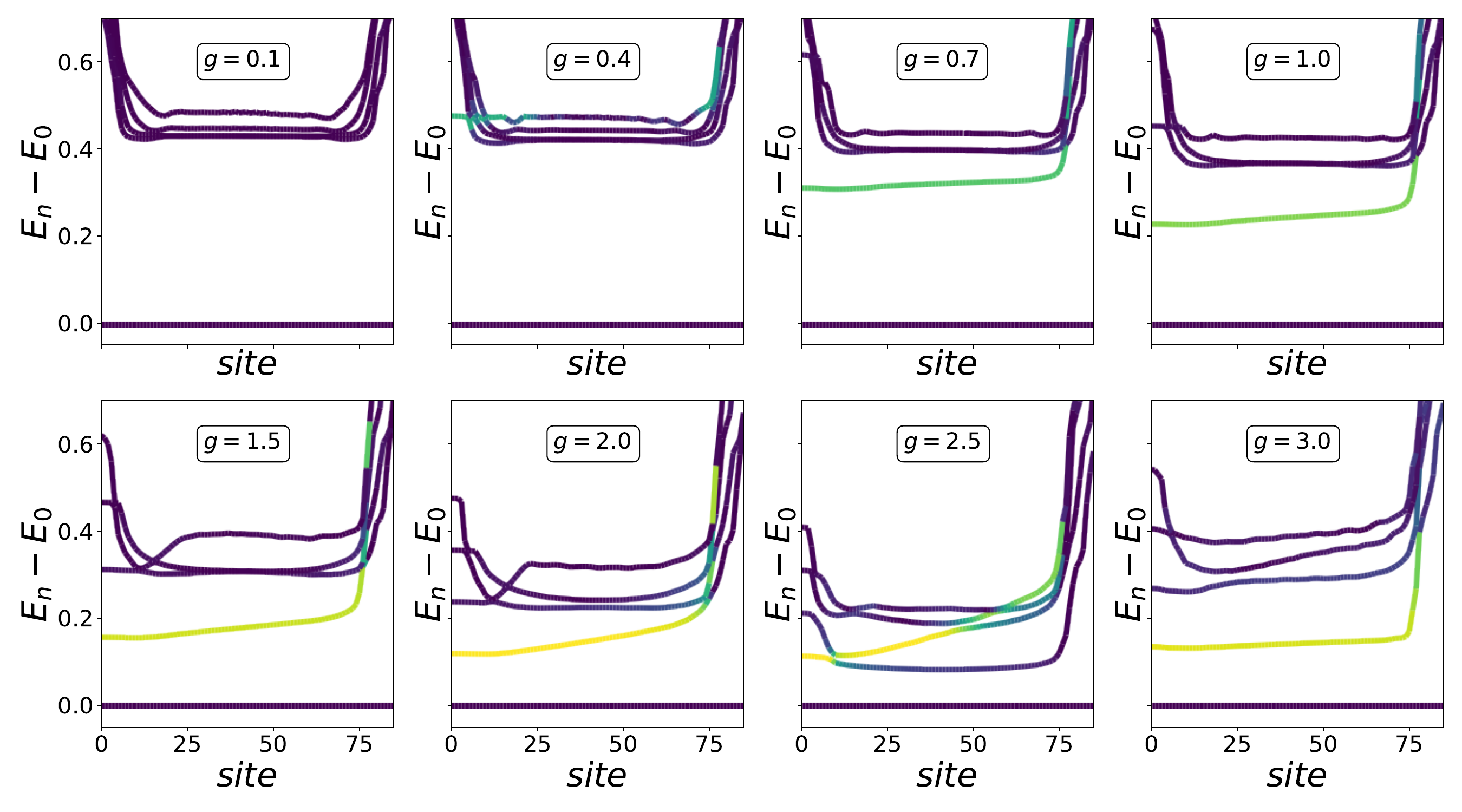}\put(10,57){(b)}
     \end{overpic}}}$
    \caption{Excited state targettig with DMRG local effective Hamiltonians. Panel (a) shows the middle-chain value of the excited state as a function of coupling $g$ for a system size $N_e=30 $ and $L_y=16$. Orange dashed line indicate the lower polariton of the effective model. Panel (b) shows the site dependency of the excited states for different values of $g$, marked by vertical gray dotted lines in panel (a). The colorbar indicate the one photon matrix element with the ground state.}
\label{fig:excited_conv}
\end{figure*}
\section{Excited state targeting}\label{app:excited_appendix}
In this section we present additional numerical result on the DMRG excited state targetting used in the main text (Fig. \ref{fig:magnetoroton}). As discussed in Ref. \cite{Mila_prb2017_excited}, excited states of local effective Hamiltonians constructed during DMRG sweeps are good approximations to excited states of the full Hamiltonian. By construction they are variational estimates, orthogonal to lower energy states at a fixed position during the sweep. The persistence of a certain excited state during the sweep was found to be a good indication of a converged excited state \cite{Mila_prb2017_excited}. In Fig. \ref{fig:excited_conv} we show the excited state spectrum obtained during a sweep as a function of the updating sites for different values of the coupling $g$. A three site update is used here. In the FQH phase $g\lesssim 2.5$ we observe a good convergence of the lowest magnetoroton state (blu line) while the polariton state (lighter line) is flat only away from the transition. The asymmetric nature of our MPS ansatz (Sec. \ref{sec:num_meth}) with the photon on the left side clearly shows up here. Indeed the best variational estimate for the polariton state is not found in the bulk but closer to the cavity site. On the other hand magnetoroton excitations shows a phenomenology similar to Ref. \cite{Mila_prb2017_excited} and have a lower variational energy in the bulk.

\section{Large cavity frequency limit}
\label{app:large-omegac-limit}

In this appendix we briefly discuss the large cavity frequency limit $\omega_c\gg V_0$. This regime is subtle within the truncated model, as higher Landau level corrections become significant when $\omega_c$ approaches the cyclotron gap $\omega_B$. Setting these refinements aside, the large $\omega_c$ regime can be understood intuitively: as $\omega_c \to \infty$, cavity photons become increasingly energetic, effectively decoupling from the low-energy electronic degrees of freedom. This behavior is consistent across dipole-gauge models and corresponds to the static limit in the path integral formulation presented in Ref. \cite{Dmytruck_prb2021}. Note however that in order to take the large cavity frequency limit one need to assume the linear interaction term $ig\omega_c (\hat{a}-\hat{a}^\dagger)\hat{P}$ to be smaller than the cost of a photon, otherwise no elimination of the cavity mode is possible. The physically motivated choice of keeping $g\sqrt{\omega_c}$ fixed, or equivalently the Rabi frequency $\Omega=2 \tilde{\gamma}_0 g\sqrt{\omega_c V_0}$ fixed, acually does automatically the job in the $\omega_c\to \infty$ limit. Perturbation theory provides a straightforward way to verify this decoupling. To outline the derivation, we divide the Hamiltonian in diagonal and off-diagonal terms:
\begin{align}
&\hat{H}_{\chi}= \hat{H}'+ \hat{H}''\\
&\hat{H}'=\omega_c\hat{a}^\dagger\hat{a}+\hat{H}_\mathrm{int}+g^2\omega_c\hat{P}^2\\
&\hat{H}''=ig\omega_c(\hat{a}-\hat{a}^\dagger)\hat{P}\;;
\end{align} 
Using a Schrieffer-Wolff transformation with a generator:
\begin{align}
    \hat{S}=-ig(\hat{a}+\hat{a}^\dagger)\hat{P}\;
\end{align}
one can successfully decouple the photon dynamics and get an effective Hamiltonian up to second order in perturbation theory:
\begin{align}
    \hat{H}_{\mathrm{eff}} &= e^{\hat{S}}\hat{H}_{\chi}e^{-\hat{S}}\simeq \hat{H}' - g^2 \omega_c \hat{P}^2 +O\left(g^3 \omega_c\right) =\\
    &=\hat{H}_\mathrm{int} +\omega_c \hat{a}^\dagger\hat{a} +O\left(g^3 \omega_c\right)\;.
\end{align}
We thus find the $\hat{P}^2$ terms cancel out, leaving us with just the decoupled theory $\hat{H}_{\mathrm{eff}}\simeq \hat{H}_0=\hat{H}_\mathrm{int}+\omega_c\hat{a}^\dagger\hat{a}$ up to second-order in $g$. Note that the corrections identified in \cite{enknerfaist_2024_fqhenhanced}, with the caveat of being obtained in a different LLL truncation scheme, scale with the gradient to the fourth power.

%%%%%%%%%%%%%%%%%%%%%%%%
\bibliography{main_v3.bbl}
\end{document}